\documentclass[twoside,12pt]{article}
\usepackage{epsfig}
\usepackage{rotating}
\usepackage{color}
\usepackage[utf8]{inputenc}
\usepackage[english]{babel}

\def\iso#1{$^{#1}$}
\def\msun{M$_\odot$}
\newcommand{\be}{\begin{equation}}
\newcommand{\ee}{\end{equation}}
\newcommand{\bea}{\begin{eqnarray}}
\newcommand{\eea}{\end{eqnarray}}

\usepackage[backend=biber,style=chem-angew,sorting=none]{biblatex}

\begin{document}

\title{ \vspace{1cm} Radioactive nuclei from cosmochronology to habitability}
\author{M.\ Lugaro,$^{1,2}$ U.\ Ott,$^{3,4}$ \'A.\ Kereszturi$^1$\\
\\
$^1$Konkoly Observatory, Research Centre for Astronomy and Earth Sciences,\\
Hungarian Academy of Sciences, H-1121 Budapest, Hungary\\
$^2$Monash Centre for Astrophysics, Monash University, VIC3800, Australia\\
$^3$Atomki Institute for Nuclear Research,\\
Hungarian Academy of Sciences, H-4026, Debrecen, Hungary\\
$^4$Max-Planck Institute for Chemistry, D-55128 Mainz, Germany\\
}
\maketitle

©2018. This manuscript version is made available under the CC-BY-NC-ND 4.0 license\\ 
http://creativecommons.org/licenses/by-nc-nd/4.0/ \\

{\em This paper is dedicated to the memory of Gerald~J.~ Wasserburg, who pioneered, built up, and inspired the science presented here.} 

\begin{abstract} 


In addition to long-lived radioactive nuclei like U and Th isotopes, which have been used to measure the age of the Galaxy, also radioactive nuclei with half-lives between 0.1 and 100 million years (short-lived radionuclides, SLRs) were present in the early Solar System (ESS), as indicated by high-precision meteoritic analysis. We review the most recent meteoritic data and describe the nuclear reaction processes responsible for the creation of SLRs in different types of stars and supernovae. We show how the evolution of radionuclide abundances in the Milky Way Galaxy can be calculated based on their stellar production. By comparing predictions for the evolution of galactic abundances to the meteoritic data we can build up a time line for the nucleosynthetic events that predated the birth of the Sun, and investigate the lifetime of the stellar nursery where the Sun was born. We then review the scenarios for the circumstances and the environment of the birth of the Sun within such a stellar nursery that have been invoked to explain the abundances in the ESS of the SLRs with the shortest lives -- of the order of million years or less. Finally, we describe how the heat generated by radioactive decay and in particular by the abundant \iso{26}Al in the ESS had important consequences for the thermo-mechanical and chemical evolution of planetesimals, and discuss possible implications on the habitability of terrestrial-like planets. We conclude with a set of open questions and future directions related to our understanding of the nucleosynthetic processes responsible for the production of SLRs in stars, their evolution in the Galaxy, the birth of the Sun, and the connection with the habitability of extra-solar planets. 

\end{abstract}
\tableofcontents
\section{Introduction}
\label{sec:intro}

More than a century has passed since Marie Skłodowska Curie\footnote{The 150$^{\rm th}$ anniversary of her birthday was recently celebrated on the 7$^{\rm th}$ of November 2017.} coined the term Radioactivity to indicate the emission of radiation and particles from peculiar nuclei. Since then, the role and applications of radioactivity have had a profound impact in many fields of science and technology. The role of radioactive nuclei in the field of astrophysics has been long recognised and described. For example, radioactive nuclei power the light of supernovae and the radiation they emit can be mapped throughout the Galaxy by satellite observatories \cite{radiobook}. Here we focus on the most recent advances in the research directions that relate the process of short-lived (half-lives\footnote{See Table~\ref{table:acronyms} for a list of all the symbols and the acronyms used throughout the paper.} T$_{1/2}$ $\sim$ 0.1 to 100 million years, Myr) radioactivity to the concept of {\it cosmochronology}, and on the relatively more recent link between short-lived radioactivity and {\it habitability}. We consider in particular the applications of radioactivity in the field of {\it cosmochemistry}, i.e., the study of the composition of meteorites and other solid Solar System samples aimed at explaining the origin of chemical matter in the Solar System and in the Universe. Due to extensive technological advances in the laboratory analysis of the isotopic composition of terrestrial and extraterrestrial materials, the amount of information and constraints that can be derived from such studies are expanding at a very fast rate. Much effort on the theoretical interpretation is needed to keep up with the experimental data. In this landscape, the connections between radioactivity, cosmochronology, and habitability are becoming more relevant than ever, and the implications of these connections are quickly becoming far reaching. The aim of this paper is to illustrate and discuss these connections and their implications.

{\it Cosmochronology} is intrinsically linked to radioactivity, being defined as the use of the abundances of radioactive nuclei to compute either the age of the elements themselves, or the age of astronomical objects and events. The first aim typically relies on very long-lived radionuclides with half-lives T$_{1/2}$ of the order of billions of years (Gyr), such as \iso{238}U, \iso{232}Th, \iso{187}Os, \iso{87}Rb; an introduction to this topic can be found, for example, in Chapter 1 of \cite{radiobook}. Here we address the second aim: to use radioactive nuclei to calculate the age of astronomical objects and events, specifically in relation to the birth of our Sun and Solar System, with the ultimate aim to compare the birth of our Sun to the birth of other stars and their extra-solar planetary systems. To such aim we use short-lived radionuclides (SLRs, T$_{1/2}$ $\sim$ 0.1 to 100 Myr), which provide us with a range of chronometers of the required sensitivity.  

It is well known that radioactive decay can be used as an accurate clock because the rate at which the abundance by number of a radioactive nucleus $N_{\rm SLR}$ decreases in time due to its radioactive decay is a simple linear function of the abundance itself, where $\lambda$ is the time-independent constant of proportionality referred to as the {\it decay rate}:

\begin{equation}
\label{eq:basicdiff}
\frac{dN_{\rm SLR}}{dt} = - \lambda \,N_{\rm SLR}. 
\end{equation}

\noindent A quick integration between two set times $t_1$ and $t_0$ delivers:  

\begin{equation}
\label{eq:basic1}
N_{\rm SLR}(t_1) = N_{\rm SLR}(t_0) e^{- \lambda (t_1 - t_0)},
\end{equation}

\noindent which can also be written as 

\begin{equation}
\label{eq:basic2}
t_1 - t_0 = \tau [{\rm ln}(N_{\rm SLR}(t_0)) - {\rm ln}(N_{\rm SLR}(t_1))],
\end{equation}

\noindent where $\tau = 1/\lambda ={T_{1/2}/ ln(2)}$ is the mean-life, i.e., the time interval required to decrease $N_{\rm SLR}$ by a factor $1/e$ (instead of a factor $1/2$, as for the half-life).  

\begin{figure}[tb]
\begin{center}
\begin{minipage}[t]{8.5 cm}
\includegraphics[width=8.5cm,angle=0]{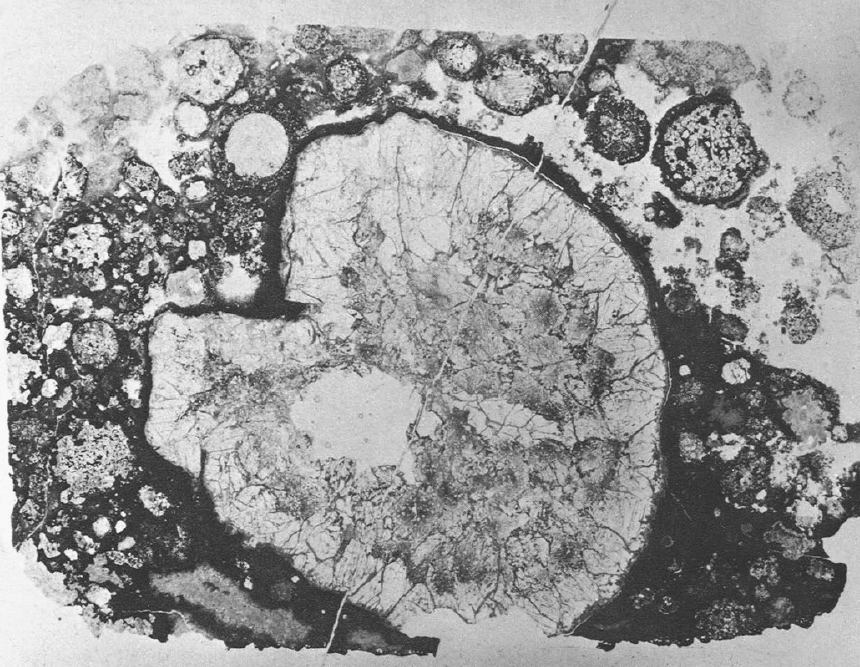} \\
\end{minipage}
\begin{minipage}[t]{16.5 cm}
\caption{Photomicrograph produced in 1977 \cite{clayton77} of the CAI named Al3S4 from the Allende meteorite. The field of view is 22 mm $\times$ 17 mm. In 2014, the initial \iso{146}Sm/\iso{144}Sm ratio in the ESS was derived from  analysis of this CAI \cite{marks14} (bottom left panel of Fig.~\ref{fig:data}). \label{fig:CAI}}
\end{minipage}
\end{center}
\end{figure}

Radioactive clocks have been used extensively to measure a large variety of time intervals. The decay of \iso{14}C, a nuclide with a half-life of 5730 yr, allows us to measure timescales related to human history; and the age of our Milky Way Galaxy of approximately 13 Gyr has been estimated also based on the ages of some of the oldest observed stars inferred from their U and Th abundances \cite{cayrel01,frebel07}. Thanks to the SLRs considered here, it has become possible to investigate in detail the early history of the Solar System and build a chronology of planetary growth from micrometer-sized dust to terrestrial planets \cite{dauphas11}. For example, the solidification of the lunar magma ocean has been dated to about 200 Myr after the birth of the Sun also thanks to the $\alpha$ radioactive decay of \iso{146,147}Sm into \iso{142,143}Nd, respectively \cite{borg11}. The age of the oldest solids in the Solar System, the calcium-aluminium-rich inclusions (CAIs) found in primitive meteorites (Fig.~\ref{fig:CAI}), is 4567-4568 Myr (see Table 3 of \cite{tissot17}) as measured from the radioactive decay chain starting at the U isotopes and ending into the Pb isotopes. CAIs are believed to be among the first solids to have formed in the protosolar nebula, thus, their age is taken also as indicative for the age of the Sun.  

Unlike cosmochronology, {\it habitability} has been linked to short-lived radioactivity only recently. Here we use the concept of habitability in the following sense: whether or not an astronomical object can support the formation or the maintenance of life forms partly similar to those we have on Earth \cite{Gargaud2011}. Formation and maintenance, however, are two different processes, both related to habitability. It should be  kept in mind that life forms elsewhere in the Universe could be fundamentally different from those we know from Earth. However, the definition of life as a system based on chemicals, built on organic material, and supported by liquid water as a solvent is generally accepted by the astrobiological community and thus is also used here. 

The paper is structured as follows. Section~\ref{sec:background} introduces some basic methodology and considerations and is separated into four sections: Sec.~\ref{sec:abundances} presents the methods by which the initial SLR abundances in the early Solar System are inferred from meteoritic analysis. Section~\ref{sec:stars} presents a broad overview of stellar evolution and nucleosynthetic processes in stars.  Section~\ref{sec:galaxy} describes the processes that have built up the Solar System chemical matter, from galactic chemical evolution to the formation of the Sun itself. Section~\ref{sec:habit} presents how, in general, radioactivity may influence habitability in several direct and indirect ways. Section~\ref{sec:list} discusses in more detail each SLR, from its meteoritic abundance to the nuclear path of its stellar production. The 19 SLRs considered here are grouped into 9 subsections, according to their nucleosynthetic production processes. In Sec.~\ref{sec:GCE} we deal with Galactic evolution: Sec.~\ref{sec:GCEmodels} presents the simple analytical models used so far to describe the evolution of SLRs in the Galaxy, and Sec.~\ref{sec:times} shows how the SLR galactic abundances can be used to establish the timing of specific events related to the birth of our Sun. In Sec.~\ref{sec:birth} we discuss inferences derived from the presence of SLRs in the ESS concerning the circumstances of the Sun's birth. For sake of clarity, we distinguish three different questions related to the general problem: the stellar sources, the injection mechanism, and the plausibility and probability of the possible scenarios (covered in Sec.~\ref{sec:Q1}, \ref{sec:Q2}, and \ref{sec:Q3}, respectively). In Sec.~\ref{sec:radioheat} we describe the potential sources of radioactive heat in the ESS and the implications on planet formation and habitability: first, we analyse all the possible radioactive heat sources (Sec.~\ref{sec:heat}), then we consider carrier minerals (Sec.~\ref{sec:carriers}), and finally the specific, important case of \iso{26}Al (Sec.~\ref{sec:26Alhabit}). Section~\ref{sec:conclusions} summarises the main points of the paper and presents a final set of open questions and future research directions.

The topic of the present paper covers a range of research fields, from nuclear physics, via astronomy and astrophysics, to planetary sciences, from both the experimental and the theoretical perspective. We focus here on the interdisciplinary connections between these topics. As such the paper has been written keeping in mind different audiences and with the broad aim to foster and enhance the efficiency of the knowledge transfer required to answer the currently open questions.

\section{Background information}
\label{sec:background}

\begin{table}
\begin{center}
 \centering
 \caption{List of acronyms and symbols used throughout the paper.}
 \label{table:acronyms}
\vspace{0.3cm}
 \begin{tabular}{ll}
\hline
\multicolumn{2}{l}{General} \\
Myr & Millions of years \\
SLR & Short-lived radionuclide \\
$N_{\rm SLR}$ & Abundance by number of a SLR \\
$N_{\rm stable}$ & Abundance by number of a stable reference isotope \\ 
$\lambda$ & Decay rate \\
$\tau = 1/\lambda$  & Mean-life  \\
T$_{1/2}=\tau\,{\rm ln} 2$ & Half-life  \\
ESS & Early Solar System \\
CAI & Calcium-aluminium rich inclusion \\
$\delta/\epsilon(N_1/N_2)$ & Per mil/per ten thousands variation of the abundance ratio $N_1/N_2$ \\
\hline
\multicolumn{2}{l}{Stars and supernovae} \\
\msun\ & Solar mass \\
$Z$ & Stellar metallicity \\
AGB star & Asymptotic giant branch star \\
CCSN & Core-collapse supernova \\
SNIa & Type Ia supernova \\
WD & White dwarf \\ 
NSM & Neutron star merger \\
WR star & Wolf-Rayet star \\
(G)MC & (Giant) molecular cloud \\
CRs & Cosmic rays \\
\hline
\multicolumn{2}{l}{Galaxy} \\
GCE & Galactic chemical evolution \\
ISM & Interstellar medium \\
$k$ & Infall parameter in GCE analytical models \\
$K$ & GCE parameter in analytical granularity equation \\
$\delta$ & Recurrence time between stellar additions from the same source \\ 
$T_{\rm Gal}$ & Age of the Galaxy up to the formation of the Sun \\
$T_{\rm isolation}$ & Isolation time of the (G)MC where the Sun was born \\
$T_{\rm LE}$ & Time of a last nucleosynthetic event \\
\hline
\multicolumn{2}{l}{Nucleosynthesis processes} \\
NSE & Nuclear statistical equilibrium process \\
$s$ process & $Slow$ neutron-capture process \\
$r$ process & $Rapid$ neutron-capture process \\
$p$ process & Process responsible for the production of p-rich isotopes heavier than Fe \\
$\gamma$ process & Photodisintegration process \\
$\nu$ process & Neutrino process \\
\hline
 \end{tabular}
\end{center}
 \end{table}

\subsection{The derivation of the SLR abundances in the ESS}
\label{sec:abundances}

\begin{figure}[tbh]
\begin{center}
\begin{minipage}[t]{16.5 cm}
\includegraphics[width=8cm,angle=0]{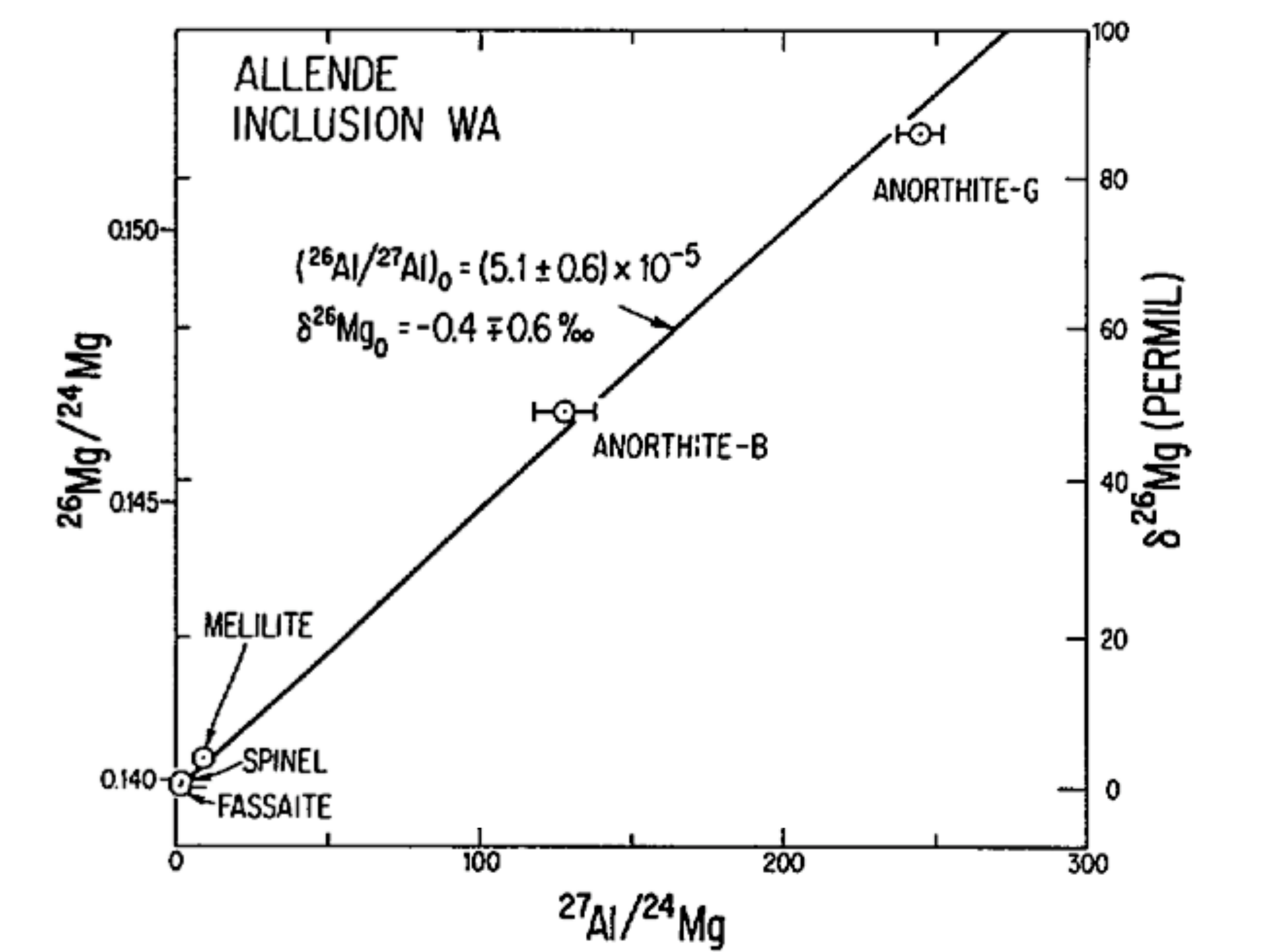} 
\includegraphics[width=8cm,angle=0]{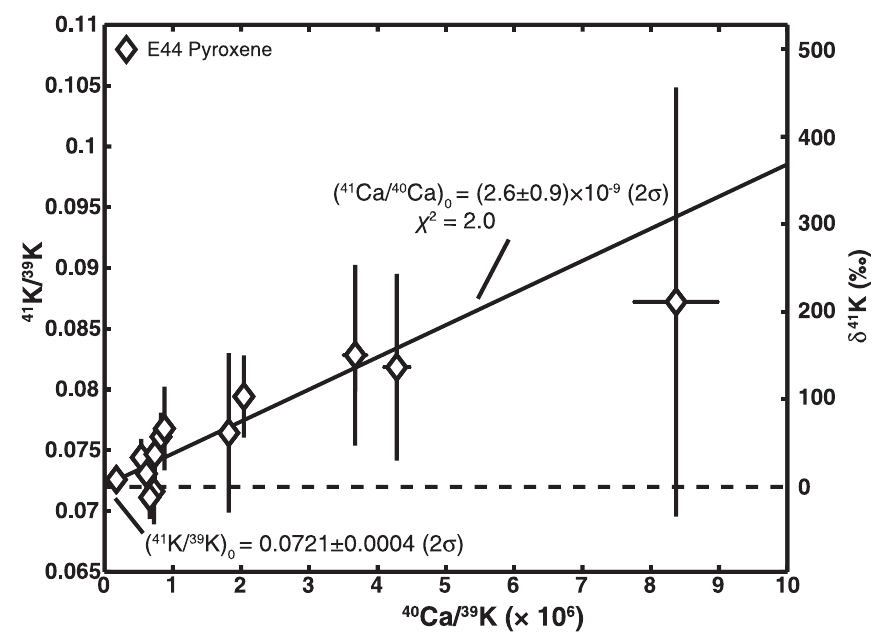} \\
\vspace{0.3cm}
\includegraphics[width=8cm,angle=0]{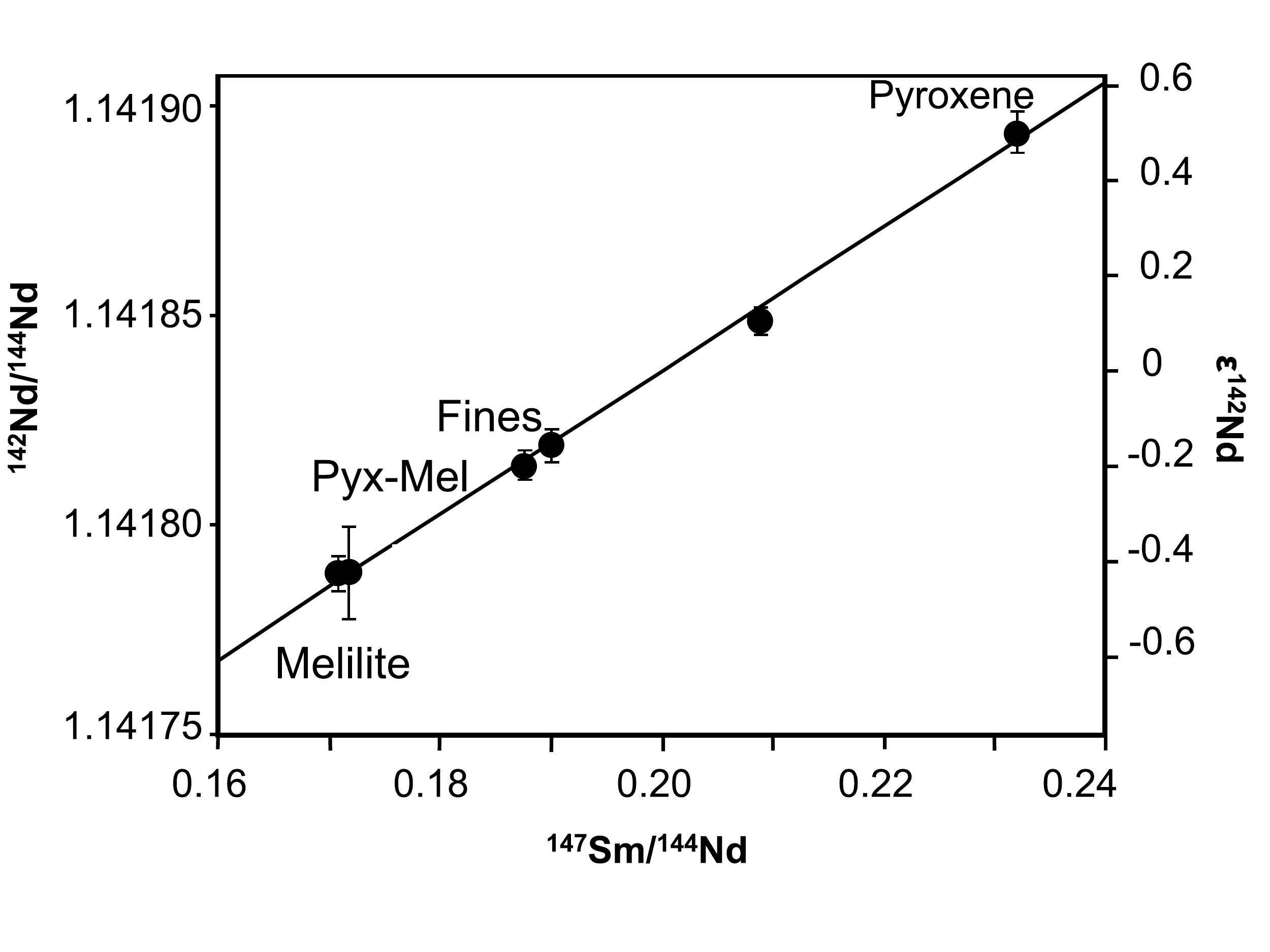} 
\includegraphics[width=8cm,angle=0]{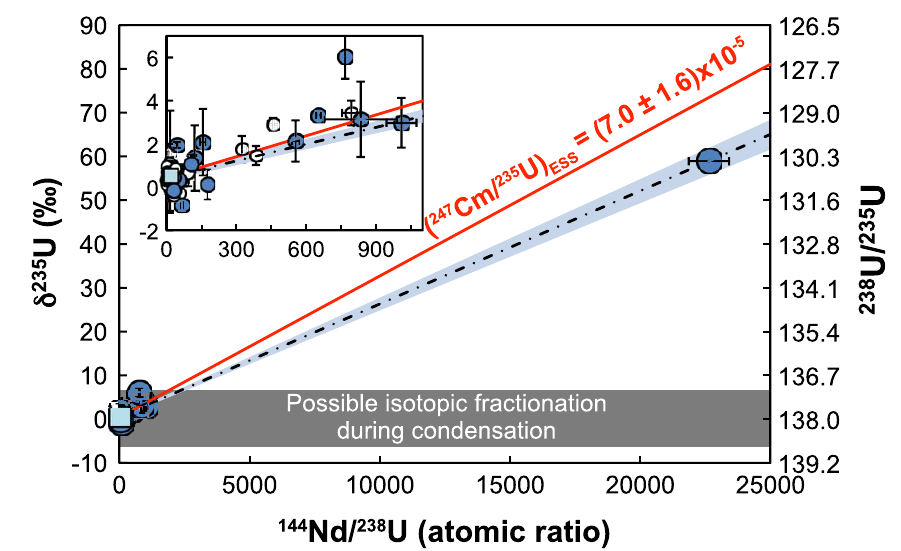} \\
\end{minipage}
\caption{Four typical examples of derivation of the ratio of a SLR relative to its stable (or long-lived) reference isotope from
excesses in the daughter nucleus of the SLR. The excess with respect to one of the 
most abundant isotopes of the same element is plotted on the y-axis, both as ratio and as $\delta$-value or $\epsilon$-value, i.e., per mil or per ten thousand, respectively, variation with respect to the laboratory standard (see Eq.~\ref{eq:delta}).
The x-axis reports the isotopic ratio of two isotopes taken to represent the relative
abundances of the two elements involved, which is controlled by the chemistry and mineralogy of the sample.
{\it Top left}: Measurements of different minerals with varying Al/Mg ratios in the inclusion WA
from the Allende meteorite from Lee et al. ``Aluminum-26 in the early solar system - Fossil or fuel'' \cite{lee77} ©AAS. Reproduced with permission. The linear correlation between the \iso{26}Mg excess and the elemental ratio 
represented the first clear evidence that  \iso{26}Al was incorporated live in these solids. If
\iso{26}Al was incorporated extinct instead, i.e., already fully decayed into \iso{26}Mg, the \iso{26}Mg excess
would be constant as function of Al/Mg. {\it Top right}: Inferred  \iso{41}Ca/\iso{40}Ca ratio in CAI E44 from the Efremovka meteorite from Liu et al. ``A Lower Initial Abundance of Short-lived $^{41}$Ca in the Early Solar System and Its Implications for Solar System Formation''( https://doi.org/10.1088/0004-637X/761/2/137)\cite{liu12} ©AAS. Reproduced with permission. This is an example of a case of a weaker evidence (see Table~\ref{table:SLRs}) due to the large error bars. 
{\it Bottom left}: Derivation of the initial \iso{146}Sm/\iso{144}Sm ratio in CAI Al3S4 (Fig.~\ref{fig:CAI}) from very high-precision data \cite{marks14}. {\it Bottom right}:
Derivation of the ESS \iso{247}Cm/\iso{235}U ratio based on analysis of the peculiar U-depleted CAI {\it Curious Marie} also from the Allende meteorite \cite{tissot16}. In this case Nd is used as chemical proxy for Cm, since Cm does not have stable isotopes. 
The blue line represents the isochrone obtained from the data, the red line represents the isochrone shifted an assumed age of 5 Myr.
However, \cite{tang17} reported a much shorter age, and  
the ESS ratio reported in Table~\ref{table:SLRs} is essentially the same as the blue line. Reprinted from Tissot et al. (2016) Science Advances, 2, e1501400 © The Authors, some rights reserved; exclusive licensee American Association for the Advancement of Science. Distributed under a Creative Commons Attribution Non Commercial License 4.0 (CC BY-NC) http://creativecommons.org/licenses/by-nc/4.0/. 
\label{fig:data}}
\end{center}
\end{figure}

Analysis of meteoritic whole rocks and separate inclusions is applied to derive the abundances of the SLRs as close as possible to the time when the Sun was born, i.e., in the early Solar System (ESS). The CAIs (FIG.~\ref{fig:CAI}) are one of the major components (amounting up to several \%) of the most primitive meteorites, the carbonaceous (CC) and unequilibrated ordinary (UOC) chondrites, and consist of high-temperature (refractory) solids. The other components of chondrites are chondrules -- solidified melt droplets that gave these meteorites their name -- and matrix, both of which consist largely of silicate minerals. These meteorites are ``undifferentiated'', i.e., they were not affected by major planetary/asteroidal processes like magmatism and formation of a metallic core. ``Differentiated'' meteorites, in contrast, suffered from such processes and include the rarer (by number) ``achondrites'', and the iron and stony-iron meteorites. Important among the achondrites in the context of establishing ESS abundances of SLRs (e.g., $^{244}$Pu) are the angrites, named after the type specimen {\it Angra dos Reis}. Achondrites include also eucrites (magmatic rocks likely from the asteroid Vesta) as well as meteorites from the Moon and from Mars.

Since the birth of the Sun was a process that lasted a few Myr, rather than a specific point in time, the definition of the time when the Sun was born is ambiguous. As usual in cosmochemistry we define this as the time when the first solids formed, in other words, as the age of the oldest solids found in meteorites, the CAIs. As mentioned above, the age of CAIs is very well determined using U to Pb radioactive dating.  Furthermore, it appears that CAIs, unlike chondrules, formed over a very short timescale of the order of 0.1 Myr \cite{connelly12}, similar to the median lifetimes of proto-stars hydrostatic cores surrounded by a dense accretion disk\footnote{These are referred to as proto-stars of Class 0. In the Class I objects more than 50\% of the envelope has fallen onto the central protostar, Class II objects have circumstellar disks, while Class III proto-stars have lost their disks.}. In the following we will refer to the ESS as the time when the CAIs formed. 

Given that the Sun is almost 4.6 Gyr old and the SLRs we consider here live less than 100 Myr, even if they were abundantly present when the Sun was born, today they are completely extinct and their abundances in the ESS cannot be not measured directly. They
are rather inferred from analysis of meteoritic samples via the identification of an $excess$ in the daughter nucleus into which each SLR decays. For example, 
excesses in \iso{26}Mg or \iso{60}Ni, with respect to their normal abundance ratios relative to isotopes without a possible radiogenic component such as \iso{24}Mg or \iso{58}Ni, can be the product of the radioactive
decay of \iso{26}Al or \iso{60}Fe, respectively. This is conceptually very different from observing, as done recently, live \iso{60}Fe in
the Earth's deep sea crust \cite{wallner16} (as well as \iso{244}Pu \cite{wallner15}), in fossilised bacteria \cite{ludwig16}, and
on the Moon \cite{fimiani16}. This live \iso{60}Fe is the fingerprint of a recent injection, roughly 2 Myr ago, from one or more supernova(e) resulting from the core-collapse of massive stars (core-collapse supernovae, CCSNe) \cite{breitschwerdt16}. On the other hand, an excess in \iso{60}Ni relative to \iso{58}Ni measured in meteorites represents {\it extinct} \iso{60}Fe and potentially the fingerprint of one or more CCSNe that occurred more than 4.6 Gyr ago. Also, fifteen atoms of live \iso{60}Fe have been counted in accelerated particles (cosmic rays, CRs) that reach the Earth \cite{binns16}. These live \iso{60}Fe atoms are the fingerprint of recent production events from CCSNe in the groups of massive stars (OB associations) from where the CRs are believed to originate. 

In the case of the ESS abundances, to make more evident the radiogenic origin of the observed excesses, it is necessary to analyse materials with variable amounts of the element to which the SLR isotope belongs, relative to the element
to which the daughter isotope belongs, e.g., the Al/Mg and the Fe/Ni ratios in the case of \iso{26}Al and \iso{60}Fe, respectively. True radiogenic excesses should be more evident in materials with the higher elemental ratios. These materials are advantageous in disentangling the true radiogenic excesses from other effects that may cause unusual isotopic ratios, such as statistical flukes as well as instrumental and natural mass fractionation effects. Excesses in the daughter nuclei are usually measured relative to the most abundant isotope of the same element, and to better highlight their nature as excesses, they are reported in the form of $\delta$-values or $\epsilon$-values, i.e., per mil or per ten thousand, respectively, variations with respect to a corresponding ``normal'' isotopic ratio, as defined by a laboratory standard. For example, in the case of the \iso{26}Mg/\iso{24}Mg ratio the $\delta$-value is:

\begin{equation}
\delta(^{26}{\rm Mg}/^{24}{\rm Mg}) = \left({\frac{(^{26}{\rm Mg}/^{24}{\rm Mg})_{\rm measured}}{(^{26}{\rm Mg}/^{24}{\rm Mg})_{\rm standard}}} - 1\right) \times 1,000. 
\label{eq:delta}
\end{equation}

\noindent The $\epsilon$-value is defined in the same way, except that the variation is multiplied by 10,000 instead of 1,000. A linear correlation between the excess and the elemental ratio (e.g., $\delta(^{26}{\rm Mg}/^{24}{\rm Mg})$ versus Al/Mg) proves that the SLR was incorporated in the samples while still alive (\cite{lee77}, Fig.~\ref{fig:data}).
The slope of the line gives the abundance ratio of the SLR
to the stable reference isotope at the time of {\it closure of the system}, i.e., the time after which the system was not disturbed anymore by any redistribution of isotopes or elements, the only compositional change coming from radiogenic decay.
Any alteration event after formation of a solid can be responsible for ``resetting'' the  chronometers. The line defined by the data points is referred to as an $isochrone$, since data points located on a given
line have by definition the same ratio of the SLR to its reference isotope, i.e.,
their closure time is the same. Any younger sample, i.e., one that closed after some time, would lie on a line with a shallower slope, since it would contain a lower initial abundance of the SLR due to its decay during the given time interval. Using this method, SLRs can be used to derive relative ages for Solar System samples, 
from which we can infer the history of the formation of planetesimals and planets \cite{dauphas11}. 

The intercept at x=0 represents the composition of a virtual sample that did not
include any abundance of the SLR. As such it provides the initial ratio of the daughter
nucleus to the reference isotope of the same element at the time of closure, relative to the laboratory standard. Samples that formed later from a same reservoir, 
as explained above, would present a shallower slope, at the same time, they would also have a higher intercept, since the SLR would have decayed further within the reservoir itself. 
However, different $\delta$-values at x=0 for different samples could also indicate non-radiogenic (i.e., not dependent on the decay time) heterogeneities in the initial abundance of the daughter nucleus and/or the SLR itself. For example, discussion is on-going on whether \iso{26}Al itself was distributed heterogeneously or homogeneously in the ESS (Sec.~\ref{sec:26Al}). It is crucial to determine the presence of SLR heterogeneities also because these would disturb the derivation of the isochrone-based ages for Solar System samples. 

Time differences between different samples can contribute to the uncertainties in our knowledge of the ESS abundances of the SLRs. Clearly, the best samples for this purpose are the oldest possible materials, the CAIs. In some cases analysis of a given element in CAIs is not easily possible, and other materials younger than CAIs need to be used. This is the case, for example, for \iso{60}Fe, due to the fact that not much Fe is present in CAIs. The age difference between the analysed sample and the CAIs can be measured using other radioactive
systems and then be used to extrapolate back from the abundance measured in the sample to the ESS value (see, e.g., the case of \iso{247}Cm/\iso{235}U in the bottom right panel of Fig.~\ref{fig:data}).


Two more issues should be mentioned. The first is the case where excesses in the daughter nucleus may be present, which are not related to the radiogenic decay of the SLR. Potential intrinsic heterogeneities could be produced both by natural and artificial effects. Natural effects include nucleosynthetic signatures, i.e., anomalies in the stable isotopes due to the original presence of presolar stardust, as well as mass fractionation, both mass-dependent and non mass-dependent \cite{dauphas16}. Artificial effects can occur during the laboratory chemical procedures and the measurement itself and are mostly of the mass-dependent fractionation type. These effects can be prominent relative to the true radiogenic effect, which is usually quite small (in fact, as explained above, it is measured in per mil or per ten thousand variations). The mass-depended artificial effects can be corrected by analysing at least three isotopes, and normalising the system to a chosen set of ``normal'' non-radiogenic ratios. A typical example where these issues are particularly relevant is the hotly debated case of \iso{60}Fe (Sec.~\ref{sec:60Fe}). 

The second issue is related to the derivation of useful SLR to stable isotope ratios in the ESS for the few SLRs heavier than Fe produced by the proton-capture process (the $p$ process; see Sec.~\ref{sec:stars}), and potentially for \iso{244}Pu (Sec.~\ref{sec:rpiso}). In these cases, to obtain a ratio that is possible to interpret within the framework of stellar nucleosynthesis it is necessary to re-normalise the measured ratio to a different stable isotope than that used for the measurement. This involves the use of the Solar System abundances of stable isotopes and their associated uncertainties, which can be relatively large when different elements are involved. A main example is \iso{92}Nb, whose ESS abundance is measured relative to \iso{93}Nb, which is the only stable isotope of Nb and is produced by neutron-capture processes. The abundance of \iso{92}Nb needs to be re-normalised instead to \iso{92}Mo, a neighbouring nucleus that is produced by the $p$ process like \iso{92}Nb (see Sec.~\ref{sec:92Nb}). 

In Table~\ref{table:SLRs} we present an update of Table 1 of Dauphas \& Chaussidon (2011) \cite{dauphas11} for 19 SLRs. 
The half-lives are taken from the National Nuclear Data Center website (www.nndc.bnl.gov, including errors in brackets), except for \iso{10}Be and \iso{146}Sm, for which references are given in the table footnotes.
Roughly a dozen new measurements and estimates have become available since 2011, improving the accuracy and precision of our knowledge of the initial ESS abundances of roughly half of the listed nuclei. The number of nuclei with three stars in the quality ranking (last column of Table~\ref{table:SLRs}) has increased by one since 2011 because of the more precise determination of the \iso{107}Pd/\iso{108}Pd ratio \cite{brennecka16}. Further, the \iso{247}Cm/\iso{235}U ratio is now much more solidly determined, thanks to the discovery of the peculiar U-depleted CAI {\it Curious Marie} named after Marie Skłodowska Curie (\cite{tissot16}, bottom right panel of Fig.~\ref{fig:data}). On the other hand, we have downgraded the estimate of the \iso{244}Pu/\iso{238}U ratio from three- to two-star quality due to the fact that two different values are reported from two different types of experiments. The value given by \cite{lugmair77} is roughly half of that listed in the table from \cite{hudson89} (see discussion in Sec.~\ref{sec:rpiso}). The number of ratios with one-star quality has decreased from five to three with respect to Table~1 of \cite{dauphas11} due to the upgrade of the \iso{247}Cm/\iso{235}U ratio, as well as the recently improved upper limit of the \iso{135}Cs/\iso{133}Cs ratio \cite{brennecka17a}. This is now more than two orders of magnitude lower than the previous estimate, providing a more significant constraint. 
For three of the SLRs produced by the $p$ process (\iso{92}Nb and \iso{97,98}Tc) we provide both the experimental ratio and the ratio re-normalised to a different stable isotope using the most recent Solar System abundances of the stable isotopes \cite{lodders09,burkhardt11}.

Most of the uncertainties listed in Table~\ref{table:SLRs} are statistical only and given at 2$\sigma$, however, several exceptions are present, which are discussed in detail within the subsections of Sec.~\ref{sec:list} dedicated to the different isotopes. Systematic uncertainties, on the other hand, are not included since they derive from specific suppositions and cannot be evaluated quantitatively. An indication of the magnitude of such uncertainties can only be derived by comparing the results from different experiments, approaches, and assumptions. For example, in the case of the ESS abundance of \iso{107}Pd, the main current uncertainty is related to a potential systematic error related to the age of the considered sample \cite{matthes18}.

Three more SLRs exists with half-lives in the range of interest here: \iso{81}Kr (0.23 Myr), \iso{93}Zr (1.5 Myr), and \iso{99}Tc (0.21 Myr). They are not included in Table~\ref{table:SLRs}, however, for various reasons: \iso{81}Kr is a noble gas isotope, and as such was virtually absent from the solid materials with which we deal here. Even if it was introduced therein by ion implantation, as in the case of noble gas trapped in meteoritic components such as stardust nanodiamond and SiC, as well as Phase Q \cite{ott14}, its abundance would still be very low compared to the neighbouring less volatile elements and not reflect the abundance produced in a stellar source. In addition, its daughter nucleus \iso{81}Br is also  volatile and thus prone to secondary loss, complicating matters further. The daughter of \iso{93}Zr is \iso{93}Nb; for this nucleus it is not possible to observe an excess relative to other isotopes of Nb since it is the only stable isotope of Nb. Finally, \iso{99}Tc decays into \iso{99}Ru. Only upper limits are available for the similar 
case of \iso{98}Tc decaying into \iso{98}Ru, but  
\iso{99}Tc is even more challenging \cite{becker03} due to the 20 times shorter half-life of \iso{99}Tc with respect to \iso{98}Tc, and the 7 times higher natural abundance of \iso{99}Ru with respect to \iso{98}Ru.

\begin{table}
\begin{center}
 \centering
 \caption{For the 19 SLRs we list their daughter nuclei, stable or long-lived reference isotopes, T$_{\rm 1/2}$ (and $\tau$) from the National Nuclear Data Center website (www.nndc.bnl.gov, including errors on the last digits in brackets), and ESS ratios. In the last column, following Dauphas \& Chaussidon \cite{dauphas11} a quality ranking is given: three stars
 indicate those SLRs whose ESS abundance is well determined; two stars indicate those SLRs for which there is convincing evidence for their presence in the ESS, but the initial abundance is less certain; one star indicates those SLRs for which there are reports, but the evidence is weak and awaits confirmation; $<$ means that only an upper limit on the initial abundance exists.}
\label{table:SLRs}
\vspace{0.3cm}
 \begin{tabular}{llllllll}
 \hline
SLR & Daughter & Reference & T$_{1/2}$(Myr) & $\tau$(Myr) & ESS ratio & Ref. & Quality \\
 \hline
\iso{26}Al & \iso{26}Mg & \iso{27}Al & 0.717(24) & 1.035 & $(5.23 \pm 0.13) \times 10^{-5}$ & \cite{jacobsen08} & $\star\star\star$ \\ 
\iso{10}Be & \iso{10}B & \iso{9}Be & 1.388(18)$^a$ & 2.003 & $3-9 \times 10^{-4}$ & \cite{tatischeff14}$^b$ & $\star\star\star$ \\ 
\iso{53}Mn & \iso{53}Cr & \iso{55}Mn & 3.74(4) & 5.40 & $(7 \pm 1) \times 10^{-6}$ & \cite{tissot17} & 
$\star\star\star$ \\ 
\iso{107}Pd & \iso{107}Ag & \iso{108}Pd & 6.5(3) & 9.4 & $(6.6 \pm 0.4) \times 10^{-5}$ & \cite{matthes18}$^c$ & $\star\star\star$ \\ 
\iso{182}Hf & \iso{182}W & \iso{180}Hf & 8.90(9) & 12.8 & $(1.018 \pm 0.043) \times 10^{-4}$ & \cite{kruijer14}  & $\star\star\star$ \\  
\iso{247}Cm & \iso{235}U & \iso{235}U & 15.6(5) & 22.5 & $(5.6 \pm 0.3) \times 10^{-5}$ & \cite{tang17} & 
$\star\star\star$ \\
\iso{129}I & \iso{129}Xe & \iso{127}I & 15.7(4) & 22.6 & $(1.28 \pm 0.03) \times 10^{-4}$ & \cite{ott16} & 
$\star\star\star$ \\ 
\iso{92}Nb & \iso{92}Zr & \iso{93}Nb & 34.7(2.4) & 50.1 & $(1.57 \pm 0.09) \times 10^{-5}$ & \cite{haba17} & $\star\star\star$ \\  
 &  & \iso{92}Mo$^d$ & & & $(3.2 \pm 0.3) \times 10^{-5}$ & & $\star\star\star$ \\  
\iso{146}Sm & \iso{142}Nd & \iso{144}Sm & 68$^e$/103$^f$ & 98$^e$/149$^f$ & $(8.28 \pm 0.44) \times 10^{-3}$ & \cite{marks14} & $\star\star\star$ \\ 
\iso{36}Cl & \iso{36}S, \iso{36}Ar & \iso{35}Cl & 0.301(2) & 0.434 & $2.44 \pm 0.65 \times 10^{-5}$ & \cite{tang17}$^g$ & $\star\star$ \\ 
\iso{60}Fe & \iso{60}Ni & \iso{56}Fe & 2.62(4) & 3.78 & $(1.01 \pm 0.27) \times 10^{-8}$ & \cite{tang15}$^h$ & $\star\star$ \\ 
\iso{244}Pu & $^i$ & \iso{238}U & 80.0(9) & 115 & $(7 \pm 1) \times 10^{-3}$ & \cite{hudson89} & 
$\star\star$ \\ 
\iso{7}Be & \iso{7}Li & \iso{9}Be & 53.22(6) days & 76.80 days & $(6.1 \pm 1.3) \times 10^{-3}$ & \cite{chaussidon06} & $\star$ \\ 
\iso{41}Ca & \iso{41}K & \iso{40}Ca & 0.0994(15) & 0.1434 & $(4.6 \pm 1.9) \times 10^{-9}$ & \cite{liu17} & $\star$  \\ 
\iso{205}Pb & \iso{205}Tl & \iso{204}Pb & 17.3(7) & 25.0 & $(1.8 \pm 1.2) \times 10^{-3}$ & \cite{palk18} & $\star$ \\
\iso{126}Sn & \iso{126}Te & \iso{124}Sn & 0.230(14) & 0.33 & $< 3 \times 10^{-6}$ & \cite{brennecka17b} & $<$ \\
\iso{135}Cs & \iso{135}Ba & \iso{133}Cs & 2.3(3) & 3.3 & $< 2.8 \times 10^{-6}$ & \cite{brennecka17a} & $<$ \\
\iso{97}Tc & \iso{97}Mo & \iso{92}Mo & 4.21(16) & 5.94 & $< 1 \times 10^{-6}$ & \cite{burkhardt11} & $<$ \\
& & \iso{98}Ru$^l$ & & & $< 1.1 \times 10^{-5}$ & & $<$ \\
\iso{98}Tc & \iso{98}Ru & \iso{96}Ru & 4.2(3) & 6.1 & $< 2. \times 10^{-5}$ & \cite{becker03} & $<$ \\
 &  & \iso{98}Ru$^l$ & & & $< 6. \times 10^{-5}$ & & $<$ \\
 \hline
 \end{tabular}
 \end{center}
$^a$According to \cite{chmeleff10}. $^b$and references therein. A single CAI with a very high value of $104 \times 10^{-4}$ also exists \cite{gounelle13}. $^c$The value needs to be confirmed by Pb-Pb dating using the U isotope composition determined for the same sample, it could be lowered down to $4 \times 10^{-5}$ \cite{matthes18}. $^d$Renormalised using Solar System abundances \cite{lodders09,burkhardt11}. $^e$According to \cite{kinoshita12}. $^f$According to \cite{marks14}. $^g$ We calculated the error bar translating the age of less than 50 kyr \cite{tang17} into an age of $25 \pm 25$ kyr.
$^h$Values from $10^{-7}$ to $10^{-6}$ are also reported \cite{mishra14,telus18}.
$^i$The main (99.88\%) decay mode of \iso{244}Pu is by $\alpha$ emission. The ensuing decay chain proceeds through the very long lived \iso{232}Th (T$_{1/2}$=14 Gyr). The spontaneous fission of \iso{244}Pu, which results in measurable excesses of some Xe isotopes used to derive the ESS abundance of \iso{244}Pu, represents only 0.12\% of the decay process. $^l$Renormalised using Solar System abundances \cite{lodders09}.
\end{table}

\subsection{Stellar evolution and nucleosynthesis}
\label{sec:stars}

\begin{figure}[tb]
\begin{center}
\begin{minipage}[t]{16.5 cm}
\includegraphics[width=16cm,angle=0]{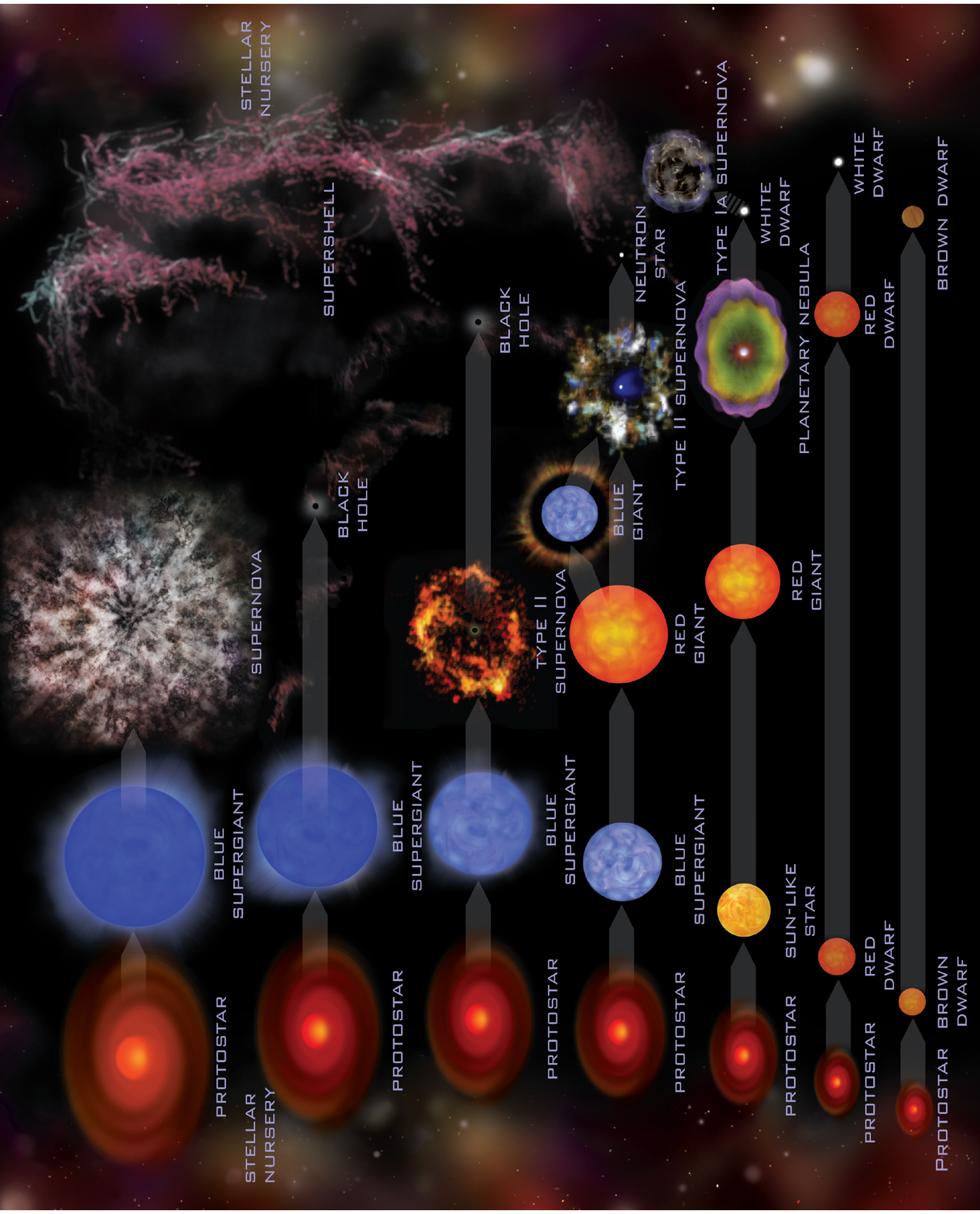} \\
\end{minipage}
\begin{minipage}[t]{16.5 cm}
\caption{Schematic illustration of stellar evolutionary phases with time (increasing on the x-axis), according to their initial mass (increasing on the y-axis). 
Image credit: NASA/CXC/M.Weiss/Public domain.
\label{fig:starevol}}
\end{minipage}
\end{center}
\end{figure}

The cosmic abundances of the vast majority of the nuclei of the elements heavier than H and He are produced 
by processes occurring during the various hydrostatic and explosive evolutionary phases of single 
stars, as well as during the interaction of two or more stars in multiple stellar systems. This applies 
to the abundances of both stable and radioactive nuclei, the only difference being, of course, that the latter,  
following production, decay according to their half-lives.
In fact, it was thanks to the discovery of the signature of the short-lived element Tc in the spectra of red giant stars that it was possible to definitely prove that nucleosynthesis occurs {\em in situ} inside stars \cite{merrill52}. Any Tc originally present would have been completely decayed -- its isotopes have half-lives of a few million years at most -- by the time of the order of billions of years that the observed low-mass stars take to reach the red giant phase. Here below we provide a brief summary of the processes of stellar evolution and nucleosynthesis. For more detailed reviews see \cite{woosley02,langer12,karakas14,demarco17}. 

In the interiors of stars matter can reach extremely high temperatures, for example, 10 million K (MK) in the core of the Sun and up to billions of K in supernovae. Under the force of gravity, high density conditions are also maintained, for example, roughly 100 gr/cm$^3$ in the core of the Sun and up to 10$^{10}$ gr/cm$^3$ in supernovae. Such conditions force nuclei to keep in a 
confined volume and to react via a huge variety of nuclear interaction channels. This complexity and 
diversity created all the variety of atomic nuclei from carbon up in the Universe. It is, however, not enough to produce 
nuclei in the hot and dense interiors of stars and supernovae. Mechanisms must also exist such that these nuclei are expelled into the surrounding 
medium and recycled into newly forming stars and planets. In stars born with mass similar to 
the Sun (solar mass, hereafter \msun) and up to roughly ten times this value, these mechanisms are identified as 
the combination of the mixing of matter from the deep layers of the star to the stellar surface and the stellar winds that peel off the external layers of the 
star. These processes are active most efficiently during the final phases of the 
lives of these stars, the so-called asymptotic giant branch (AGB) phase. During the AGB, efficient dredge-up episodes of matter from the hot core of the star occur together with strong stellar winds, driving mass-loss rates up to $10^{-4}$ \msun/yr. The winds are powered by variations in the stellar radius, and thus the luminosity and the surface temperature, as well as by the presence of large amounts of dust that form in the cool ($\sim$2000 K) external layers of the star. When most of the original stellar mass is lost, the matter expelled by the wind can be illuminated by UV photons coming from the central star, producing what we observe as a colourful planetary nebula. Eventually the core of the star, rich in C and O produced by previous He burning, is left as a white dwarf (WD, Fig.~\ref{fig:starevol}). The evolutionary timescales of such low-mass stars are relatively long, from 1,000 Myr for stars of mass around that of the Sun, down to several tens of Myr for stars of mass around 7 times larger.  

More massive stars live much shorter lives, from a few Myr to a few tens of Myr, and end their lives due to the final collapse of their core (Fig.~\ref{fig:starevol}). Once nuclear fusion processes have turned all the material in the core into Fe,
neither fusion nor fission processes can 
release enough energy anymore to prevent the core collapse.
As the core collapses, matter starts falling onto it, which results in a bounce shock and a final CCSN explosion. The exact mechanism of the explosion is not well known although remarkable progress has been made in the past decade \cite{janka12}. The supernova ejecta are responsible for carrying out into the interstellar medium the fraction of 
synthesised nuclei that does not fall back into the neutron star or black hole remnant. In the earlier phases of the evolution of these massive stars, also stellar 
winds can play the important role of shedding freshly synthesised material into the stellar surroundings. In fact, in some cases, the winds can be so strong that layers previously affected by nuclear burning are exposed, and the ashes of the burning of H and He in the stellar core can be observed directly at the stellar surface. These rare, peculiar stars are known as Wolf-Rayet (WR) stars \cite{langer12} and the strong winds that characterised them are driven by radiation, when the mass of the star is so high (roughly $>$ 40 \msun) that its luminosity can push matter away. Binary interaction can also result in significant loss of matter from stars, if the presence of a companion results in gravitational pull, enhanced mass loss, and non-conservative mass transfer when mass is lost from the system. 
Another interesting consequence of binary interaction is when accretion of mass from a stellar companion onto a WD is followed by explosive thermo-nuclear burning on the surface of the WD, which results in what we observe as nova explosions. These explosion events also shed matter into their surroundings. An even more extreme case of thermo-nuclear explosions are the supernovae classified as Type Ia (SNIa). In contrast to supernovae classified as SNII, which are rich in H and originate from CCSNe, SNIa are characterised by the absence of H in their spectra. In this case C-burning initiated within a WD made mostly of C and O produces a detonation or a deflagration that tears 
the whole WD apart (Fig.~\ref{fig:starevol}). Even though the light from these events is used as a standard candle to measure the expansion of the Universe (e.g., \cite{riess98}), their origin is still mysterious. Two major binary channels are
currently proposed: a WD accreting matter from a stellar companion, and the collision of two WDs.

Stellar nucleosynthesis was first systematically organised by Cameron \cite{cameron57} and Burbidge et 
al. \cite{burbidge57} -- of which an update can be found in Wallerstein et al. \cite{wallerstein97}. Hydrogen burning is mostly
responsible for the production of N by conversion of C and O into it, as well as a 
large variety of minor isotopes: from \iso{13}C produced via proton captures on \iso{12}C, to the only stable 
isotope of Na (\iso{23}Na) produced by proton captures on \iso{22}Ne, and the SLR \iso{26}Al, produced by proton 
captures on \iso{25}Mg. Typical temperatures are from 10 to 100 MK, depending on the stellar site. Helium burning is 
mostly identified with the triple-$\alpha$ (\iso{4}He particle) reaction producing \iso{12}C, and the 
\iso{12}C($\alpha$,$\gamma$)\iso{16}O reaction, producing \iso{16}O. Many other secondary channels of burning open 
as the temperature increases above 100 MK, for example, conversion of  already present \iso{14}N nuclei 
into \iso{22}Ne via double $\alpha$-captures. Also reactions that produce free neutrons are 
typically associated with He-burning, the most famous being \iso{13}C($\alpha$,n)\iso{16}O and 
\iso{22}Ne($\alpha$,n)\iso{25}Mg. In stars with mass below roughly 10 \msun, nuclear burning processes do not typically proceed past He burning. When He is exhausted in the core these stars enter the AGB phase with a degenerate, inert C-O core. In more massive stars, instead, the temperature in the core increases further. A large variety of reactions can 
occur. These processes involve C, Ne, and O  
burning, and include many channels of interactions, with free protons and neutrons driving a large number of possible nucleosynthetic paths. The cosmic abundances of the ``intermediate-mass'' elements, roughly from Ne to Cr, are mainly the 
result of this burning. Once the temperature reaches billions of degrees, the probabilities of fusion 
and photodisintegration reactions become comparable and the result is nuclear statistical equilibrium (NSE). This process favours the production of nuclei with the highest binding energy per 
nucleon, resulting in a final composition predominantly characterised by high abundances of the nuclei around the Fe peak in the Solar System abundance distribution.

Beyond the Fe peak, charged-particle reactions are not efficient anymore due to the large Coulomb barrier around these heavy nuclei (with number of protons greater than 26). Neutron captures, in the form of {\it slow} neutron captures, the $s$ process (see \cite{kaeppeler11} for a review), and {\it rapid} 
neutron captures, the $r$ process (see \cite{thielemann11} for a review), are instead the main channels for the production of the atomic nuclei up to 
the actinides. Traditionally, these two neutron capture processes stand as the two extreme cases: 
during the $slow$ process, neutron captures are always slower than  decays, during the $rapid$ process, 
neutron captures are always faster than decays. However, intermediate cases do also occur in 
nature, ranging from the mild case of the operation of {\it branching points} on the $s$-process 
path (as discussed below in relation to a variety of SLRs, such as \iso{60}Fe), to the neutron burst in CCSNe (again, possibly affecting the abundances of many SLRs), to the $intermediate$ neutron-capture process, the $i$ process, 
identified so far mostly in low-metallicity environments and post-AGB stars \cite{herwig11,hampel16}. 

The $s$ process 
requires relatively low neutron densities ($\sim 10^{7}$ cm$^{-3}$) and is at work during He and C 
burning in low-mass AGB stars (producing most of the $s$-process elements, from Sr to Pb) and the hydrostatic burning phases of massive stars (producing the $s$-process abundances from Fe to Sr). The  neutrons are provided by the neutron source reactions on \iso{13}C 
and \iso{22}Ne mentioned above \cite{kaeppeler11}. The $r$ process 
requires much higher neutron densities ($> 10^{20}$ cm$^{-3}$) and is at work in explosive 
neutron-rich environments. The stellar site of the $r$ process has been one of the most uncertain and highly 
debated topic in astrophysics. Currently, neutron star mergers (NSMs) are being favoured due to new constraints from the discovery of the gravitational wave source GW170817
and its counterparts in $\gamma$-rays (NSMs are believed to be the origin of short $\gamma$-ray bursts) and in the optical and infrared, where the source is a {\it kilonova} resulting from the radioactive decay of heavy $r$-process nuclei \cite{kilpatrick17,cote18}.
Measurements of \iso{244}Pu in the Earth's crust as compared to its ESS abundance also support rare events such as NSM as the site of the $r$ process \cite{hotokezaka15}. Peculiar flavours of CCSNe (with strong magnetic fields, jets, as well as accretion disks around black holes) could also contribute to $r$-process production in the Galaxy \cite{thielemann11}. Another problem with the modelling of the $r$ process is the fact that the nuclei involved are extremely unstable and it is difficult to experimentally determine their properties, even their mass. The coming up large Facility for Antiproton and Ion Research (FAIR) at GSI (Germany) is one of the facilities promising future improvements on this problem, together with the Facility for Rare Isotope Beams (FRIB) at MSU (USA) and the RI Beam Factory at RIKEN (Japan).

A few tens of nuclei heavier than Fe are located on the proton-rich side of the valley of $\beta$-stability and cannot be produced via neutron captures. These nuclei have typically very low abundances, i.e., they represent at the very most a few percent of the total Solar System abundance of the element they belong to. To account for their production a so-called $p$ process is invoked, whose mechanism 
and astrophysical site is still debated. One popular flavour of the $p$ process is the 
$\gamma$ process \cite{pignatari16}, where heavier, abundant nuclei are photodisintegrated in an explosive 
environment to produce lighter $p$-process nuclei. Other possibilities 
are related to the inverse case, where lighter nuclei capture charged particles to reach some heavier 
$p$-process nuclei, typically the lightest, and most abundant up to Mo and Ru at atomic mass around 90-100. There are several proposed options for this modality, from the $rp$ process (rapid $p$ process), for example, occurring in 
X-ray bursts from explosive burning due to accretion of matter from a stellar 
companion onto a neutron star, to explosive nucleosynthesis during CCSNe, in particular when matter cools down from NSE and 
$\alpha$ particles becomes available ($\alpha$-rich freeze out), as well as the neutrino winds from a nascent neutron star (the $\nu$ process).

Finally, the bulk of the production of B, Be, and Li\footnote{For Li a contribution from Big Bang and stellar nucleosynthesis is also present \cite{travaglio01}.} does not occur in stars. The abundances of these nuclei are produced via spallation reactions in the interstellar medium (ISM). Spallation reactions occur when material is hit by accelerated particles, i.e., cosmic rays (CRs). This process can be also referred to as non-thermal nucleosynthesis, given that it does not occur within a Maxwellian plasma as nucleosynthetic processes in stars.

The list of SLRs present in the ESS (Table~\ref{table:SLRs}) cover almost the whole range of atomic masses, from approximately 10 to 250. As such, their production mechanisms cover the whole range of nucleosynthetic processes that occur in stellar production sites, as will be discussed in detail in Sec.~\ref{sec:list}.

\subsection{Galactic chemical evolution and the build-up of Solar System matter}
\label{sec:galaxy}

As stars end their life polluting their environment via winds or explosions with atomic nuclei freshly synthesised in 
their interiors, new stars are born in the ISM, collecting the gas and dust 
expelled by the dying stars. In this way, the chemical composition of the
Galaxy evolves with time and results in stars of different ages located in different regions of the Galaxy to 
present different chemical compositions \cite{tinsley80,matteucci12}. This process is referred to as galactic chemical 
evolution (GCE) and is specifically responsible for the fact that stars have different 
{\it metallicities} ($Z$), i.e., amounts of ``metals''. Traditionally, in astronomy $Z$ refers to all the elements heavier
than H and He; the Fe abundance is often used as a proxy for it. For the Sun $Z=0.014$ \cite{asplund09},
but we also find stars in our Galaxy with $Z$ varying from six orders of magnitude lower than for the Sun to 
more than a factor of two higher, depending on both the time and place where they were born. A simple GCE model predicts that metallicity  increases as the Galaxy evolves with time. The consequence is that younger (relative to present day) stars should show higher metallicity than older stars, since the younger stars would have formed later during the evolution of the Galaxy and collected material from more previous generations of stars. However, the most recent observations of large stellar galactic populations show that for each stellar age there is a large spread of metallicity \cite{casagrande11,bensby14}. This is interpreted as the result of stellar migration from different regions of the Galaxy \cite{spitoni15}, where different star formation rates produce different numbers of stellar generations and in turn different metallicities. In this respect the field of GCE is now moving towards a more complete picture of galactic ``chemo-dynamical'' evolution. 

\begin{figure}[tb]
\begin{center}
\begin{minipage}[t]{8.5 cm}
\includegraphics[width=8.5cm,angle=0]{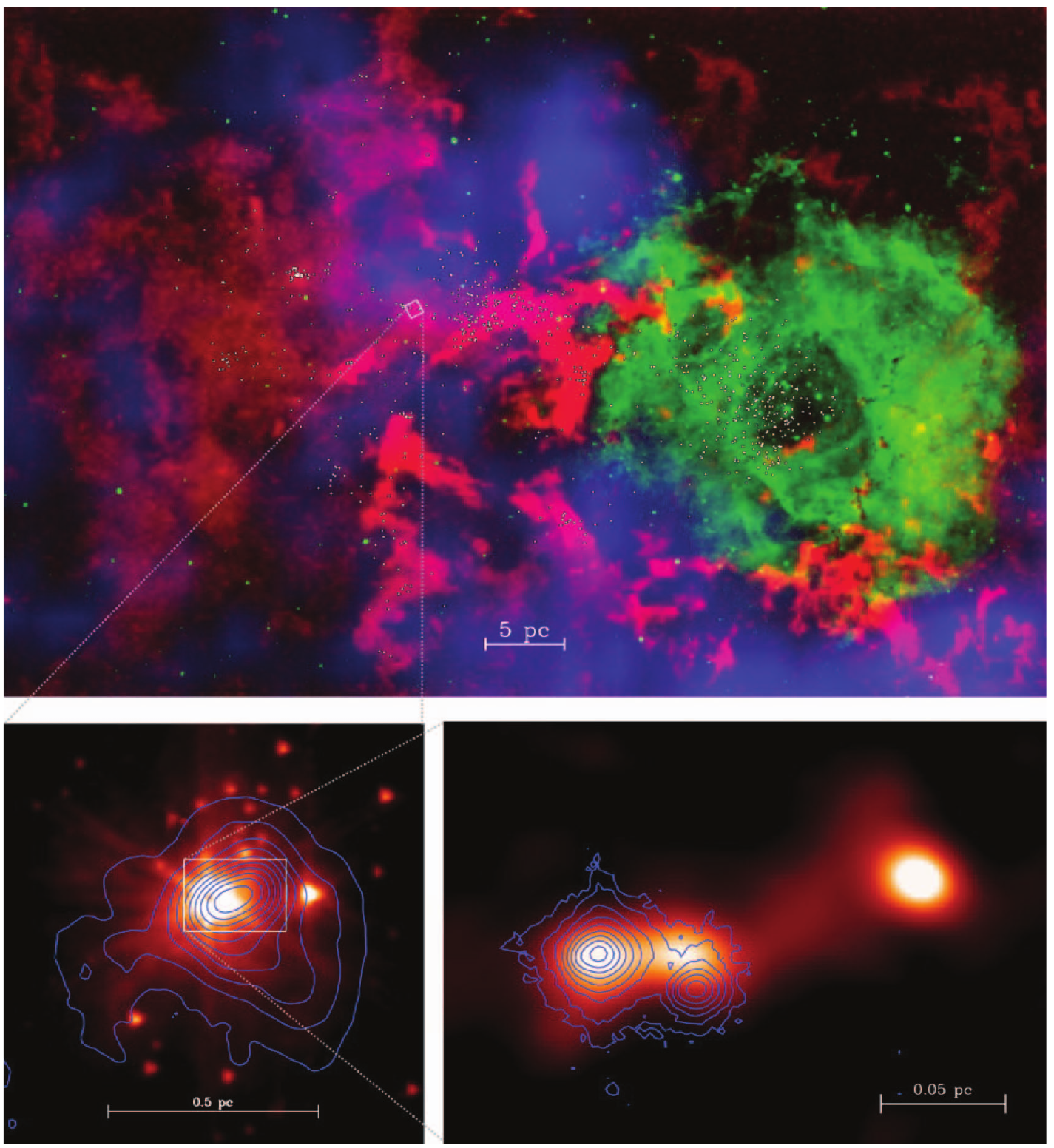} \\
\end{minipage}
\begin{minipage}[t]{16.5 cm}
\caption{The scales of star formation, as illustrated by \cite{williams10}. The upper panel shows a composite view of the Rosette nebula and accompanying GMC. The nebula (in green) is powered by a collection of massive stars at the centre of a large cluster. The red is emission from the CO molecule, indicating the GMC. The lower panels zoom in on the star formation process: the left panel shows contours of emission from a cold dusty envelope around a deeply embedded stellar cluster; the right panel shows contours from embedded protostars. Figure reproduced with permission from J. Williams, Contemporary Physics 2010, 51, 381–396 Taylor \& Francis Ltd www.tandfonline.com. \label{fig:williams}}
\end{minipage}
\end{center}
\end{figure}

Within the ISM, star formation occurs within hierarchical structures (see \cite{williams10} for an accessible review). Stellar nurseries are the coldest and denser regions of the ISM and are 
referred to as molecular clouds (MCs, named molecular because of the presence of molecules, in particular hydrogen molecules) or giant molecular clouds (GMCs), depending on their size, which is of the order of 50 parsec (pc) for GMCs (top panel of Fig.~\ref{fig:williams}). Molecular clouds in the solar neighbourhood appear to be relatively short-lived, of the order of a few Myr \cite{hartmann01}, instead, molecular clouds further away have been observed to have lifetimes in the range 20 to 40 Myr \cite{murray11}. Such differences have been attributed to their larger masses. 
It is now well established that the vast majority of stars are born in MCs large enough to produce at least a group of stars, referred to as a stellar cluster. GMCs potentially host a number of stellar clusters. The clusters have sizes on the order 0.5 pc (left bottom panel of Fig.~\ref{fig:williams}) and the number of stars can vary largely, from a few tens to tens of thousands. Within clusters, the star formation process is relatively fast, on the order of a few Myr at most (see \cite{dib13} and references therein).
Within the 0.05 pc scale, a dense core (the protosolar nebula in the case of the Sun) collapses to form a single star or a multiple stellar system of typically two or three stars. The star itself is first observed as embedded within a thick envelope from which it accretes matter. Since the nebula rotates, a protoplanetary disk forms (the protosolar disk in the case of the Sun). Within a few Myr, possibly up to 10 Myr based on statistical observations \cite{haisch01,williams11}, only solids are left in the disk as all the gas is dispersed.
This complex, hierarchical structure of star formation results in the possibility of stars within a given stellar cluster or GMC to evolve and pollute the gas from which new stars are born or the already formed protoplanetary disks, with some SLRs, as will be discussed in relation to the Sun in Sec.~\ref{sec:Q3}.

\begin{figure}[tb]
\begin{center}
\begin{minipage}[t]{16.5 cm}
\includegraphics[width=16cm,angle=0]{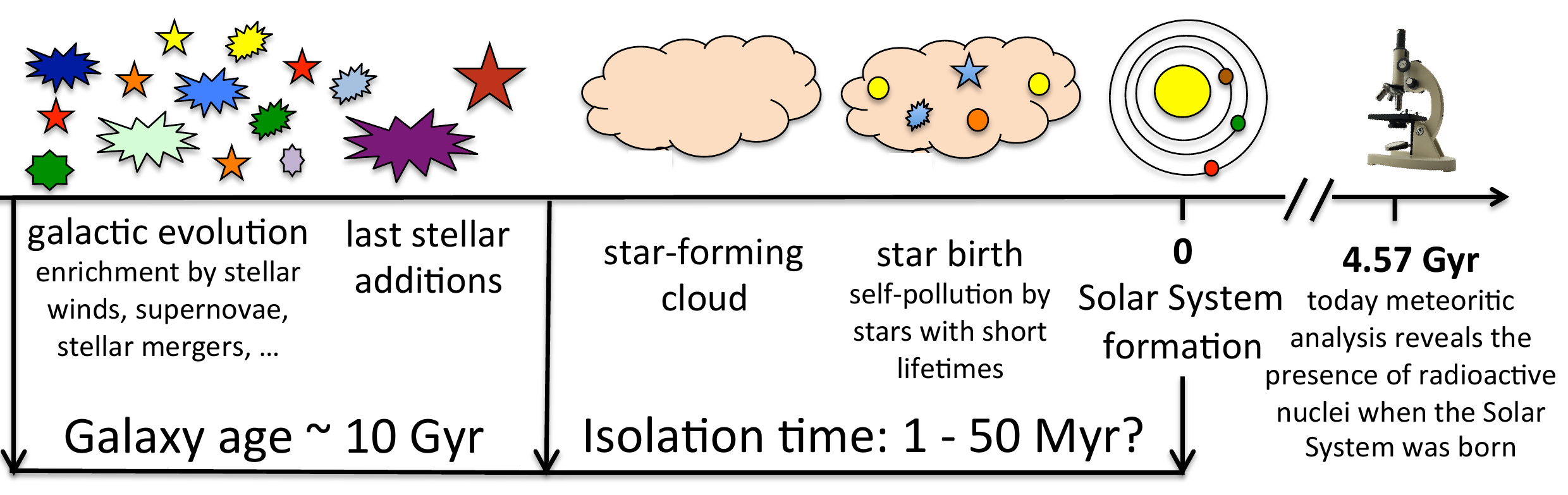} \\
\end{minipage}
\begin{minipage}[t]{16.5 cm}
\caption{Schematic illustration of the presolar history of the Solar System matter. \label{fig:Sun}}
\end{minipage}
\end{center}
\end{figure}

Within this global picture we can identify two phases for the presolar history of Solar System matter (Fig.~\ref{fig:Sun}). The first phase is related to the evolution of the Galaxy on the relatively long time interval from the formation of the Milky Way Galaxy to the birth of the Sun of $\sim$9 Gyr (equal to the age of the Milky Way of $\sim$13 Gyr minus the age of the Sun of 4.6 Gyr). The bulk of the composition of our Sun and its planets was constructed by generations of hundreds to thousands of stars in the Galaxy, each contributing their parcel of atomic nuclei to the build-up of the matter that ended up in the Solar System.
The elemental and isotopic composition of the Sun has been used as one of the
fundamental benchmarks for GCE models because it is very well known, thanks both to
spectroscopic observations interpreted using sophisticated models of the atmosphere of the Sun \cite{asplund09}, and to
laboratory analyses of pristine meteorites \cite{lodders09,lodders10}. Specifically, GCE models are required to match the Sun's
composition for stars born at the time (4.6 Gyr ago) and place (roughly 8 kpc from the galactic centre, under the assumption that the Sun did not migrate from its birth place) when and where the Sun was born. 

The end of the GCE contribution is marked by the incorporation of such presolar matter into a (G)MC. At this point in time the second phase of the presolar history of Solar System matter begins: its residence time in the stellar nursery where it was born. This phase lasted of the order of few to tens of Myr, i.e., roughly three to four orders of magnitude less than the GCE timescale. In relation to the investigation of SLRs in the ESS, the time when such an incorporation occurred has been referred to as the isolation time ($T_{\rm isolation}$). The reason is that the mixing between material in the hotter ISM and in the colder (G)MC is relatively slow, i.e., the time scale to achieve complete mixing is long ($\sim$100 Myr \cite{deavillez02}), thus, during the isolation time the presolar matter was isolated from stellar contributions in the GCE regime. In other words, $T_{\rm isolation}$ is the time the ESS matter spent inside a (G)MC before the formation of the Sun, isolated from the evolution of the ISM matter driven by GCE. It can also be described as the time interval between the birth of the parent (G)MC and the birth of the Sun itself, and called  an ``incubation'' time. During $T_{\rm isolation}$ a number of SLRs were only affected by radioactive decay, which thus can be used as a clock to measure $T_{\rm isolation}$ (as will be presented in detail in Sec.~\ref{sec:times}). This method gives us the most accurate way to investigate the lifetime of the specific (G)MC where the Sun was born. As mentioned above, molecular clouds are observed to live between a few to a few tens of Myr, probably depending on their size and mass, however, we do not know 
which side of this range is applicable to the particular case of the Sun. 

It is important to highlight here the difference between stable nuclei and SLRs in the context of the build-up of Solar System matter. In relation to stable nuclei, GCE is the most significant process and the contributions from all previous generations of stars count, given that the abundances of these nuclei continue to increase as the Galaxy evolves. Furthermore, for stable nuclei potential additions from
one or a few more short-lived, massive stars within the (G)MC or the stellar cluster where the Sun was born would not have
made a noticeable difference since their abundances produced by the GCE in the ISM at the time and place of the birth
of the Sun were already relatively high\footnote{Some care is still needed in specific cases, e.g., if the ESS was polluted by a nearby star
or supernova, this could have affected the O isotopic ratios to the level of
per mil variations, which is within the resolution of measurements of
meteoritic samples \cite{gounelle07,lugaro12a}.}. Long-lived radioactive nuclei such as Th and U
behave in this respect in a very similar way to stable nuclei, while the situation for
SLRs is highly dependent on their specific half-lives. The longer the half-life, the more the
abundance of the nucleus in the ESS carries the imprint of its production by GCE, as in the
case of stable nuclei. The shorter the half-life, the more the abundance of the SLR
nucleus in the ESS carries the imprint of its production within the (G)MC or stellar cluster
where the Sun was born, simply because the isolation time is more likely to have erased its GCE contribution. In this case the SLR cannot be used as a clock to measure the isolation time, but acts instead as an indicator for the circumstances 
of the birth of the Sun within its stellar nursery, i.e., it indicates that the Sun was born close enough in time and space to a production event. The SRLs are thus the fingerprint of the stellar
objects that populated the environment where the Sun's birth happened. There are many different
scenarios and hypotheses on the circumstances and the environment of the birth
of the Sun based on such shortest-lived SLR fingerprints, particularly \iso{26}Al; they
are discussed in more detail in Sec.~\ref{sec:birth}.

\subsection{Radioactivity and habitability}
\label{sec:habit}

\begin{figure}[tb]
\begin{center}
\begin{minipage}[t]{15 cm}
\includegraphics[width=15cm]{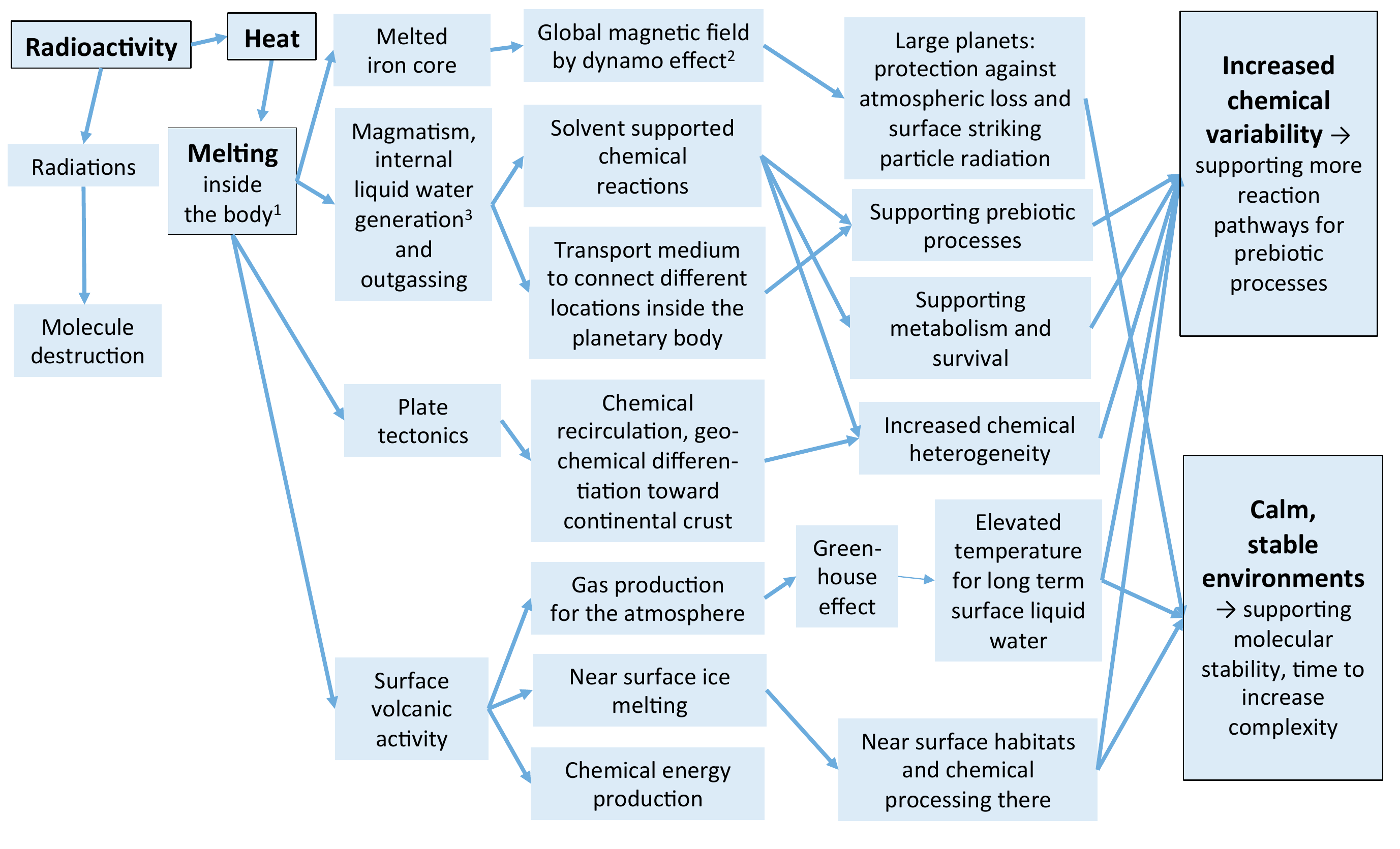}
\caption{Possible connections between radioactivity and various factors that influence habitability of solid planetary bodies. Notes according to numbers: 1.  Ancient melting was also supported by other processes, including exothermic heat generated by serpentinization, i.e, the addition of water into the crystal structure of minerals. 2. A very early differentiated iron core is expected to have been present in many planetesimals, based on the paleomagnetic signature of internal magnetic dynamos even in carbonaceous chondrites, where the primitive chondritic material accumulated on the surface of an already differentiated core. 3. Magmatic activity and  internal liquid water generation were supported by the SLRs only in the first periods. Later on, long-lived radionuclides became the more important to enable continued activity.  \label{fig:habit}}
\end{minipage}
\end{center}
\end{figure}


Here, we briefly list connections and interactions related to how radioactivity may influence habitability. It does so in complex ways, with many lines of often intricate interaction between different factors. These factors can be grouped into two classes. 
Direct influences of radioactivity include internal heat generation of planetary bodies.
Beside the radioactive sources, the relict heat from the accretion process contributes significantly to the temperature of the mantle and the core. There is an ongoing debate on which factor dominates among these two on the Earth today \cite{herzberg10}. Some researchers consider them equally important \cite{jaupart15,andrault16}, however, in the longer term radioactive heat may dominate over accretionary heat. Measurements of geologically produced antineutrinos may help to settle this question, however, current uncertainties are very large \cite{araki05}. Volcanism is a possible consequence of internal heat generation, which may then be responsible for melting of ice, additions to the atmosphere (which in turn may lead to protection of the planetary surface from UV irradiation), increasing chemical heterogeneity, and generation of additional chemical energy sources from heat driven chemical reactions. Plate tectonics is also driven by internal heat and supports planetary scale chemical circulation, while increasing geochemical diversity by producing granitic crust and continents (on the Earth). Weathering then produces an even wider variety of materials that differ from those that would have been present if the whole surface were covered by water. Sufficient rates of internal heat production can also lead to the formation of a (partially) molten iron core, which can generate a global magnetic field on a rotating planet, which in turn protects the atmosphere against erosion by stellar wind and the surface against ionising charged particle bombardment.

Indirect influences are related to the formation of molecules essential for life. Radioactivity affects the characteristics of the environment, which in turn determines whether such molecules could form or not (because of temperature, volcanic activity, and presence or absence of liquid water). Not only the organic materials produced by chemical reactions matter here, but more indirect effects are also important, like the generation of phyllosilicate minerals by the action of water. Phyllosilicates help in molecular polymerisation and increase the stability of organic molecules. Such molecules might then support prebiotic processes, and the origin of life as well. They could also support the maintenance of life after its origin by serving as nutrients and building blocks for the already emerged life.


The effects listed above and their consequences are linked to each other, creating a complex system, which influences habitability in a variety of ways. The possible connections between radioactivity and various factors that influence habitability of solid planetary bodies are shown in Fig.~\ref{fig:habit}. Note that the figure is applicable only to bodies with a solid surface (including rocky planets, icy moons or asteroids, comets), while gaseous planets and brown dwarfs are different cases. Radioactive heat sources are considered, but other heat sources could be also present or even dominate over radioactivity, and may produce similar consequences as those that are listed here. Two main causal branches are visible: the heat production from radioactivity that has far reaching consequences, and the radiation itself that has a smaller number of consequences with less complexity. For example, radioactive heat driven melting causes differentiation of a planetary body, which in turn affects volcanic activity, material circulation, as well as chemical and atmospheric characteristics. In this respect it is relevant to note that the duration of radioactivity as well as its level differ between shorter- and longer-lived radionuclides. While short decay times lead to early activity on a planetary body, the presence of radionuclides with longer decay time may be essential in supporting long duration habitability – however, here the thermal budget of the body also matters: losing the continuously generated heat too efficiently may keep the given body in a frozen state.

Without the heat generated by radioactivity the conditions for habitability would be quite different and often much less favourable. In the cases where heating leads to internal melting, differentiation of the planetary body interior could contribute to liquid water and magnetic field generation, volcanic activity, as well as contribute to the generation of an atmosphere, and in general result in mineral diversity, where the latter may have a complex but poorly known connection with habitability \cite{Hazen2008b}. Without such radioactive heat-generated differentiation and melting, liquid iron cores may be much less abundant among terrestrial planets, allowing - due to lack of a magnetic field - cosmic radiation to bombard the surface \cite{Lazio2016}. The bombardment by cosmic ray particles probably reduces the chance of the origin of life on the surface and also the survival of organisms there. While in the subsurface region both origin and survival of life is possible even in such a case \cite{Fisk1999}, subsurface niches seem unlikely to be sufficient for supporting the emergence of more advanced life, and radiation in such cases does not allow surface organisms to exploit stellar irradiation -- which is a much larger energy source than subsurface chemical sources, therefore opening the way for faster evolution \cite{Trevors2002}. In the case of a missing magnetic field, habitability may still be possible but in restricted and limited form. 

Within the context of this paper we will mostly discuss the effects on habitability of the specific case of SLRs as heat sources in the ESS (Sec.~\ref{sec:26Alhabit}). We will see that the most interesting case is that of \iso{26}Al ($T_{\rm 1/2}$=0.7 Myr), simply because this SLR was so abundant. Potentially, also \iso{60}Fe and \iso{36}Cl can be of interest as heat sources in the ESS, depending on their initial abundances, which for the ESS are still debated (see Secs.~\ref{sec:36Cl} and \ref{sec:60Fe}). In relation to the case of longer-lived radionuclides as long-term sources of heating, \iso{232}Th (T$_{\rm 1/2}$=14 Gyr), \iso{235}U (T$_{\rm 1/2}$=0.703 Gyr), \iso{238}U (T$_{\rm 1/2}$=4.5 Gyr), and \iso{40}K (T$_{\rm 1/2}$=1.2 Gyr) are still alive today in the Solar System and are of paramount importance in relation to the internal energy budget of the Earth. As mentioned above, their decay currently provides possibly half of the total heat budget of the solid Earth (the other half being the primordial heat left over from its formation \cite{turcotte02}), with implications on its surface habitability. Th and U are actinides produced by the $r$ process, while \iso{40}K is produced together with the two stable isotopes of K (at masses 39 and 41) in CCSNe via O burning. Interestingly, from stellar observations and GCE models it is possible to determine the abundances of some of these isotopes in extra-solar planets. 

Since U and Th are refractory, we can assume that their abundances, relatively to Si, observed or predicted in stars should be close to the abundances present in rocky planets around the stars.
Recent observations of solar twins, with and without planets, have shown that most of these stars have larger Th abundances than the Sun \cite{unterborn15}, with a spread of almost a factor of 3. This difference probably has important implications on the habitability of extra-solar terrestrial-like planets.
Galactic chemical evolution modelling of the elements produced by the $r$ process are needed to establish the reason for the spread in the abundance of Th. Because of its long half-life, Th can almost be treated as a stable isotope with respect to GCE and as such its abundance should be intrinsically less prone to inhomogeneities in the ISM as opposed to the SLRs (Sec.~\ref{sec:GCE}). However, since it is likely that the creation of the $r$-process elements occurs in rare nucleosynthetic events associated with NSMs \cite{cote18}, it seems qualitatively feasible that the abundances of $r$-process elements may show a relatively large spread, even for stars very similar in age and metal content (i.e., the recurrence time of the additions to a particular parcel of the ISM may be actually comparable with the half-life, see Sec.~\ref{sec:GCE}). This was already demonstrated using models of inhomogeneous GCE for the typical $r$-process element Eu \cite{wehmeyer15}. More observations of Th and U in stars with planets should be feasible in the future and will provide more information on the internal heat budget from long-lived radioactivity in extra-solar terrestrial planets. 

The abundance of \iso{40}K, on the other hand, cannot be disentangled from stellar spectra from that of the much more abundant \iso{39}K. In this case, we will need to rely on GCE models to predict its abundance in  stars. We may expect a smaller spread than in the case of Th and U since its CCSNe sources are much more common than NSMs in the Galaxy. A further problem, however, is that K is moderately volatile (with a 50\% condensation temperature in the ESS of 1006 K as compared to 1610 for U \cite{lodders03}), and presents abundance variations in the Solar System, for example, between the Earth, Mars, and chondrites. In this case, model predictions for stars cannot be directly translated into predictions for the planets around them. 

\begin{figure}[tb]
\begin{center}
\begin{minipage}[t]{16.5 cm}
\includegraphics[width=8.5cm,angle=0]{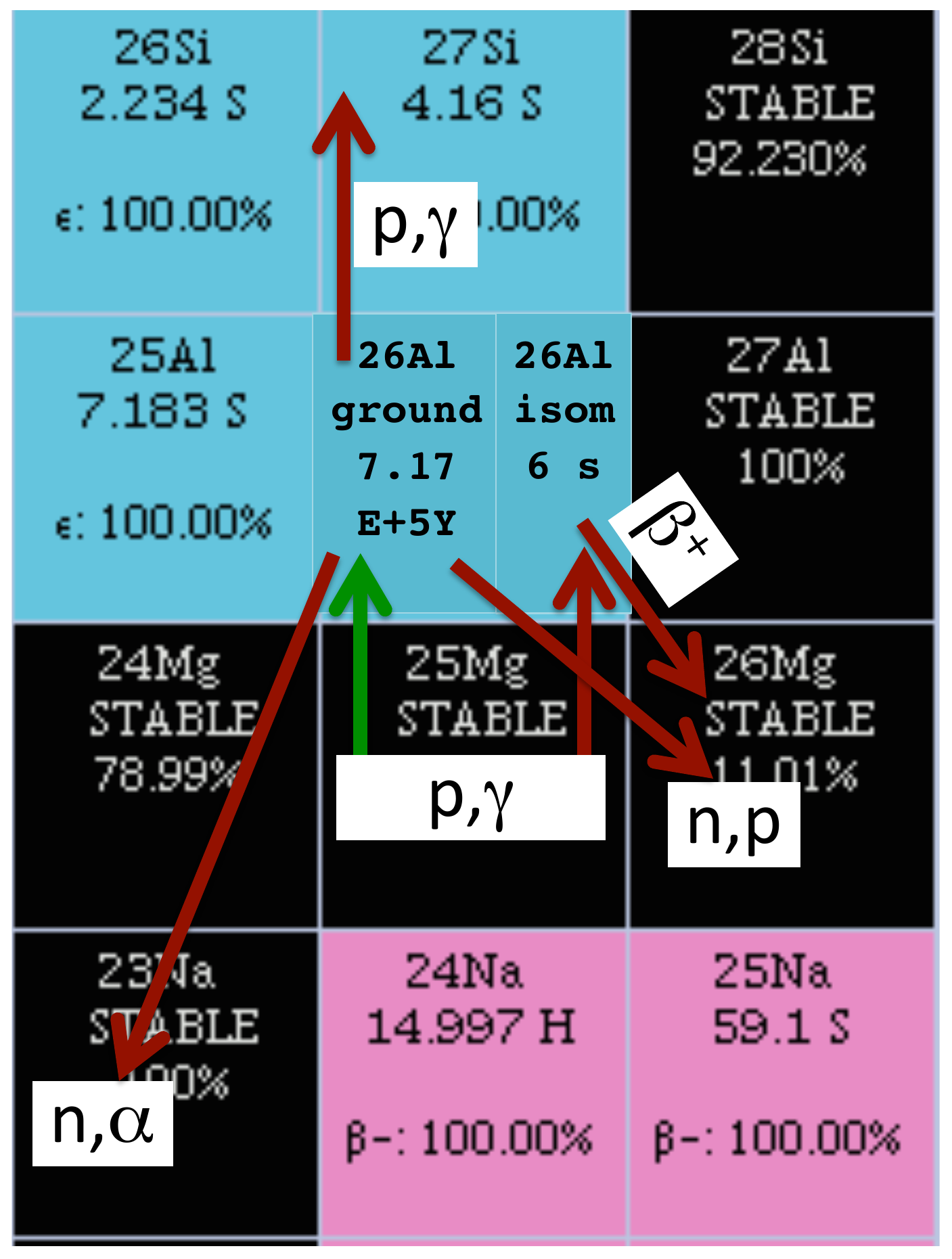} 
\includegraphics[width=8.5cm,angle=0]{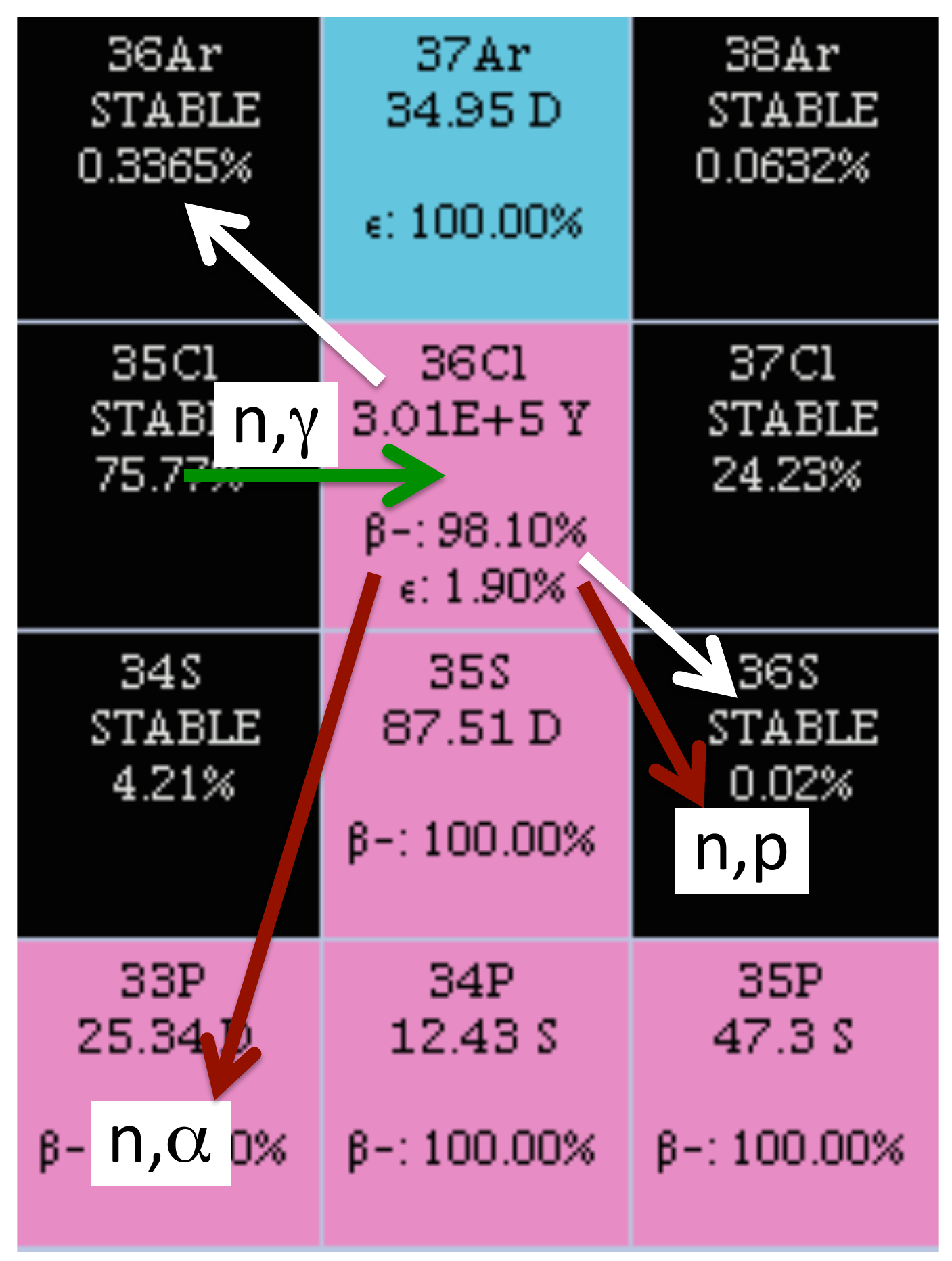} \\
\includegraphics[width=8.5cm,angle=0]{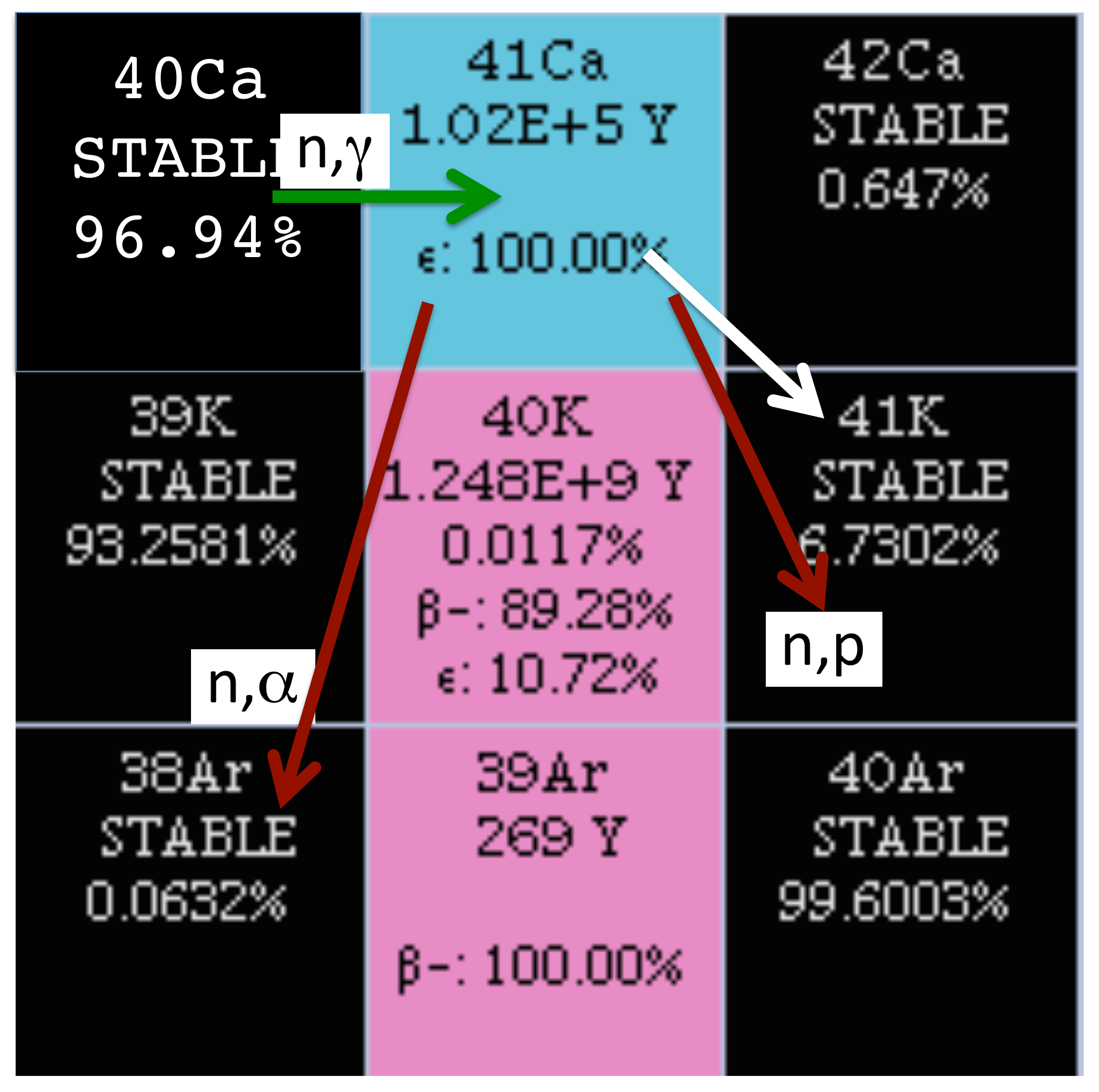} 
\includegraphics[width=8.5cm,angle=0]{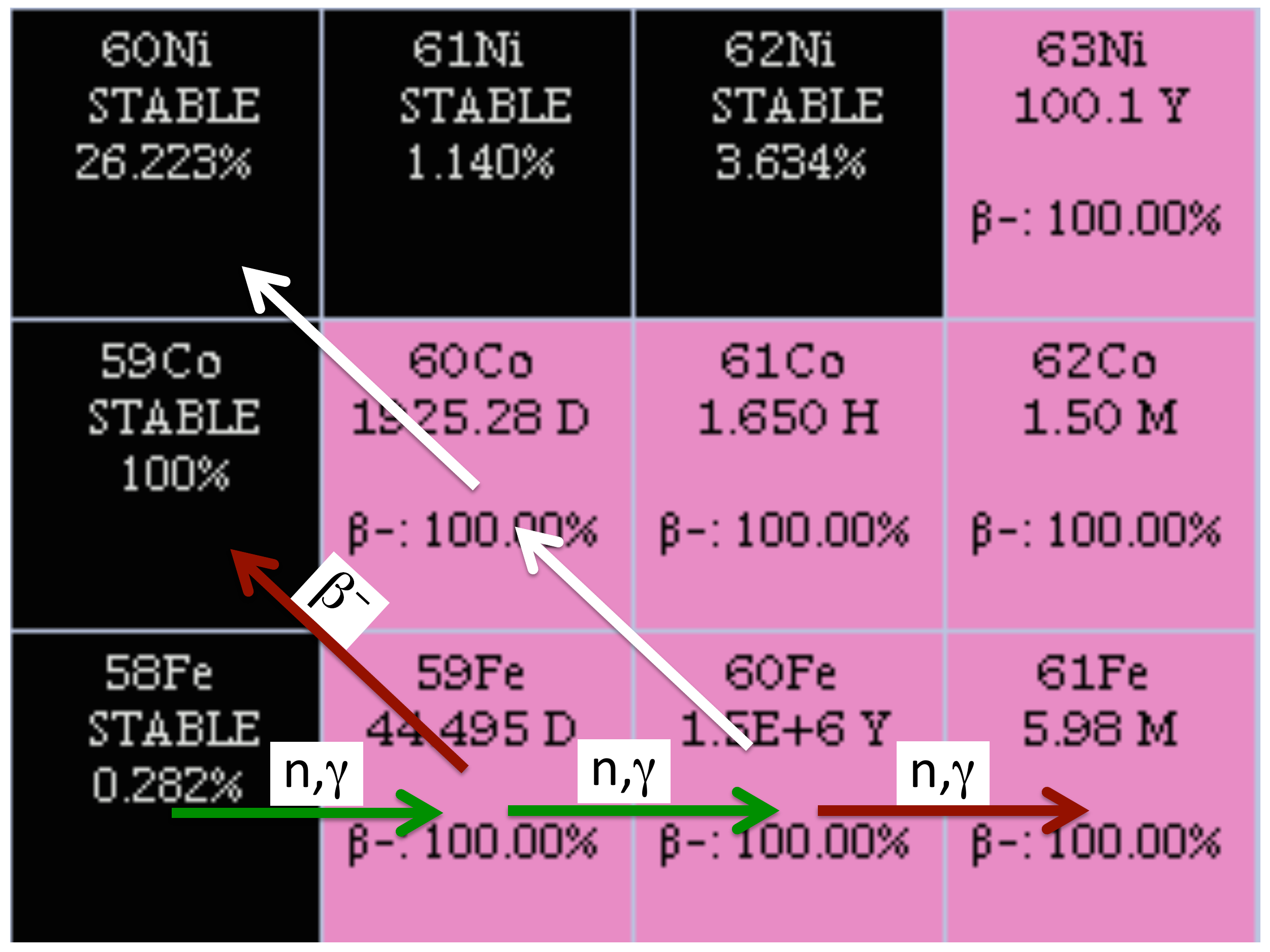} \\
\end{minipage}
\begin{minipage}[t]{16.5 cm}
\caption{Sections of the nuclide chart (modified from those obtained from the National Nuclear Data Center web site, www.nndc.bnl.gov) illustrating the nuclear-reaction paths favouring (green arrows) or inhibiting (red arrows) the production in stellar objects of four SLRs lighter and up to Fe: \iso{26}Al, \iso{36}Cl, \iso{41}Ca, and \iso{60}Fe. The remaining three SLR lighter than Fe are not plotted because \iso{7}Be and \iso{10}Be are not produced in stars, while \iso{53}Mn is produced by NSE, rather than by a defined nuclear reaction pathway. White arrows represent the radiogenic decay of each SLR, except for  \iso{26}Al, where the decay to \iso{26}Mg is not overlaid onto the nuclide chart to avoid the plot being too busy.
\label{fig:NCn1}}
\end{minipage}
\end{center}
\end{figure}

\begin{figure}[tb]
\begin{center}
\begin{minipage}[t]{16.5 cm}
\includegraphics[width=8.5cm,angle=0]{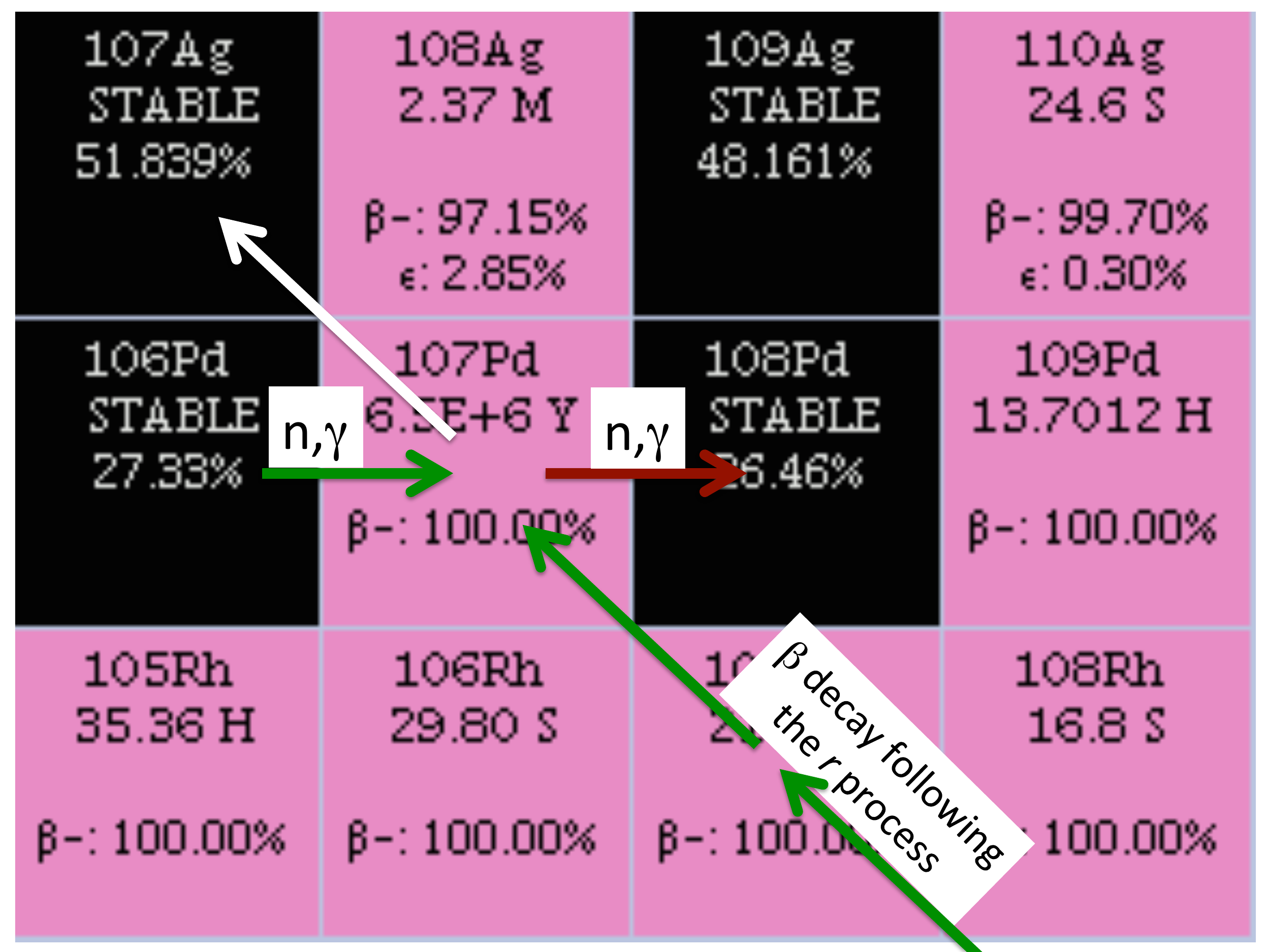} 
\includegraphics[width=8.5cm,angle=0]{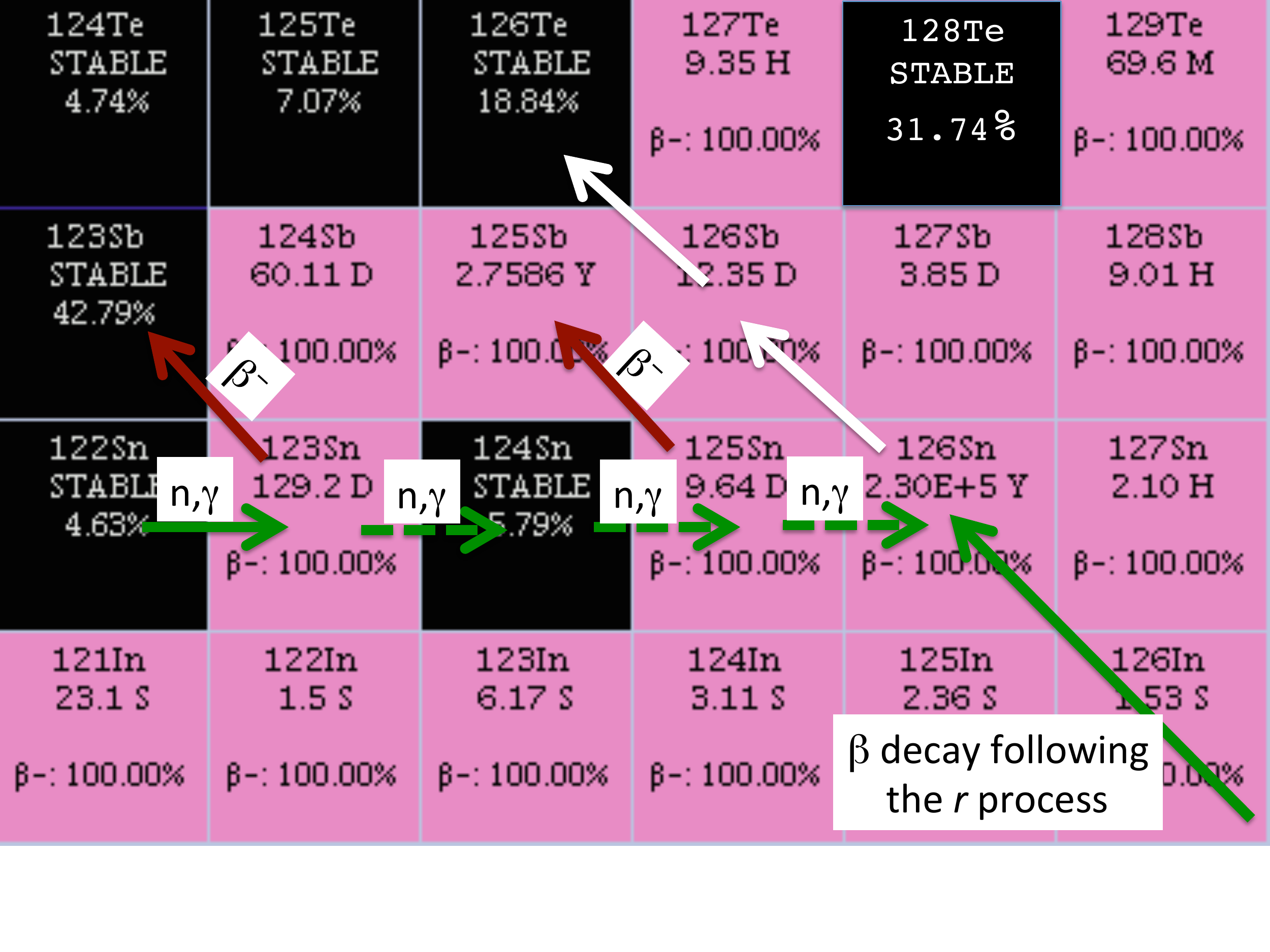} \\
\includegraphics[width=8.5cm,angle=0]{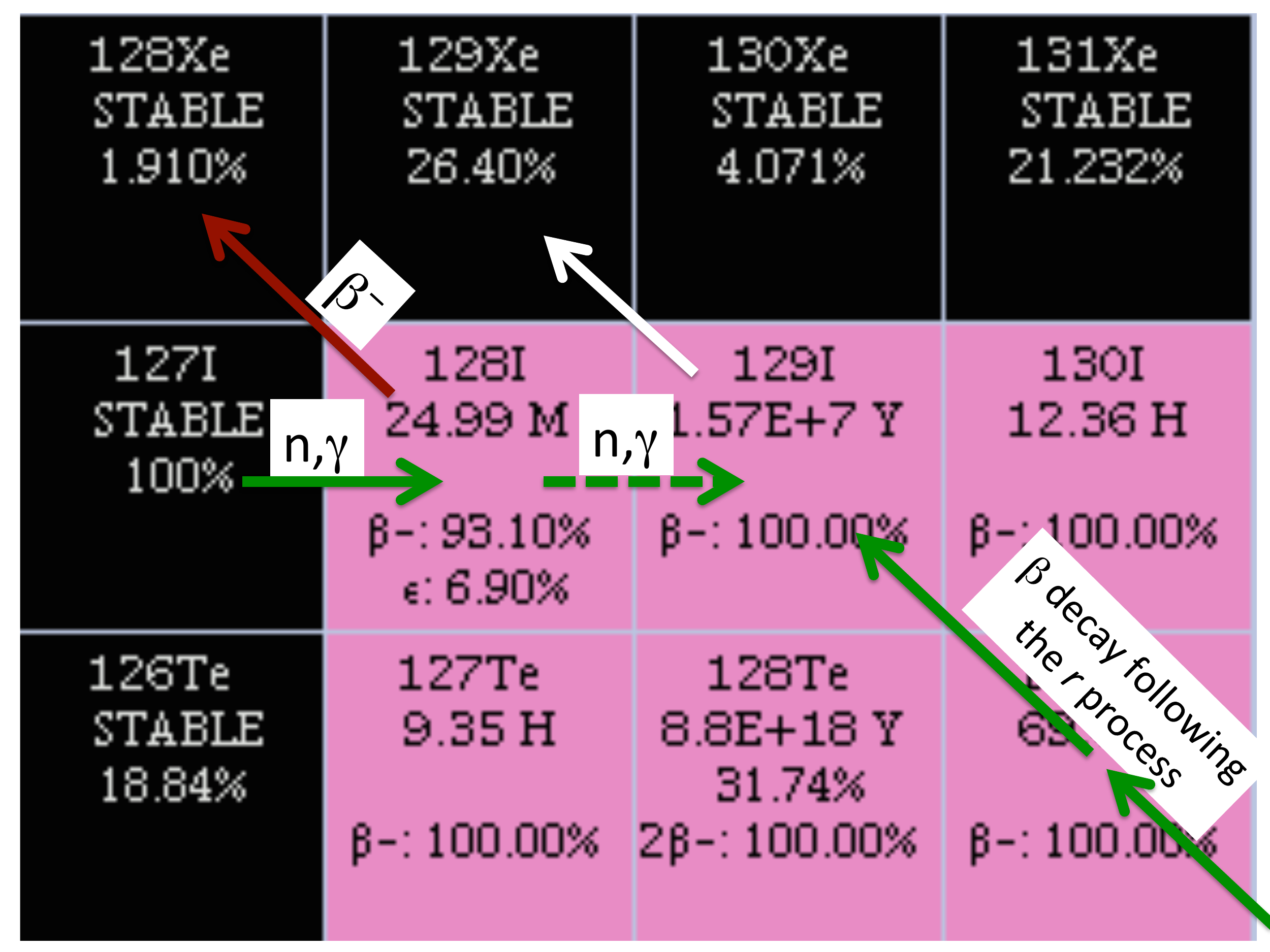} 
\includegraphics[width=8.5cm,angle=0]{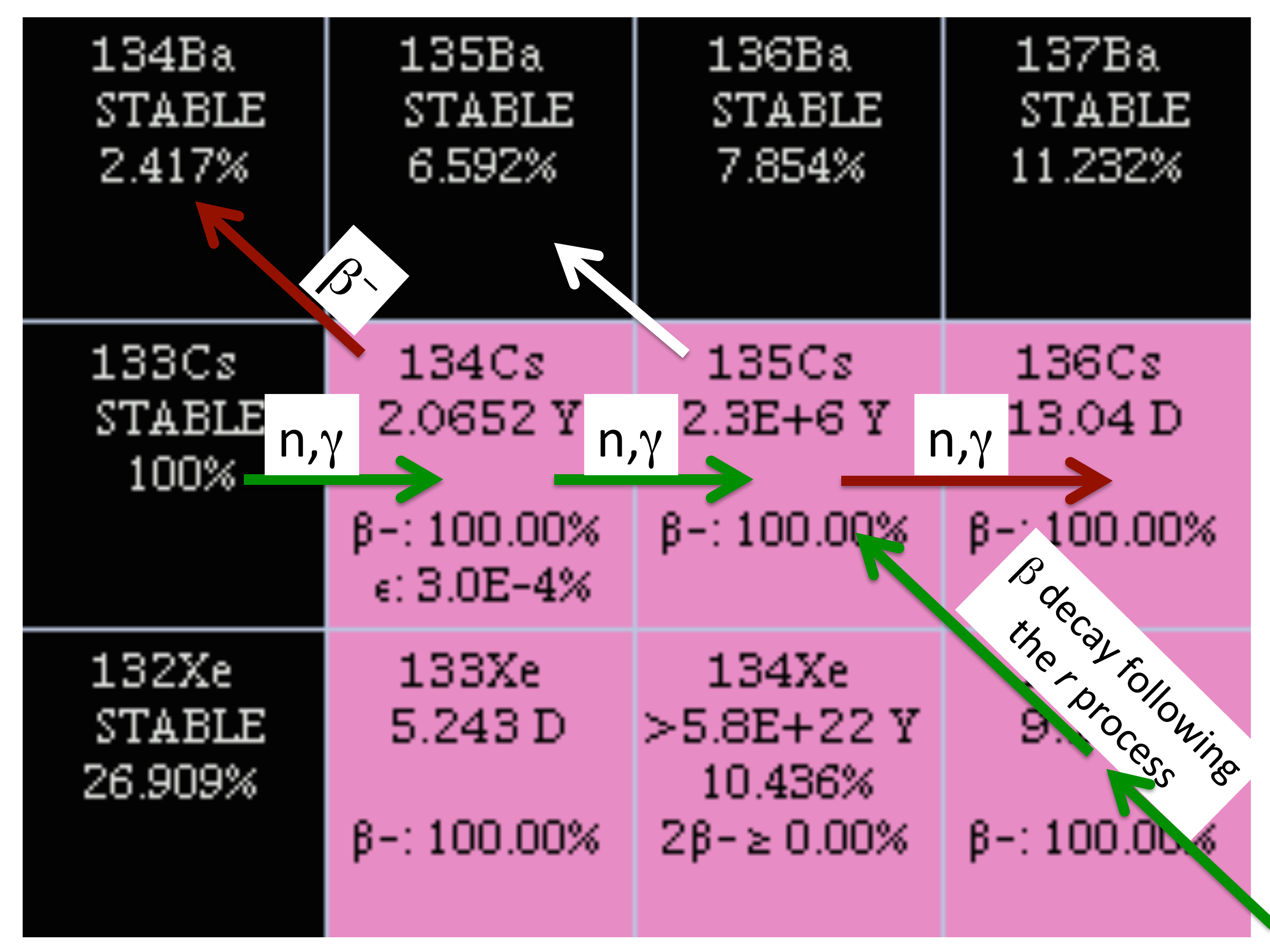} \\
\includegraphics[width=8.5cm,angle=0]{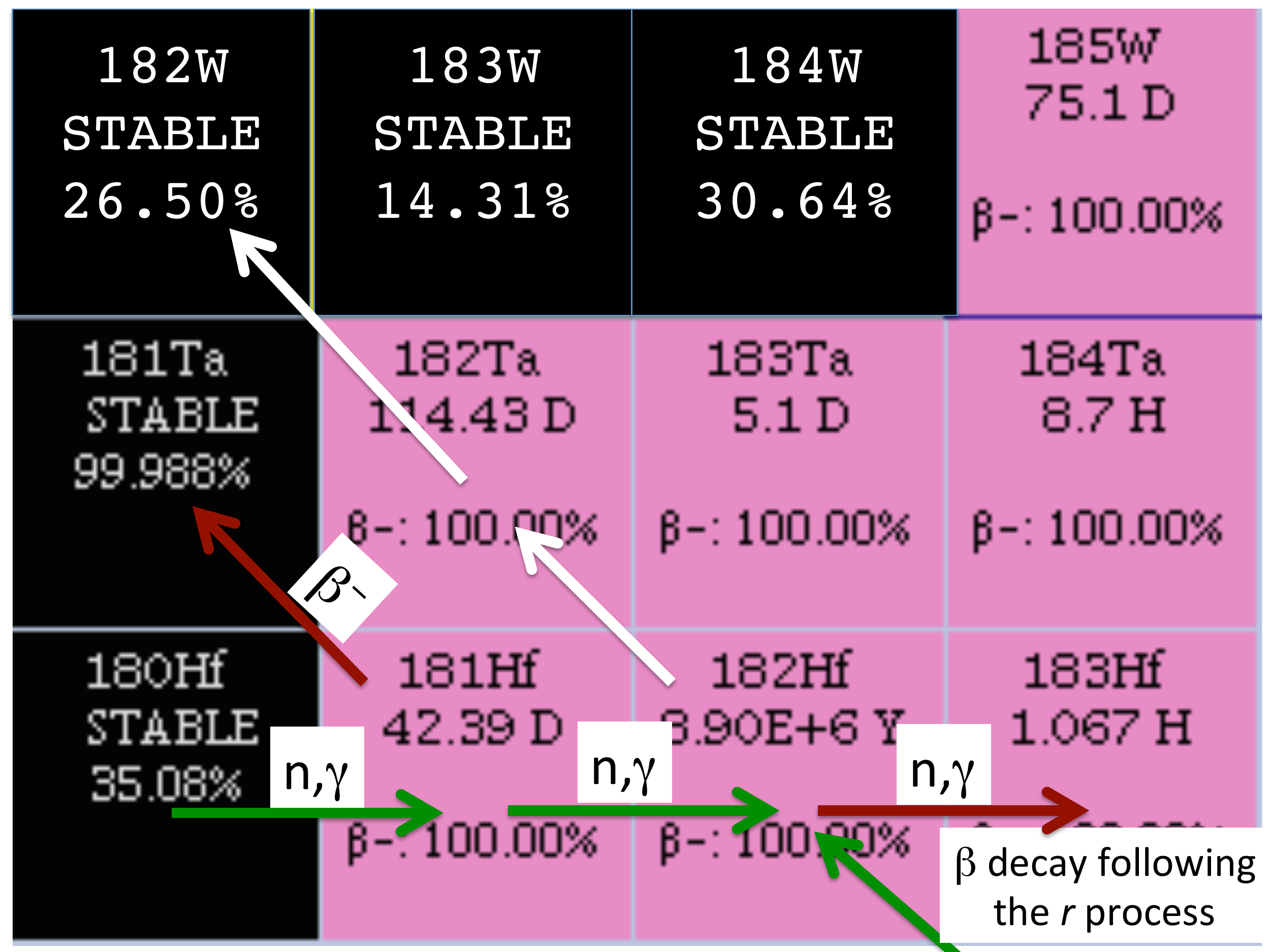} 
\includegraphics[width=8.5cm,angle=0]{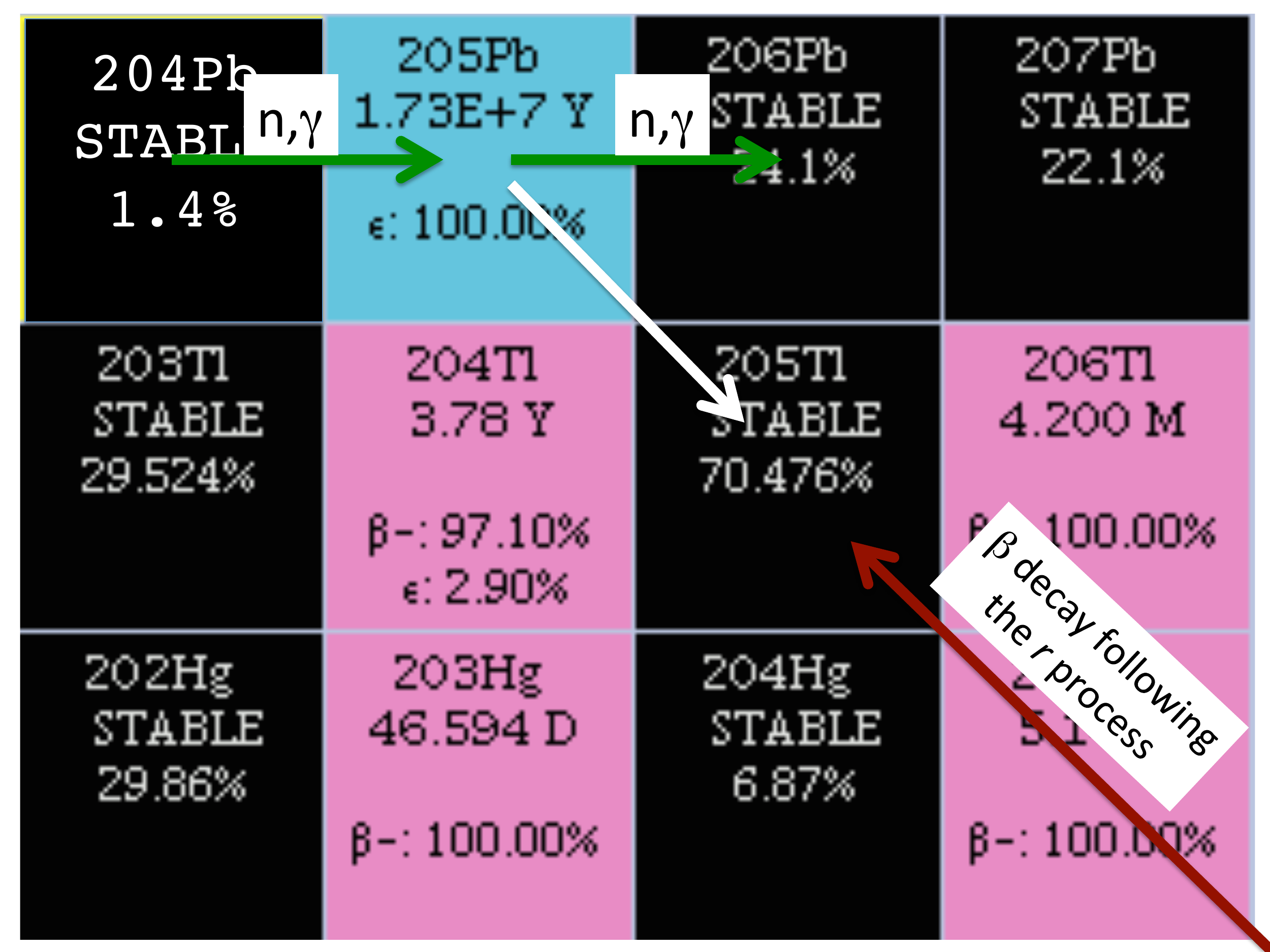} \\
\end{minipage}
\begin{minipage}[t]{16.5 cm}
\caption{Sections of the nuclide chart (modified from those obtained from the National Nuclear Data Center web site, www.nndc.bnl.gov) illustrating the nuclear-reaction paths favouring (green arrows) or inhibiting (red arrows) the production of the SLRs heavier than iron whose cosmic abundances are attributed to neutron-capture processes (in order of increasing mass: \iso{107}Pd, \iso{126}Sn, \iso{129}I, \iso{135}Cs, \iso{182}Hf, and \iso{205}Pb). The actinides \iso{244}Pu and \iso{247}Cm are not plotted, since their production is solely due to the $r$ process. Solid green arrows represent major production paths, while the dotted green arrows represent minor production paths. White arrows represent the radiogenic decay of each SLR. 
 \label{fig:NCn2}}
\end{minipage}
\end{center}
\end{figure}

\section{The SLR variety: ESS abundances and stellar origins}
\label{sec:list}

\begin{table}
\begin{center}
 \centering
 \caption{List of stellar nucleosynthesis sites and the nucleosynthetic processes occurring within them that are responsible for the production of the SLRs and stable reference isotopes listed in Column 3. Column 4 indicates if the site of production is important in terms of GCE ({\bf M}=Major) or not ($m$=minor); {\bf M}/$m$ indicates that it is still debated whether the site is major or minor. Indicative references are listed in Column 5.}
\label{table:stars}
\vspace{0.3cm}
 \begin{tabular}{lllll}
 \hline
Stellar site & Process & Products & Relevance & Ref. \\
\hline
Low-mass AGBs & $s$ process & \iso{107}Pd, \iso{108}Pd & {\bf M} & \cite{wasserburg06,lugaro14} \\
                &  & \iso{135}Cs, \iso{133}Cs & {\bf M} & \\
                &  & \iso{182}Hf, \iso{180}Hf & {\bf M} & \\
                &  & \iso{205}Pb, \iso{204}Pb & {\bf M} & \\
Massive and & p captures & \iso{26}Al & $m$ & \cite{trigo09,lugaro12a,lugaro14,wasserburg17} \\
Super-AGBs & n captures & \iso{41}Ca, \iso{36}Cl, \iso{60}Fe & $m$ &  \\
& $s$ process & \iso{107}Pd, \iso{135}Cs, \iso{182}Hf & $m$ &  \\
WR stars & p captures & \iso{26}Al & {\bf M} & \cite{arnould97,arnould06} \\
         & n captures & \iso{41}Ca, \iso{36}Cl & $m$ &  \\
         & n captures & \iso{97}Tc, \iso{107}Pd, \iso{135}Cs, \iso{205}Pb & $m$ &  \\
CCSNs & p captures+explosive & \iso{26}Al, \iso{27}Al & {\bf M} & \cite{limongi06} \\
      & n captures & \iso{60}Fe & {\bf M} & \cite{limongi06} \\
      & n captures & \iso{36}Cl, \iso{41}Ca & {\bf M} & \cite{takigawa08,lugaro14} \\
      & C/Ne/O burning & \iso{35}Cl, \iso{40}Ca & {\bf M} & \cite{rauscher02} \\
      & NSE & \iso{53}Mn, \iso{55}Mn, \iso{56}Fe & {\bf M}/$m^a$ & \cite{rauscher02} \\
      & n captures & \iso{107}Pd, \iso{126}Sn, \iso{135}Cs & $m$ & \cite{meyer00} \\
      &            & \iso{129}I, \iso{182}Hf, \iso{205}Pb & $m$ & \\
      & $\alpha$-rich freezeout & \iso{92}Nb, \iso{92}Mo, \iso{97}Tc, \iso{98}Tc & {\bf M}/$m$ & \cite{lugaro16} \\
      & $\gamma$ process & \iso{144}Sm, \iso{146}Sm & {\bf M}/$m$ & \cite{rauscher13,lugaro16} \\
      & $\nu$ process & \iso{10}Be, \iso{92}Nb & $m$ & \cite{banerjee16,hayakawa13} \\
SNIa & NSE & \iso{53}Mn, \iso{55}Mn, \iso{56}Fe & {\bf M} & \cite{travaglio04} \\
            & $\gamma$ process & \iso{92}Nb, \iso{93}Nb, \iso{146}Sm, \iso{144}Sm & {\bf M}/$m$ & \cite{travaglio14} \\
            &  & \iso{97}Tc, \iso{98}Tc, \iso{98}Ru & {\bf M}/$m$ & \\
NSM/special CCSN & $r$ process & \iso{107}Pd, \iso{108}Pd, \iso{126}Sn, \iso{124}Sn & {\bf M} & \cite{bisterzo11}$^b$\\
& & \iso{135}Cs, \iso{133}Cs, \iso{129}I, \iso{127}I & {\bf M} & \\
& & \iso{182}Hf, \iso{180}Hf & {\bf M} & \\
& & \iso{247}Cm, \iso{235}U, \iso{244}Pu, \iso{238}U & {\bf M} & \cite{goriely01,goriely16} \\ 
novae ejecta & p captures & \iso{26}Al & $m$ & \cite{jose07}\\
CRs & non-thermal & \iso{7}Be, \iso{10}Be, \iso{9}Be & {\bf M} & \cite{tatischeff14} \\
& & \iso{26}Al, \iso{41}Ca, \iso{36}Cl, \iso{55}Mn & $m$ & \cite{gounelle06} \\
\hline
 \end{tabular}
 \end{center}
$^a$The current understanding is that roughly 1/3 of the abundances of the Fe-peak elements in the Galaxy are produced by CCSNe, with the rest coming at later times from SNIa.
$^b$Abundances to be derived using the $s$-process predictions provided in the reference via the $r$-residual method, where the $r$-process abundance is given by the Solar System abundance minus the $s$-process abundance.
\end{table}

The possible nucleosynthetic production sites for the SLRs and their stable reference isotopes are summarised in Table~\ref{table:stars}. All the processes listed in the table occur in stars, except for non-thermal nucleosynthesis. As mentioned above, spallation typically occurs in the ISM, however, it could also have had an important role in the ESS, with CRs coming from the Galactic background \cite{desch04}, the young, active Sun \cite{gounelle06}, or resulting from the interaction of one or more nearby CCSN remnant(s) with the ISM \cite{tatischeff10}. A number of SLRs can be produced by this process, clearly \iso{7}Be and \iso{10}Be, but also \iso{26}Al, \iso{36}Cl, \iso{41}Ca, and \iso{53}Mn. However, there are several arguments against a major contribution the ESS as models have difficulties in providing a self-consistent solution that matches the abundances of all these isotopes \cite{desch10}. Another difficulty is that a homogeneous distribution of the SLRs is not expected in this method of production, given the variability of the CR flux, but it is observed for \iso{26}Al and \iso{53}Mn. Furthermore, for the widely discussed model of irradiation by cosmic rays from the young Sun, it appears that not enough energy was available to produce all the \iso{26}Al if this SLR was homogeneously distributed throughout the ESS at the abundance level listed in Table~\ref{table:SLRs} \cite{duprat07}. Experimental data for the relevant nuclear reaction rates involved are scarce, but we note that there are recent new data on the \iso{33}S($\alpha$,p)\iso{36}Cl \cite{anderson17} and   \iso{24}Mg(\iso{3}He,p)\iso{26}Al reactions \cite{fitoussi08}, which are important in the context of solar cosmic ray irradiation. The latter further disfavours this production channel for \iso{26}Al. 

Column 4 of Table~\ref{table:stars} clarifies if the listed site is a major ({\bf M}) or a minor ({\it m}) site of production of the cosmic abundances of the listed isotopes. If the site is major, it means that not only the ratio of the SLR to the stable isotope of reference is significant, but also that the absolute abundance produced is large enough to impact the evolution of the SLR abundance in the Galaxy. To measure this, one can compare the mass fraction of the stable isotope in the stellar ejecta (i.e., the mass expelled in form of the given isotope divided by the total mass lost) to its mass fraction in the Solar System abundance distribution. The ratios of these two numbers can be referred to as ``production factors'', and values roughly above 10 are needed to make the site under consideration a potentially important site.
With respect to the presence of SLRs on the ESS, the distinction between major versus minor site is crucial. 
Major sites of production must be included in the analysis of the evolution of SLRs in the Galaxy described in Sec.~\ref{sec:GCE} and they affect the use of SLRs as clocks to measure the isolation time. Minor sites of production are irrelevant in this context. On the other hand, if we consider the environment of the birth of the Sun, and a potential nearby stellar source of SLRs, then also minor production sites could have played a role in polluting the ESS with SLRs. In this case the stellar yields are diluted according to the distance from the star to the Sun, and given that such local sources are supposed to have been relatively close to the early Sun ($\sim$ 0.5 - 5 pc, see Sec.~\ref{sec:birth}), pollution even from a source that provides a relatively low absolute abundance can result in noticeable variations.

In Column 2 we also list a process referred to as ``n captures'', which was not included in the list of the traditional nucleosynthesis processes described in Sec.~\ref{sec:stars}. We use this label when we are in the context of neutron capture reactions, but the $s$- or the $r$-process labels do not apply. There are two possibilities for this: first, in relation to the SLRs up to Fe, \iso{36}Cl, \iso{41}Ca, and \iso{60}Fe. The traditional $s$ or the $r$ processes were introduced specifically for the production of the elements heavier than Fe, hence, it is not strictly appropriate to use these terms for neutron-capture reaction that produce nuclei up to Fe. The second instance involves the production of SLRs heavier than Fe, however, the neutron-capture process does not produce a significant abundance of the elements heavier than iron. Only a small number of neutrons are released in these cases, and the production of SLRs relies on the original presence of stable nuclei belonging to the same element. In line with this, the n-capture process in the case of SLRs heavier then Fe is always indicated as a minor ({\it m}) site of production in the table.

Expanding on the information given in Tables~\ref{table:SLRs} and \ref{table:stars}, in the 
following subsections we group the SLRs according to their nucleosynthetic production processes and for each of them we discuss in more detail their ESS abundances and nucleosynthetic origins. 

\subsection{\iso{10}Be and \iso{7}Be}
\label{sec:10Be}

As shown in Table~\ref{table:SLRs}, there is a large range of values observed for the abundance of \iso{10}Be in the ESS, and no compelling evidence exists for choosing one specific value over the others \cite{chaussidon06}. The different values probably do not indicate time differences, but are the result of an inhomogeneous distribution. This in line with production by CR irradiation, since the particle flux driving the spallation reactions is likely to vary with time and location within the disk. Furthermore, the \iso{10}Be abundance does not correlate with that of  \iso{26}Al. This is expected if they were produced by different processes: \iso{10}Be via CR irradiation and \iso{26}Al via stellar nucleosynthesis. 

Data reported for FUN-CAIs show \iso{10}Be/\iso{9}Be in the range 3-4 $\times 10^{-4}$ \cite{wielandt12}. FUN-CAIs show large mass-dependent fractionation effects and have much larger anomalies in stable isotopes than other CAIs (hence the name FUN, which stands for Fractionated and Unknown Nuclear anomalies). FUN-CAIs also show much lower abundances of \iso{26}Al than the value given in Table~\ref{table:SLRs}. Due to these properties, they are believed to be among the oldest CAIs, formed before \iso{26}Al was injected or homogenised in the disk, and before the dust carriers of the stable isotope anomalies were  efficiently homogenised. Hence, the \iso{10}Be variations shown by the FUN-CAIs may be taken as the range of values produced by CRs that did not originate from the Sun, but from the galactic background or from the interaction with one or more nearby CCSN remnants \cite{tatischeff14}. An alternative explanation for this baseline value was proposed by considering a model of a CCSN with low mass and explosion energy, which predicts production of \iso{10}Be via neutrino interactions \cite{banerjee16}. This CCSN on the other hand does not produce enough \iso{26}Al to explain the ESS data, so a different source must be invoked for this SLR. The highest value of 104 $\times 10^{-4}$ for \iso{10}Be/\iso{9}Be  was observed in one specific CAI only \cite{gounelle13}. This value must clearly be due to irradiation within the ESS. All the other CAIs are only moderately higher than the baseline value. 

The case of \iso{7}Be, which stands out from all the other SLRs for having an extremely short half-life of 53 days, is controversial: only one measurement (given at 2$\sigma$ in Table~\ref{table:SLRs}) is available and awaits confirmation. While \iso{7}Be can be made in some stars on the pathway to \iso{7}Li production \cite{cameron71}, given its short half-life the only possible origin for a potential presence of \iso{7}Be in the ESS is that of solar CR irradiation. 

\subsection{\iso{26}Al}
\label{sec:26Al}

Aluminium-26 is probably the most famous SLR. Not only was it one of the first discovered to have been present with a high abundance in the ESS \cite{lee77}, but its presence was predicted more than two decades before its discovery on the basis of the need for a heat source in the early Solar System \cite{urey55}. The first hypothesis on the circumstances of the birth of the Sun, the collapse of the protosolar cloud triggered by a nearby CCSN, was also based on the discovery of \iso{26}Al \cite{cameron77}. Furthermore, for \iso{26}Al there is some consensus that its abundance in the ESS was homogeneously distributed \cite{villeneuve09}, and thus the value reported in Table~\ref{table:SLRs} is defined as a ``canonical'' value for the ESS \cite{jacobsen08}. This allows us to use the decay of \iso{26}Al as a sensitive chronometer for the very early history of the Solar System \cite{dauphas11}. 

However, some inhomogeneities exist also in case of this SLR. As mentioned above, FUN-CAIs are well known to contain \iso{26}Al in variable amounts (e.g., \cite{park17}). Another case are micro-corundum (Al$_2$O$_3$) grains extracted from meteorites, which represent very early Solar nebula condensates since corundum is one of the first minerals predicted to condense in a cooling gas of solar composition. These grains show a bimodal distribution in \iso{26}Al: half of the grains belong to a \iso{26}Al-rich population, with \iso{26}Al/\iso{27}Al close to the canonical value, and the other half belong to a population of \iso{26}Al-poor grains with more than 20 times lower ratios \cite{makide11}. Corundum-bearing CAIs also show large variations \cite{makide13}. Interestingly, the \iso{26}Al-rich and \iso{26}Al-poor grains show the same O isotopic composition, close to that of the Sun\footnote{The O isotopic composition of the Solar System is not uniform, with the Sun being more rich in \iso{16}O by 6\% with respect to planets and bulk meteoritic rocks. This difference is typically interpreted as the effect of self-shielding of CO molecules from UV radiation in the ESS, but the exact mechanism is a matter of debate, see  \cite{ireland12} for an accessible review.}, and typical of CAIs. This observation poses strong constraints on the origin of \iso{26}Al, since a successful pollution model should avoid predicting a correlation between the presence of \iso{26}Al and modification of the O isotopes \cite{gounelle07}.

Furthermore, it has also been pointed out that the  \iso{26}Al/\iso{27}Al ratio may have been heterogeneous not only in relation to micro-corundum and special CAIs, but also at large scale in the protoplanetary disk. For example, Larsen et al. \cite{larsen11} conclude that the canonical value is representative only of the CAI forming region, while the rest of the disk was characterised by \iso{26}Al/\iso{27}Al roughly half of the canonical value. This is based on high-precision determination of the initial \iso{26}Mg/\iso{24}Mg ratio (i.e., at the Al/Mg=0 intercept, see Sec.~\ref{sec:abundances}) and the fact that the decay of a canonical abundance of \iso{26}Al should have modified the global abundance of \iso{26}Mg by a larger amount than observed. Other interpretations are also possible, however. For example, heterogeneities in the Mg isotopes themselves, unrelated to the decay of \iso{26}Al. A similar conclusion of a lower ESS \iso{26}Al abundance was reached on the basis of comparing ages based on the Pb-Pb system and the Al-Mg system \cite{schiller15a}. On the other hand, recently derived concordant Hf-W and Al-Mg ages for angrites and CV chondrules provide evidence for an homogeneous distribution of \iso{26}Al in the ESS
\cite{budde18}. 
In case the finding of a lower canonical value for \iso{26}Al will be confirmed and consensus achieved, the discussion on the origin of \iso{26}Al in the ESS will need to be revised, as well as its implications as a heat source \cite{schiller15a,larsen16}.

The production of \iso{26}Al in stars and supernovae is due to proton captures on the stable \iso{25}Mg (see top left panel of Fig.~\ref{fig:NCn1}) occurring in various kinds of environments. The crucial reaction is \iso{25}Mg(p,$\gamma$)\iso{26}Al, which has been recently measured in the Laboratory for Underground Nuclear Astrophysics (LUNA, at the Italian National Laboratories of the Gran Sasso, LNGS). Thanks to the background suppression provided by the km-thick rock of the Gran Sasso mountain, the reaction is now known to high accuracy and better precision than before \cite{straniero13}. However, a major problem is the fact that the reaction can feed both the ground state of \iso{26}Al and its isomeric state, which immediately decays into \iso{26}Mg with a half-life of just 6 seconds. The feeding factor to the ground state is not very well known, with large error bars and inconsistent data from different experiments (see discussion in \cite{straniero13}). This still hampers a precise knowledge of the rate of the reaction channel leading to the ground state of \iso{26}Al. 

In massive and Super-AGB stars\footnote{Super-AGB stars differ from AGB stars in that they experience C burning in their core, which result in a degenerate, inert core made mostly of O and Ne. They derive from the highest values of the AGB initial mass range, roughly $>$ 7-8 \msun.} (of initial mass $>$ 5 \msun), H burning can occur at the base of the convective envelope, when the temperature reaches of the order of 60-100 MK. At such temperatures, the Mg-Al chain of proton captures is established, which results in the production of \iso{26}Al \cite{trigo09,lugaro12a,wasserburg17}. In this environment, the main destruction channel for \iso{26}Al is also proton captures, via the \iso{26}Al(p,$\gamma$)\iso{27}Si reaction. The rate of this reaction is not very well determined because it is controlled by the strength of low-energy resonances at 68, 94, 127, and 189 keV, which are difficult to measure. Indirect methods have been used to gather more information, but have not been applied yet to a revision of the rate and its uncertainties. As for \iso{27}Al, relatively little production occurs in AGB and Super-AGB stars, with production factors barely above unity.

In low-mass AGB stars (of initial mass $<$ 5 \msun) the base of the convective envelope is too cold to allow production of \iso{26}Al. Extra-mixing mechanisms have been invoked to drive material from the base of the convective envelope into the hotter region lying below it, and boost the production of \iso{26}Al \cite{wasserburg06}. The idea of extra-mixing in low-mass AGB stars was proposed on the basis of observations of stardust oxide grains, and specifically those classified as Group II \cite{nittler97} showing the signature of H-burning via the CNO cycle and at the same time excesses in \iso{26}Al higher than the other oxide grain populations \cite{palmerini11}. However, a new measurement of the rate of the \iso{18}O(p,$\alpha$)\iso{15}N reaction performed by LUNA \cite{bruno16} resulted in a rate more than twice the one previously recommended \cite{iliadis10}. This has allowed to attribute the origin of Group II grains to massive AGB stars instead, whose base of the convective envelope is hot enough to drive H burning \cite{lugaro17}. Furthermore, the existence of extra-mixing during the AGB phase of low-mass stars is currently not supported by the direct observations of these stars \cite{abia17}. 

In massive stars (of initial mass $>$ 10 \msun), large amounts of both \iso{26}Al and \iso{27}Al are produced particularly during the CCSN phase. The mechanisms at play have been previously analysed and described in detail \cite{timmes95a,limongi06}. In brief, during the pre-CCSN phases, WR stars can be strong producers of \iso{26}Al due to peeling of the H-burning ashes from the convective envelope by strong winds. The same reaction chain as in AGB and Super-AGB stars applies under these circumstances, albeit activated at slightly lower temperatures (30-50 MK) and higher densities. During the CCSN explosion, further production of \iso{26}Al and \iso{27}Al occurs in the O/Ne shells, where destruction is mainly wrought by neutron captures, in particular the \iso{26}Al(n,$\alpha$)\iso{23}Na and \iso{26}Al(n,p)\iso{26}Mg reactions. These have relatively large cross sections, of the order of 100 mbarn \cite{desmet07al,oginni11}, whereas the (n,$\gamma$) channel cross section has a cross section of approximately 4 mbarn\footnote{Neutron-capture cross sections are quoted from the KADoNiS database kadonis.org \cite{dillmann06}, unless indicated otherwise.}. Studies on the impact of nuclear uncertainties on the production of \iso{26}Al in massive stars have indicated its sensitivity not only to reactions directly related to its path of production and destruction but also indirectly to a number of other reactions (see \cite{iliadis11} for details). 

Aluminium-26 is of high interest also in the field of $\gamma$-ray spectroscopic observations performed, e.g., by the COMPTEL and INTEGRAL satellites, because the $\gamma$-ray photon at 1.8 MeV produced by its decay can be detected \cite{diehl13}. This has allowed to establish that roughly 2 to 3 \msun\ of \iso{26}Al are currently present in the Galaxy. Also, it has been possible to spatially map the emission line from the \iso{26}Al decay, which has allowed us to identify its main production regions as being in the mid-plane of the Galaxy, where we expect more massive stars to be present. Furthermore, regions of higher \iso{26}Al abundance correlate with associations of massive stars (OB associations, see below). These observations provide important constraints in relation to the origin of \iso{26}Al in the Galaxy, and also in relation to its ESS abundance via comparison to meteoritic data. To translate  
the total mass of \iso{26}Al in the Milky Way ISM given from the $\gamma$-ray observation of 1.5 to 3.6 \msun\ \cite{diehl13} into a \iso{26}Al/\iso{27}Al ratio, it is necessary to normalise it to the total mass of gas and dust in the Milky Way, of 8.1 $\pm$ 4.5 $\times 10^{9}$ \msun \cite{kubryk15a}. 
This results in approximately $10^{-10}$ - $10^{-9}$ of \iso{26}Al in the Galaxy by mass fraction. To calculate the \iso{26}Al/\iso{27}Al ratio the abundance of \iso{27}Al is also required. This may differ from the solar value because $\gamma$-ray observations sample the ISM today, while the Solar System abundances sample the ISM 4.6 Gyr ago. However, as mentioned in Sec.~\ref{sec:galaxy} in recent years it has become clear that the evolution of the ISM is dominated by the effect of stellar migration \cite{spitoni15,kubryk15a}, which results in a large spread of metallicity, as traced by the abundance of Fe, for any given stellar age \cite{casagrande11,bensby14}. For example, the increase in the abundance of Fe in the past 4.6 Gyr is predicted to be less than 25\%, while the observed spread for stars in this age range is roughly a factor of 4. 
The evolution of Al in the Galaxy for the metallicity around solar of interest here approximately follows that of Fe \cite{bensby14}. This is because Al is a secondary element produced more efficiently in stars of higher metallicities, and much of the production of Fe is also delayed in the Galaxy as it occurs in SNIa from WD, the product of long-living low-mass stars.  
Using the Solar System abundance of \iso{27}Al to normalise the current day $\gamma$-ray data, a \iso{26}Al/\iso{27}Al ratio in the ISM between $2 \times 10^{-6}$ and $1.7 \times 10^{-5}$ is derived. This is 3 to 25 times lower than the canonical ESS value. Taking into account an isolation time would further decreases the ISM ratio that might be inherited by the ESS. Even an isolation time of only 1 Myr would increase the lower bound of the discrepancy from 3 to 8 times lower. Thus it appears difficult to reconcile the high abundance of \iso{26}Al in the ESS with its current ISM abundance, and an extra source has been invoked, as will be discussed in detail in Sec.~\ref{sec:birth}. In this context it should be noted that the large-scale emission observed from galactic \iso{26}Al is quite irregular \cite{wang09}, indicating clumpy distribution of massive stars. Localised \iso{26}Al emission has been reported for regions of OB associations of massive stars, such as Cygnus \cite{martin09} and Scorpius-Centaurus \cite{diehl10}. This suggests that (G)MCs in the neighbourhood of OB associations may in fact be more enriched in \iso{26}Al than the average ISM. Whether and how this enriched, hot material can find its way into cold clumps of star formation, however, still needs to be determined (see Sec.~\ref{sec:birth}).

Finally, the initial abundance of \iso{26}Al in meteoric stardust grains recovered from meteorites \cite{zinner14} at the time of their formation in stellar outflows can also be inferred using excess \iso{26}Mg. The presence of \iso{26}Al has been reported both for C-rich grains (silicon carbide SiC and graphite) and for O-rich grains (in particular, corundum Al$_2$O$_3$ stardust). The derivation of the initial \iso{26}Al/\iso{27}Al for stardust grains is not based on an isochrone, as done for Solar System materials (Sec.~\ref{sec:abundances}) because carbonaceous and corundum grains contain much larger amounts of Al than Mg, hence, it can be assumed that all the measured \iso{26}Mg excess results from the original presence of \iso{26}Al\footnote{Magnesium is not a main component
of SiC, corundum (Al$_3$O$_2$), and hibonite (CaAl$_{12}$O$_{19}$) grains, however, it is a main component of spinel (MgAl$_2$O$_4$). Stoichiometric spinel would contain two atoms of Al per each atom of Mg, which corresponds roughly to 25 times a higher ratio than in the average Solar System material. However, in single stardust spinel grains this proportion may vary.}.  
The grains believed to be originating from CCSNe show very high abundances of \iso{26}Al, with inferred \iso{26}Al/\iso{27}Al ratios in the range 0.1 to 1 \cite{groopman15}, and higher than theoretical predictions. They need to be used to further constrain the nucleosynthesis models \cite{pignatari13c}. The grains that originated in AGB stars show somewhat lower abundances, with \iso{26}Al/\iso{27}Al in the range $10^{-3}$ to $10^{-2}$, which can also be used for comparison and constraints to the nucleosynthesis models \cite{vanraai08,palmerini11,lugaro17}. 

\subsection{\iso{36}Cl and \iso{41}Ca}
\label{sec:36Cl}

These two SLRs are among the shortest lived in Table~\ref{table:SLRs}, with half-lives of the order of a few 10$^5$ yr. As such they can help disentangle the events that occurred  closest to the birth of the Sun. 
The issue of the ESS abundance of \iso{36}Cl has been debated for some time. Potentially, this SLR can have both an initial ESS contribution resulting from stellar pollution, and a late contribution from irradiation by solar CRs in the disk. 
A difficulty arises from the fact that the main (98\%) decay channel of \iso{36}Cl is via $\beta^-$ decay to \iso{36}Ar, a noble gas that easily escapes from solid material. Instead, estimates of the ESS value of \iso{36}Cl rely on measurements of excesses in \iso{36}S, the daughter of the electron capture channel.  
In support of the case for a potential stellar source of \iso{36}Cl in the ESS,  
recent analysis of the {\it Curious Marie} CAI has revealed the presence of this SLR together with \iso{26}Al in sodalite (a mineral that contains Cl) probably produced by the aqueous alteration event that depleted the CAI in U (see also Fig.~\ref{fig:data} and Sec.~\ref{sec:rpiso} in relation to \iso{247}Cm in {\it Curious Marie}). An estimate of the time of occurrence of this event, after which the CAI can be considered as a closed system, is less 50 kyr, as inferred from the \iso{26}Al-\iso{26}Mg system \cite{tang17}. Other studies have found very high levels of \iso{36}Cl in some refractory inclusions, which were not correlated to the presence of \iso{26}Al \cite{hsu06}, as well as large heterogeneities \cite{nakashima08}, providing a case for also a late irradiation contribution, as for \iso{10}Be.  

The case of \iso{41}Ca poses a difficult measurement because of its very low abundance. The latest data on a handful of CAIs \cite{liu12,liu17} demonstrate the presence of this very short-lived isotope in the ESS, with a relatively low inferred ESS value. However, also heterogenities appear to be present in its distribution since one CAI (out of four analysed) did not show a resolvable excess in the daughter \iso{41}K. Also in this case irradiation in the ESS can be responsible for the observed variations, although more data is needed to ascertain values and distributions.

The nucleosynthetic paths for the production of \iso{36}Cl and \iso{41}Ca are  similar (see Fig.~\ref{fig:NCn1}, top right panel and bottom left panel, respectively). Both SLRs are produced by neutron capture on an abundant stable isotope (\iso{35}Cl and \iso{40}Ca, respectively). \iso{35}Cl has a neutron-capture cross section roughly twice as large as that of \iso{40}Ca, however, it is roughly 15 times less abundant. The major destruction channels of \iso{36}Cl and \iso{41}Ca in the presence of neutrons are the (n,$\alpha$) and (n,p) reactions \cite{desmet07cl}, whose cross sections are not very well determined. Overall, it appears plausible to produce the two SLRs in stellar sources to the level required by their ESS ratios (see, e.g., Sec.~\ref{sec:Q1} and Fig.~\ref{fig:scipaper}). However, the large uncertainties both in their ESS abundances and distributions as well in the nuclear paths of production and destruction prevent us  from drawing strong conclusions on their sources at this time.

\subsection{\iso{53}Mn}
\label{sec:53Mn}

The value of the \iso{53}Mn/\iso{55}Mn ratio in the ESS is considered very well known since several estimates are available and in good agreement with each other (e.g., \cite{trinquier08,gopel15}, see Table 9 of \cite{tissot17}). The number reported in Table~\ref{table:SLRs} is the value recommended by \cite{tissot17} on the basis of all the available data. 
The half-life of \iso{53}Mn, on the other hand, is still a topic of debate \cite{dressler12}. The currently recommended value of 3.7 Myr is based on three different, concordant experiments from the early 1970s, however, a higher value of 4.8 Myr has been proposed in order to explain apparent discrepancies with  \iso{26}Al and Pb-Pb ages that exist for some (however not all) meteoritic samples \cite{nyquist09}. 

The bulk of the abundance of Mn in the Galaxy, both the stable \iso{55}Mn and the SLR \iso{53}Mn, is the result of the decay of \iso{53}Fe and \iso{55}Fe produced by explosive Si burning and standard NSE. The following decay of \iso{53}Mn provides much of the cosmic abundance of \iso{53}Cr. The astrophysical site where the majority of such production occurs are probably SNIa that reach the Chandrasekhar mass. Actually, the existence of such a SNIa channel appears to be required by the need to provide a significant galactic source of Mn and the other iron peak elements at late times \cite{seitenzahl13,hitomi17}. The production of \iso{53}Mn is only very marginally affected by nuclear reaction uncertainties given the nature of the NSE process, which depends more on the stability of the nuclei themselves, rather than presenting a path of production and destruction reactions, although some abundances can be affected by the properties of the freeze-out phase after the NSE. According to \cite{parikh13} the \iso{53}Mn/\iso{55}Mn production ratio in SNIa could increase by up to a factor of two due to nuclear uncertainties. Model uncertainties are of the same order, e.g., the multi-D simulations of Chandrasekhar mass SNIa by \cite{travaglio04} produce \iso{53}Mn/\iso{55}Mn ratios in the range 0.09 to 0.13 and the most recent results from \cite{seitenzahl13b} range from 0.06 to 0.13. 
The \iso{53}Mn/\iso{55}Mn ratio produced in CCSNe is higher, up to 0.2 \cite{lugaro14,lugaro16}, however, CCSNe produce a less significant absolute amount of Mn, due to the type of freeze-out that follows the NSE process. As such they are listed as a minor site in terms of cosmic abundances in Table~\ref{table:stars} (see Fig. 1 of \cite{lugaro16}).

\subsection{\iso{60}Fe}
\label{sec:60Fe}

Among the inferred abundances of the SLRs in the ESS,
that of \iso{60}Fe is the most controversial. This is not only because it represents an analytical challenge, but also because a high \iso{60}Fe/\iso{56}Fe would represent a smoking gun for stellar nucleosynthesis and specifically for a potential contribution of CCSNe to the ESS (Sec.~\ref{sec:Q1}). An extensive review of the data up to 2012 can be found in \cite{mishra12}. Since then, the situation has not been clarified: the most recent estimates of the initial \iso{60}Fe/\iso{56}Fe ratio range from $10^{–8}$ from  measurements of bulk meteorites and bulk chondrules using inductively coupled plasma mass spectrometry (ICP-MS) \cite{tang15,tang12}, to $10^{–7}$-$10^{–6}$ from in-situ measurements of high Fe/Ni phases using secondary-ion mass spectrometry (SIMS) \cite{mishra14,telus18}. Crucially, while the former value rules out a CCSN source, the latter range requires it. There is the possibility that the SIMS analyses are compromised by stable isotope anomalies in Ni and/or unrecognised mass fractionation effects \cite{boehnke17}, hence, while the debate is ongoing, in Table~\ref{table:SLRs} we have recommended the lower value. 
On the other hand, the debate related to the half-life of \iso{60}Fe can be now considered resolved, with two recent experiments \cite{wallner15a,ostdiek17} confirming within uncertainties the half-life of 2.62 Myr presented by \cite{rugel09} and roughly 75\% longer than the previous estimate. 

The reason why \iso{60}Fe is a clear signature of stellar nucleosynthesis is the fact that its production requires a chain of double neutron captures (see bottom right panel of Fig.~\ref{fig:NCn1}). When neutrons are available, the stable \iso{58}Fe suffers a (n,$\gamma$) reaction, which produces \iso{59}Fe. This isotope is unstable, with a half-life of 44.5 days. The fact that \iso{59}Fe can either decay or capture another neutron to produce \iso{60}Fe results in the possibility of a splitting along the path of neutron captures, i.e., a {\it branching point}. 
To calculate the fraction of the neutron-capture flux that branches off the main $\beta$-decay 
path at any given branching point, a {\it branching factor} is used and defined as: 

\begin{equation}
 f_{branch}=\frac{p_{branch}}{p_{branch}+p_{main}},  
\label{eq:branch1}
\end{equation}

\noindent where $p_{branch}$ and $p_{main}$ are the
probabilities per unit time associated with the nuclear reactions suffered by the branching point
nucleus and leading onto the branch or onto the main path, respectively.
The case of the \iso{59}Fe branching point is quite typical, i.e., $p_{main}$ corresponds to the
$\beta^-$ decay rate, and $p_{branch}$ corresponds to $p_{n}$,
i.e., the probability per unit time of \iso{59}Fe to capture a neutron, so that Eq.~\ref{eq:branch1} becomes:

\begin{equation}
  f_{branch}= \frac{<\sigma> v_{thermal} N_{\rm n}}{<\sigma> v_{thermal} N_{\rm n}+\lambda}, 
 \label{eq:branch2} 
\end{equation}

\noindent where $N_{\rm n}$ is the
neutron density in n/cm$^3$, $v_{thermal}$ is the thermal velocity $\sqrt{2 k_B T/m}$ ($k_B$ is Boltzmann constant, $T$ the temperature and, $m$ the mass), and $<\sigma>$ is the Maxwellian-averaged neutron-capture cross section. This is is typically inversely proportional to the velocity, so that $<\sigma> v_{thermal}$ is relatively constant. The 
$<\sigma>$ values of \iso{59}Fe and \iso{60}Fe have been determined to be $\simeq$6 mbarn via indirect experiments \cite{uberseder14} and $\simeq$24 mbarn via direct experiments \cite{uberseder09}, respectively, both with only a mild temperature dependence of 
the $<\sigma> v_{thermal}$. 
For the probability of producing \iso{60}Fe to be above a few percent, neutron densities above $5 \times 10^{9}$ n/cm$^{3}$ are required. These are produced via the \iso{22}Ne($\alpha$,n)\iso{25}Mg reaction in the He-shell burning regions of relatively massive AGB stars, Super-AGB stars, and during the pre-CCSN phase (here including also C-shell burning) of massive stars. During the CCSN blast wave \iso{60}Fe is further produced in the same region where \iso{26}Al is produced \cite{timmes95a,limongi06}. In these conditions of very high temperature (above 500 MK), the half-lives of both \iso{59}Fe and \iso{60}Fe can decrease significantly \cite{li16}, affecting the calculation of the branching point and the survival of \iso{60}Fe itself. In the case of AGB stars, \iso{60}Fe can be expelled into the surrounding medium by the stellar winds if it was previously mixed to the stellar surface by dredge-up episodes. 
Other proposed sites of neutron-rich nucleosynthesis leading to the production of \iso{60}Fe 
are electron-capture supernovae \cite{wanajo13} and carbon deflagration SNIa \cite{woosley97}. 
Overall, while some of the nuclear inputs related to the activation of the \iso{59}Fe branching point are still uncertain, it would be difficult to radically change this current production picture for \iso{60}Fe.

As for \iso{26}Al, galactic $\gamma$-rays indicate significant levels of global \iso{60}Fe production in the Galaxy, with a ratio of the flux originating from \iso{60}Fe to that originating from \iso{26}Al of 0.15 $\pm$ 0.05 \cite{diehl13}, and thus an abundance ratio of 0.55 $\pm$ 0.18. Using the \iso{26}Al abundance for the ISM derived in Sec.~\ref{sec:26Al} and the Solar System abundance of \iso{56}Fe, a \iso{60}Fe/\iso{56}Fe ratio in the ISM from 0.8 to 13 $\times 10^{-7}$ is derived. This is 8 to 130 times higher than the value of $10^{-8}$ reported for the ESS in Table~\ref{table:SLRs}, which allows for an isolation time of at least 8 Myr, and would not require an extra source for the ESS abundance of this SLR \cite{tang12}. If the highest value reported for the ESS of approximately $10^{-6}$ \cite{mishra14} is considered instead, no isolation time would be allowed, and a local source would need to be invoked. 

It is useful also to consider the \iso{60}Fe abundance relative to the \iso{26}Al abundance because the flux ratio \iso{60}Fe/\iso{26}Al is directly determined from $\gamma$-rays and thus the \iso{60}Fe/\iso{26}Al abundance ratio in the Galaxy of approximately 0.55 is better determined than the absolute abundances. In the ESS, the \iso{60}Fe/\iso{26}Al ratio corresponds to 0.00178 and 0.178, when using \iso{60}Fe/\iso{56}Fe=$10^{-8}$ and $10^{-6}$, respectively, i.e., it is roughly 300 to 3 times lower than the $\gamma$-ray ratio. This shows that whichever ESS \iso{60}Fe abundance is considered, the source of \iso{26}Al in the ESS, under-produced  \iso{60}Fe, mildly or strongly, relatively to the Galactic average. A strong under-production would require 
complete decoupling of the origin \iso{26}Al and of \iso{60}Fe in the ESS, likely excluding CCSNe as potential sources of \iso{26}Al in the ESS. 
A mild under-production could represent the detailed, specific signature of the particular CCSN sources present at the birth of the Sun.


Finally, we note that on top of the $\gamma$-rays, the other independent constraints from Earth, Moon and CR samples already mentioned in Sec.~\ref{sec:abundances} also indicate that significant levels of \iso{60}Fe are required to be produced by CCSNe in the Galaxy. For these cases the \iso{26}Al abundance is not available for comparison of the relative abundances, but may become possible to consider in the future.

\subsection{The $r$-process SLRs: \iso{129}I, \iso{244}Pu, and \iso{247}Cm}
\label{sec:rpiso}

The presence of \iso{129}I was the first among all SLRs to be revealed in the ESS by excesses in \iso{129}Xe \cite{reynolds60} given the relatively easy opportunity to analyse samples poor in the noble gas Xe. The value in the ESS presented in Table~\ref{table:SLRs} is based on the estimate by \cite{ott16} given with an uncertainty of 1$\sigma$. It is derived by combining the experimental value from analysis of the Shallowater meteorite given by \cite{gilmour06} at 1$\sigma$, with the age of that meteorite (also given at 1$\sigma$), compared to the age of CAIs given at 2$\sigma$ by \cite{connelly12}.

The difficulty of determining the initial ESS abundances of \iso{244}Pu and \iso{247}Cm is that for these nuclei there are no stable isotopes of the same element. Therefore, \iso{244}Pu and \iso{247}Cm have to be referenced to other elements with similar nucleosynthetic and chemical properties. Conventionally, \iso{244}Pu is referenced to \iso{238}U and \iso{247}Cm to \iso{235}U. However, Pu and Cm are chemically more analogous to the light rare earth elements than to U. 

For \iso{244}Pu there exist contradictory data based on two different approaches. The value reported 
in Table~\ref{table:SLRs} (given at 1$\sigma$ as usual in work related to noble gas experiments) is the result of the analysis of the Xe composition samples of the St Severin ordinary chondrite irradiated with neutrons to induce the fission of \iso{235}U \cite{hudson89}. After this treatment, the composition of Xe in the sample is a mixture of the Xe produced by the fission of the \iso{244}Pu present at its formation, and the fission induced on \iso{235}U, together with a small amount from the spontaneous fission of \iso{238}U and some ``trapped'' xenon. The relative abundance of the fission components is a function of \iso{244}Pu/U. However, the chemistry of St Severin may not be representative of the bulk ESS as pieces of this meteorite show highly variable U contents and Th/U ratios. Results from the  analysis of Xe in ancient terrestrial zircons from Western Australia \cite{turner07} yields data in agreement with this value, but the authors could not rule out fractionation between Pu and U during magmatic/mineral formation processes. In a different approach, mineral separates were analysed from the Angra dos Reis angrite meteorite, and \iso{150}Nd, an r-process-only isotope of Nd (a light rare earth element) was used as the reference isotope \cite{lugmair77}. In order to obtain the abundance of \iso{244}Pu with reference to \iso{238}U, which is closer in mass to \iso{244}Pu thus providing a ratio better predictable by $r$-process models, this must be converted using the Solar System Nd/U ratio 4.57 Ga ago \cite{lodders09}. This results in a value\footnote{The error here would a combination of statistical and systematic errors from the measurements \cite{lugmair77}, as well as from the renormalisation from Nd to U \cite{hagee90}.} of $(4.4 \pm 1.00) \times 10^{-3}$ \cite{peto08,hagee90}, significantly lower than the value given in Table~\ref{table:SLRs}. However, the absolute abundance of Nd was measured separately from the noble gases, and since angrites are differentiated meteorites the original Pu/U and Pu/Nd ratios may have been modified during melting. New efforts are ongoing to derive a better estimate of the ESS \iso{244}Pu/\iso{238}U ratio on a variety of materials \cite{peto17}. 

For \iso{247}Cm the situation is more favourable than for \iso{244}Pu thanks to the discovery of the peculiar {\it Curious Marie} CAI \cite{tissot16}. This CAI is extremely depleted in U, providing an extreme data point in terms of the Nd/U ratio (see Fig.~\ref{fig:data}). With  this new data \cite{tissot16} it has been possible to derive a clear isochrone and hence a precise \iso{247}Cm/\iso{235}U ratio. In the original work, the absolute time of the latest alteration event that depleted the U was unknown, and an age of 10 Myr was applied to derive the ESS \iso{247}Cm/\iso{235}U ratio. The latest work on {\it Curious Marie} \cite{tang17}, however, implies that the alteration occurred at most 50,000 yr after the formation of the CAI, which we have taken into account in our recommended ESS \iso{247}Cm/\iso{235}U ratio listed in Table~\ref{table:SLRs}, where the error bar is 2$\sigma$ when assuming that Cm behaved chemically exactly like Nd.

The half-life of \iso{129}I has been relatively well known since experiments in the 1970s, and the recommended value of 15.7 Myr is in agreement with the I-Xe systematics of chondrules from primitive meteorites \cite{pravdivtseva16,pravdivtseva17}. 
The half-lives of \iso{244}Pu and \iso{247}Cm are well known given that these isotopes are involved in nuclear reactor technology. 


From the point of view of nucleosynthesis, the three isotopes considered here are almost exclusively produced by the $r$ process. The two
heaviest belong to the actinide group of elements, with nuclear charges from 89 to 103 and chemical properties as rare earth elements. The $s$-process production chain ends
at Pb and Bi (with nuclear charges 82 and 83, respectively), where most of the reaction flow is trapped at \iso{208}Pb and \iso{209}Bi because of the small neutron-capture cross sections of these 
neutron magic nuclei (approximately 0.3 and 2.6 mbarn, respectively, at 30 keV). The following element on the chain of neutron captures is Po, with the isotopes \iso{210}Po and \iso{211}Po unstable against $\alpha$ decay toward \iso{206}Pb and \iso{207}Pb \cite{ratzel04}. This results in the impossibility of building elements beyond Bi with the $s$ process and the necessity for these elements to come from the $r$ process.
The $r$-process production of actinides has been studied in detail \cite{goriely01,goriely16}, also considering nuclear uncertainties. Since there are no solar abundances for these elements to compare to, the best strategy is to derive the actinide abundances from model predictions that match the Solar System $r$-process abundances. The uncertainties in the yields are large, around one order of magnitude \cite{goriely16}, however, the relative isotopic ratios that are useful to compare to ESS data are somewhat less uncertain. For example, of the 20 models presented by \cite{goriely16}, 18 models present \iso{244}Pu/\iso{238}U ratios between 0.3 and 0.5 and \iso{247}Cm/\iso{235}U ratios in the range 0.2 and 0.4. Only two models have significantly different ratios around 1.3. This indicates that the absolute production of the actinides are typically correlated to each other when changing the model parameters.

The case of \iso{129}I is
different: the reason why the $s$ process cannot produce it is the very short half-life of
\iso{128}I. As shown in the middle left panel of Fig.~\ref{fig:NCn2}, in the presence of neutrons the only stable isotope of I,
\iso{127}I captures a neutron to produce \iso{128}I. This is  unstable with a half-life of
25 minutes, too short to allow any further neutron captures, at least in $s$-process conditions.
Similarly to the case of \iso{59}Fe discussed above, the
case of the \iso{128}I branching point is quite typical. Here, $\lambda \sim 3.4 \times 10^{-4}$ 1/s and $<\sigma> \sim 1500$ mbarn. The
branching factor reaches above 1\%, 50\%, and 90\% only when $N_{\rm n}$ is above $10^{11}$,
$10^{12}$, and $10^{13}$ n/cm$^3$, respectively. In general, a branching factor also depends on
the temperature since both $\lambda$ and $<\sigma> v_{thermal}$ can have a temperature dependence, and also a density dependence for $\lambda$. In the case of \iso{128}I, both $\lambda$ and $<\sigma> v_{thermal}$ have a small dependence
on the temperature, with a variation of less than 60\% for typical $s$-process temperatures between 100 and
300 MK \cite{takahashi87}\footnote{It should be noted that there is a third path of the branching
point because \iso{128}I can also electron capture into \iso{128}Te. This channel however is
roughly 20 times less likely than the $\beta^{-}$, so we did not consider it in the calculation of
the branching point. However small, the branching factor of \iso{128}I in $s$-process conditions
has been carefully investigated because it affects the accuracy of the $s$-process contribution to
\iso{128}Xe, in principle an isotope produced mostly by the $s$-process, but with a potential $\sim$10\%
$p$-process contribution \cite{reifarth04}.}. 

The neutron densities required to produce \iso{129}I are not achieved in the typical AGB
stars that produce the bulk of the $s$ process in the Galaxy. Consequently, the bulk galactic
production of \iso{129}I has been attributed to the $r$ process. Still, in other sites such as
CCSNe, neutron ``bursts'' can occur with neutron densities up to the values that allow
\iso{129}I production (Table~\ref{table:stars}). For accurate predictions in these cases, the contribution of higher energy levels in \iso{129}I to the total neutron-capture cross section should be considered \cite{rauscher12}, which may modify the measured cross sections in stellar plasma conditions.  
These sites do not produce the bulk of \iso{129}I in the Galaxy because there are not enough neutrons released to convert Fe into heavier elements, but only small numbers of neutrons are released iin total, 
which allow some capture by the initial inventory of \iso{127}I itself.

Also the reference isotope, \iso{127}I, is a major $r$ process product.
Thus, in spite of the issues and problems currently related to the modelling of the $r$ process, both from the stellar site and the nuclear physics point of view, the \iso{129}I/\iso{127}I ratio produced by the $r$ process is well constrained. This is because it can be derived on the basis of the $r$-process {\it residuals} method, whereby the $r$-process component of the Solar System abundance distribution is obtained by subtracting from the total abundance of each isotope (derived from meteoritic analysis) the $s$-process component, which is relatively well known, often controlled mostly by the neutron-capture cross section of the stable isotope (e.g., \cite{arlandini99,bisterzo11}). In the case of the radioactive \iso{129}I, its $r$-process abundance can be simply derived from the $r$-process residual of \iso{129}Xe, since all of the $r$-process abundance of \iso{129}Xe is first produced as \iso{129}I. Due to all these advantages, the \iso{129}I/\iso{127}I ratio represents a textbook case to be used for the derivation of cosmic timescales, as will be described in more detail in Sec.~\ref{sec:times}. Furthermore, any constraints from it can be cross checked with those from \iso{244}Pu and \iso{247}Cm, providing possibly three different independent evaluations of a given time interval. 

\subsection{The SLRs with an $s$-process contribution: \iso{107}Pd, \iso{182}Hf, and \iso{205}Pb}
\label{sec:107Pd}

The ESS abundance of \iso{182}Hf has been relatively well determined for some time, and the value we provide in Table~\ref{table:SLRs} is in agreement within error bars with the value given by \cite{burkhardt08}. For \iso{107}Pd there are still problems related to the determination of the age of the iron meteorite on which the \iso{107}Pd/\iso{108}Pd ratio has been determined with a precision of better than 3\% \cite{matthes18}. In Table~\ref{table:SLRs} we have considered the higher value given by \cite{matthes18} for the initial ratio, which is in agreement with the value of (5.9 $\pm$ 2.2) $\times 10^{-5}$ based on the analysis of carbonaceous chondrites \cite{schonbachler08}. The case of \iso{205}Pb is more 
problematic because both the elements involved in the analysis (Pb and Tl) are somewhat volatile and Pb in particular is prone to contamination so that a mixing line produced by contamination could be incorrectly interpreted as an isochrone. Furthermore, Tl has only two stable isotopes, which means that it is not possible to correct directly for possible artificial or natural mass fractionation effects, and it is more difficult also to recognise potential nucleosynthetic anomalies. To address this problem, for the mass spectrometric analysis, \cite{andreasen12} added an element of similar mass (Pt) in order to infer the instrumental mass fractionation for Tl. These authors obtained a value for the \iso{205}Pb/\iso{204}Pb ratio in agreement with the previous value \cite{baker10}, but an order of magnitude higher than that given by the older analysis \cite{nielsen06}. The value in Table~\ref{table:SLRs} is that reported by \cite{palk18} and obtained combining the results of \cite{andreasen12} and \cite{baker10}. 

From the point of view of stellar production, the galactic abundances of these three nuclei have significant contributions from the $s$ process. Specifically,
\iso{107}Pd and \iso{205}Pb are produced directly by neutron-captures onto the stable \iso{106}Pd
and \iso{204}Pb, respectively (top left and bottom right panels of Fig.~\ref{fig:NCn2}). On the other hand, production of \iso{182}Hf
requires the activation of a branching point at \iso{181}Hf (bottom left panel of Fig.~\ref{fig:NCn2}). The half-life of this nucleus was
believed to greatly decrease from roughly 42 days to 30 hours in stellar conditions due to the
population of an excited state at 68 keV \cite{takahashi87}. In this case the branching factor
leading to the production of \iso{182}Hf reaches above 3\% only for $N_{\rm n}$ above
$10^{9}$ n/cm$^3$ and the bulk of the production of \iso{182}Hf needs to be attributed to the $r$ process.
This was leading to a strong disagreement in the time
intervals derived from \iso{129}I and \iso{182}Hf and to solve this problem different components for the $r$-process were proposed \cite{wasserburg96}, as well as different local sources for \iso{182}Hf \cite{meyer00} (see also \cite{ott08}). However, according to more recent
experimental data \cite{bondarenko02} such a 68 keV state in \iso{181}Hf does not exist, thus the half-life of
\iso{181}Hf should not vary greatly with the temperature. In this case, the branching factor is
above 3\% already for $N_{\rm n}$ above $5 \times 10^{7}$ n/cm$^3$ and a significant production of
\iso{182}Hf occurs in the AGB stars that produce the bulk of the $s$ process in the Galaxy, with \iso{182}Hf/\iso{180}Hf ratios around 0.15. This 
removes the issue of the time discordance with \iso{129}I \cite{lugaro14}.

\iso{107}Pd and \iso{182}Hf have significant contributions from both the $s$ and the $r$
process. As in the case of \iso{129}I, their $r$-process production ratios relative to their reference
stable isotope can be derived using the $r$-residual method and considering the $r$-process
residuals of their daughter nuclei, \iso{107}Ag and \iso{182}W, respectively. Since the $r$-process residual depends on the $s$-process contribution to each specific
isotope, the $r$-process \iso{182}Hf/\iso{180}Hf ratio had to be readjusted after the discovery that
\iso{182}Hf has a significant $s$-process production, which in turn increases the $s$-process
contribution to \iso{182}W. On the other hand, \iso{205}Pb is effectively a nucleus produced only by the $s$ process, 
being shielded from $r$-process production by \iso{205}Tl. However, this does not mean that its
production can exclusively occur in AGB stars: small neutron bursts in CCSNe
and WR stars can also produce this isotope, although these are minor production sites since there
is no conversion of Fe nuclei into Pb. The main problem with the production of \iso{205}Pb is that its electron-capture half-life is predicted to vary by several orders of magnitudes in stellar conditions: from 17 Myr in terrestrial conditions down to roughly 15 years for temperatures above 50 MK, and also depending on the density, although this temperature and density dependence is uncertain by an order of magnitude \cite{goriely99}. This makes it difficult to save \iso{205}Pb to be carried to the stellar surface, but production is still be expected \cite{yokoi85}.  
Finally, we note that in both the cases of \iso{182}Hf and of \iso{205}Pb, population of higher energy levels can modify the total neutron-capture cross sections in stellar conditions \cite{rauscher12}, which needs to be considered as an additional model uncertainty.  

\subsection{The $p$-process SLRs: \iso{92}Nb, \iso{146}Sm, and \iso{97,98}Tc}
\label{sec:92Nb}

The ESS values of both \iso{92}Nb and \iso{146}Sm are well determined. In the case of \iso{92}Nb the value of \iso{92}Nb/\iso{93}Nb given in Table~\ref{table:SLRs} is derived from rutiles and zircons with well known ages and is in agreement with the less precise value of \cite{iizuka16} from angrites and eucrites. The  \iso{146}Sm/\iso{144}Sm value, on the other hand, has been measured directly in a CAI (Fig.~\ref{fig:data}), removing previous issues related to the age adjustment. For \iso{97,98}Tc, only upper limits are available.
The half-life of \iso{146}Sm is poorly determined. A recent experiment \cite{kinoshita12} shortened the previous recommended value by approximately 50\%, however there is better agreement with the suite of meteoritic data with different ages (from Pb-Pb dating) when  the older half-life is employed \cite{marks14}. The half-life of \iso{92}Nb appears to be well determined, being the weighted average of two experiments that produced similar results, in spite of different approaches.

The stellar production of all these isotopes is broadly ascribed to the $p$ process -- although a minor production process for \iso{97}Tc is also neutron capture on the relatively abundant \iso{96}Ru, followed by the electron capture decay of \iso{97}Ru \cite{arnould97}. Model predictions and the usage of these isotopes as cosmochronometers have been discussed in detail by \cite{lugaro16}. In summary, \iso{146}Sm is most likely produced by the $\gamma$ process, although the site is still debated between SNIa \cite{travaglio14} and CCSNe, with SNIa being favoured \cite{travaglio18}. The accuracy of the theoretical predictions is hampered for this isotope by the uncertain \iso{148}Gd($\gamma$,$\alpha$)\iso{144}Sm reaction, which controls the ($\gamma$,n)/($\gamma$,$\alpha$) branching at \iso{148}Gd. The current resulting uncertainty on the \iso{146}Sm/\iso{144}Sm production ratio from SNIa models is a factor of two \cite{travaglio14}, but owing to the lack of experimental data it could be even higher, up to one order of magnitude. These nuclear uncertainties, together with the half-life uncertainty, currently hamper the opportunity of using \iso{146}Sm, the longest lived of the SLRs considered here, as an accurate cosmochronometer. 

When considering the lighter $p$-process isotopes up to Ru, different flavours and sites of $p$-process nucleosynthesis need to be considered \cite{travaglio18}, particularly to explain the relatively high Solar System abundance of the $p$-process isotopes of Mo (at masses 92 and 94) and Ru (at masses 96 and 98). In particular, another source of \iso{92}Nb is required in the Galaxy because only considering the $\gamma$ process in SNIa results in inconsistent timescales when the other SLR predominantly produced by SNIa, \iso{53}Mn, is also considered \cite{lugaro16}. For example, low-mass CCSNe could be a significant cosmic source of \iso{92}Nb. In summary, also the opportunity to use \iso{92}Nb as an accurate cosmochronometer is hampered, in this case by the current large uncertainties related to the modelling of CCSNe. Finally, the upper limits available for the ESS abundance of \iso{97,98}Tc do not allow the use of these SLRs as meaningful chronometers \cite{lugaro16}.



\subsection{\iso{126}Sn and \iso{135}Cs}
\label{sec:126Sn}

Establishing and interpreting the ESS initial abundances of \iso{126}Sn and \iso{135}Cs is challenging for a number of reasons. Concerning \iso{126}Sn, measuring Te isotopic ratios precisely is especially difficult, and compounded by the fact that given the short half-life of \iso{126}Sn we may not expect a large excess signal. The most recent work \cite{brennecka17b} reports an upper limit for the \iso{126}Sn/\iso{124}Sn ratio of 3$\times 10^{-6}$. Even so, this value is significant when compared to the stellar production of \iso{126}Sn and may be used to rule out nearby stellar sources. So far, most authors have ascribed the production of \iso{126}Sn to the $r$ process in supernovae (see \cite{fehr09} and references therein), since the branching isotope \iso{125}Sn has a half-life of 9.6 days, too short to allow for significant capture of neutrons in $s$-process conditions (Fig.~\ref{fig:NCn2}). However, a non-detection of \iso{126}Sn cannot be used to rule out a nearby CCSN source of SLRs because, as discussed above, the $r$ process is not believed to occur in standard CCSNe, but rather in NSMs or peculiar CCSNe. The branching point may still open during neutron burst conditions in CCSNe. In fact, considering the theoretically calculated $<\sigma>$ of \iso{125}Sn of around 70 mbarn, and the strong temperature dependence of the half-life \cite{takahashi87}, which decreases to 2.5 hours at 200 MK, the probability of \iso{125}Sn to capture a neutron is above 10\% for neutron densities above 10$^{11}$ n/cm$^3$, which are possible during a neutron burst. A detailed analysis of the production of \iso{126}Sn in CCSNe also considering the nuclear uncertainties is still missing and urgently needed to exploit the new experimental ESS upper limit.

The situation for \iso{135}Cs is similar and also open. The difficulty in measuring its initial ESS abundance is mostly related to the fact that Cs is a volatile element, which does not easily fully condense into solids. Recently a new approach to the problem was used by \cite{brennecka17b}, who inferred a new upper limit for the \iso{135}Cs/\iso{133}Cs ratio by analysing volatile-depleted material (rather than material in which the radionuclide should be enhanced, as is usually done), which should show a {\it deficit} in \iso{135}Ba with respect to the bulk of Solar System matter, in which \iso{135}Cs fully contributed to the abundance of \iso{135}Ba. The derived upper limit is much lower than that previously proposed based on direct measurements of Ba isotopes in CAIs (e.g., \cite{bermingham14}). Similarly to the case of \iso{126}Sn, \iso{135}Cs is produced by a branching point located at an isotope, \iso{134}Cs, whose T$_{1/2}$ presents a strong theoretically estimated temperature dependence. It decreases from 2 years at laboratory temperatures, down to 12 days at 200 MK, due to the population of levels at 60 and 177 keV in stellar condition. The uncertainty in the evaluation of the rate is of an order of magnitude \cite{goriely99}. The $<\sigma>$ of \iso{134}Cs is approximately 800 mbarn, as derived from statistical model calculations aided by the experimentally determined cross section of \iso{135}Cs \cite{patronis04}. However, the population of higher energy levels can modify the total $<\sigma>$ of \iso{134}Cs in stellar conditions by $\sim$50\% \cite{rauscher12}. Using such values, the branching factor for the production of \iso{135}Cs is above 10\% already for $s$-process neutron densities above N$_{\rm n}$ $\sim 5 \times 10^8$ n/cm$^3$. In fact, the activation of the branching point at \iso{134}Cs is required to match the \iso{134}Ba/\iso{136}Ba isotopic ratio in the Solar System, where both \iso{134}Ba and \iso{136}Ba are isotopes that can be only be produced by the $s$ process \cite{bisterzo15}. This branching point is also of interest for the interpretation of Ba isotopic anomalies measured in mainstream stardust silicon carbide grains (SiC) that originated from AGB stars and show the signature of the $s$ process \cite{liu15,lugaro17}.
Overall, while the largest contributor to the cosmic abundance of \iso{135}Cs is the $r$-process, in relation to the ESS we need to consider also the minor potential production sites, both the $s$-process and the neutron burst in CCSNe. A detailed analysis also of the related nuclear uncertainties and their impact is still missing and, again, urgently required to be able to properly interpret the meteoritic data. 

\section{The galactic chemical evolution of radioactive isotopes}
\label{sec:GCE}

\subsection{General models and considerations}
\label{sec:GCEmodels}

The simplest way to compute the evolution of the abundances of 
radioactive nuclei in the Galaxy is based on the concept of 
steady-state equilibrium, for which a simple derivation can be made as follows. The rate of 
change in time of the number $N_{\rm SLR}$ of a 
given radioactive nucleus in the ISM is given by:

\begin{equation}
\label{eq:SS1}
\frac{dN_{\rm SLR}}{dt}=- N_{\rm SLR}/\tau + \frac{dP_{\rm SLR}}{dt},
\end{equation}

\noindent where the first (negative) term represents the decay, with the mean-life $\tau$ constant in time, and the second 
(positive) term represents the stellar production rate, as stars inject 
freshly produced radioactive nuclei into the ISM. This equation is a 
typical self-regulating equation because the larger the positive 
term, the larger the abundance, and the larger the 
negative term. This means that the abundance $N_{\rm SLR}$ will converge towards the asymptotic value given by the equilibrium value resulting by setting $dN_{\rm SLR}/dt=0$. How quickly the abundance approaches the asymptotic value depends on the half-life, e.g., after 4 times the half-life it will be within 5\% of the asymptotic value, and after 7 times the half-life, within less than 1\%. In general, for the SLRs considered here it is quite safe to assume that they would have reached steady-state equilibrium in the Galaxy by the time of the birth of the Sun, $T_{\rm Gal}$, roughly 7 to 9 Gyr after the birth of the Galaxy. The equilibrium value is simply given by equating the two terms of Eq.~\ref{eq:SS1}:

\begin{equation}
\label{eq:SS2}
N_{\rm SLR}=\frac{dP_{\rm SLR}}{dt} \,\tau.
\end{equation}

To be able to compare to the meteoritic data, however, it is necessary to normalise
the abundance of the SLR to that of its reference stable isotope, 
$N_{\rm stable}$. This is given by the integral of the production rate of the stable isotope over the whole galactic time $T_{\rm Gal}$ before the birth of the Sun. If the production rate is constant then simply: 

\begin{equation}
\label{eq:SS2b}
N_{\rm stable}=\frac{dP_{\rm stable}}{dt}\, T_{\rm Gal}.
\end{equation}

\noindent From which follows: 

\begin{equation}
\label{eq:SS3}
\frac{N_{\rm SLR}}{N_{\rm stable}}=
\frac{dP_{\rm SLR}/dt}{dP_{\rm stable}/dt} \, \frac{\tau}{T_{\rm Gal}}.
\end{equation}

\noindent In steady-state equilibrium the $N_{\rm SLR}/N_{\rm stable}$ ratio 
will steadily decline as time passes, since $N_{\rm SLR}$ is constant,
while $N_{\rm stable}$ increases with time.

More sophisticated, still analytical, models of the GCE \cite{clayton85} also converge into the steady-state equilibrium Eq.~\ref{eq:SS3}. However, such more accurate description of the evolution of the Galaxy results in the introduction into the equation of a multiplication term $(k+1)$, where $k$ is a free parameter used to describe infall of primordial gas into the Galaxy, 
so that the equation becomes: 

\begin{equation}
\label{eq:SS4}
\frac{N_{\rm SLR}}{N_{\rm stable}}= (k+1)
\frac{dP_{\rm SLR}/dt}{dP_{\rm stable}/dt} \, \frac{\tau}{T_{\rm Gal}}.
\end{equation}

\noindent A value of $k=0$ implies no infall, a so-called ``closed-box'' model. However, infall of primordial gas is required to solve the problem that closed-box models overestimate the number of low-metallicity stars when compared to the Milky Way \cite{tinsley80,matteucci12}. 
This infall controls the shape and magnitude of the star formation history. The abundances of stable nuclei accumulate with time, hence they depend on the integrated stellar mass, while the production of the unstable nuclei is related directly to the local star formation rate.
When allowing more infall by increasing $k$, the peak of star formation shifts to later times and the local star formation rate increases by a larger factor than the integrated stellar mass, which results in an increase of the $N_{\rm SLR}/N_{\rm stable}$ ratio.
Actually, this would result in a factor larger than $(k+1)$, however, another effect of increasing $k$ is the reduction of the fraction of stable nuclei locked inside stellar remnants, which partially counterbalances the first effect. The reason is that with more infall stable nuclei are diluted with primordial gas, which takes up some of their place inside stars. On the basis of galactic chemical evolution models, typical values of $k$ are between 2 and 4 (see, e.g., \cite{rauscher13} and references therein). 

A more detailed analysis of the analytical GCE models \cite{huss09} also demonstrates that the $k$-dependent factor in Eq.~\ref{eq:SS4} can take more complex forms than the simple $(k+1)$, depending on how the productions of the SLR and of the reference stable nucleus evolve with time. This depends on whether their abundances result from {\it primary} nucleosynthesis, meaning that they are produced in stars starting from the initial (Big-Bang origin) H and He abundances only, or from {\it secondary} nucleosynthesis, meaning that
their production in stars requires an initial amount of metals, i.e., elements heavier than H and He. In any case, it should be kept in mind that such analytical approaches only model instantaneous recycling of material from stars into the ISM, meaning that stellar lifetimes are not accounted for. Stellar lifetimes contribute to delays, which are important, for example in the case of the products of SNIa such as Mn and Fe, because these result from the explosion of WDs, the progeny of long-lived low-mass stars. This applies also in the case of $s$-process products from low-mass AGB stars. Detailed models of GCE are required to provide a more accurate description of the evolution of SLRs in the Galaxy, however, only a few are currently available, and only for a handful of SLRs  \cite{timmes95a,travaglio14}.

The steady-state $N_{\rm SLR}/N_{\rm 
stable}$ ratios in the ISM at the time of the formation of the Sun calculated with Eq.~\ref{eq:SS4} have been traditionally taken to represent the actual ratios in the ISM material from which the Sun formed after a given isolation time. However, there is 
a serious issue with this assumption: the steady-state ratios are typical ISM ratios averaged over the whole Galaxy by assuming that the stellar production rates are {\it continuous}. On the other hand, the actual ratios in the ISM material from which the Sun formed were physically produced by {\it discrete} additions from stars, to first approximation identical, but spaced in time. The steady-state approach can be validly applied to compare model predictions to the integrated amount of \iso{26}Al and \iso{60}Fe in the Galaxy as obtained by $\gamma$-ray observations\footnote{For sake of clarity, we note that the ratio of the \iso{60}Fe/\iso{26}Al fluxes $dN_{\rm SLR}/dt$ measured via the $\gamma$-ray observatories to be $\sim$0.15 \cite{wang07} needs to be multiplied by $\tau(60)/\tau(26)$ to obtain the observed abundances (Eq.~\ref{eq:basicdiff}), that can be compared to steady-state equilibrium abundances. On the other hand, the flux ratio can be directly compared to the corresponding ratio of stellar yields (after converting them from mass $M_{\rm A}$ to number $N_{\rm A}$ for any SLR of atomic mass $A$ via $N_{\rm A}=M_{\rm A}/A$). This is because the stellar yields also need to be multiplied by $\tau(60)/\tau(26)$ to obtain the steady-state equilibrium abundances (Eq.~\ref{eq:SS2}).}. On the other hand, the $granularity$ of the stellar production events cannot be ignored when considering the SLRs that ended up specifically in the local region of the galactic disk where the Solar System matter was built up, since their half-lives can be comparable to the time interval that elapses between the stellar additions \cite{meyer00,wasserburg06}. 
Effectively, $N_{\rm SLR}$ may not evolve 
smoothly in time but show large fluctuations, depending on the 
relative values of the mean-life $\tau$ of the given SLR and the recurrence time $\delta$ between the
stellar additions of matter from a given production site.  
The crucial parameter thus becomes the ratio $\tau/\delta$. Note that this ratio also
yields the steady-state value, when multiplied by the abundance $p_{\rm SLR}$ produced by each single event, since in the case of discrete events 
$\delta$ represents $dt$ in Eq~\ref{eq:SS2}. 

In this more general case of non-continuous stellar production, the dimensionless  factor describing the temporal evolution of any generic SLR is given by the sum of many terms, each of them representing one single stellar addition, exponentially decayed from such event until the present time $t$:

\begin{equation}
\label{eq:andres}
(1 + e^{-\delta/\tau}+e^{-2\delta/\tau}+...+e^{-(N-1)\delta/\tau}) \times e^{-(t-N\delta)/\tau} = 
\sum_{n=0}^{N-1}e^{-n\delta/\tau} \times e^{-(t-N\delta)/\tau} = \frac{1-e^{-(N+1)\delta/\tau}}{1-e^{-\delta/\tau}} \times e^{-(t-N\delta)/\tau}
\end{equation}

\noindent where the term $e^{-(t-N\delta)/\tau}$ describes the decay of the abundance from the last event, i.e., the first term in the sum, equal to unity.

\begin{figure}[tb]
\begin{center}
\begin{minipage}[t]{16.5 cm}
\includegraphics[width=8.5cm,angle=0]{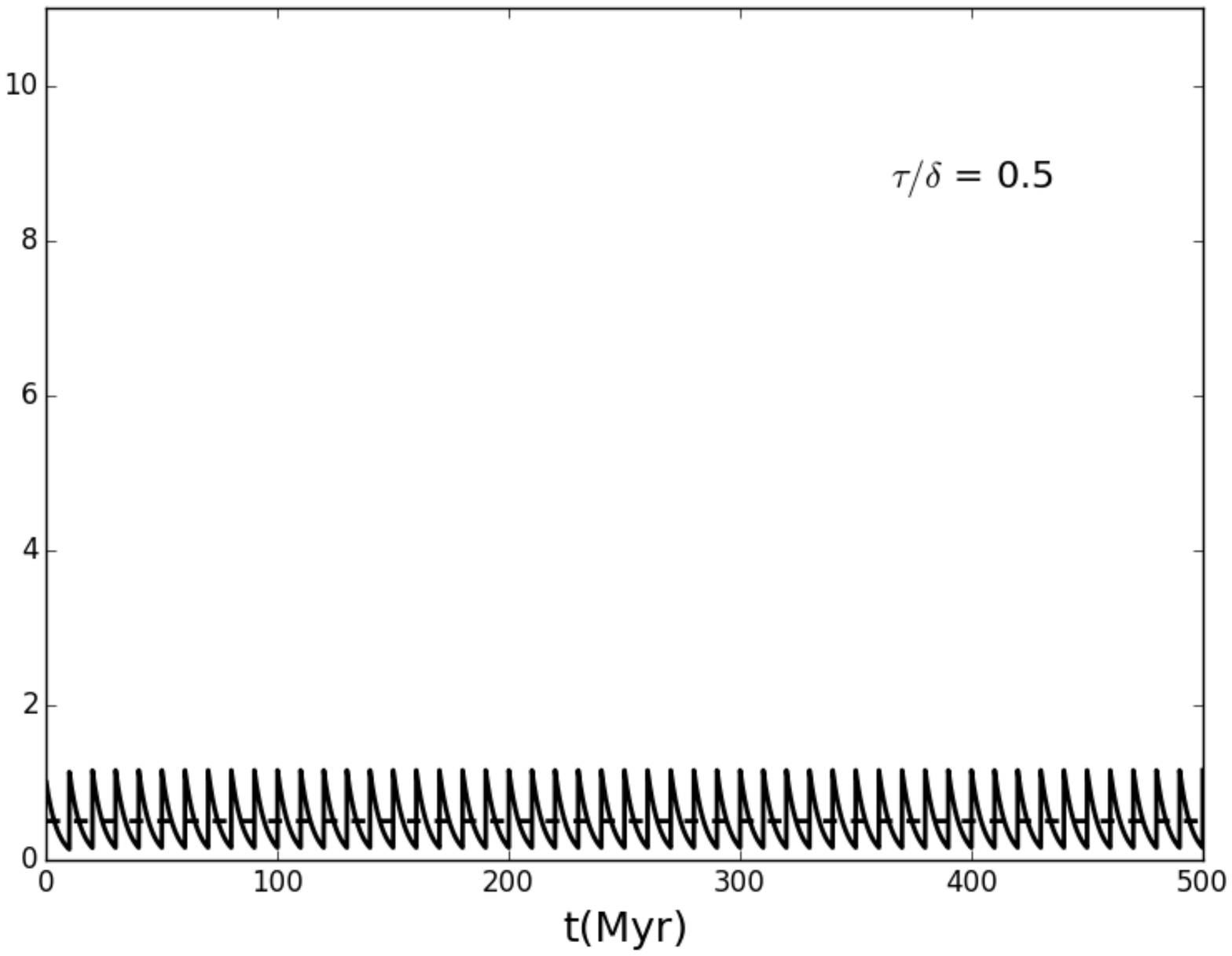} 
\includegraphics[width=8.5cm,angle=0]{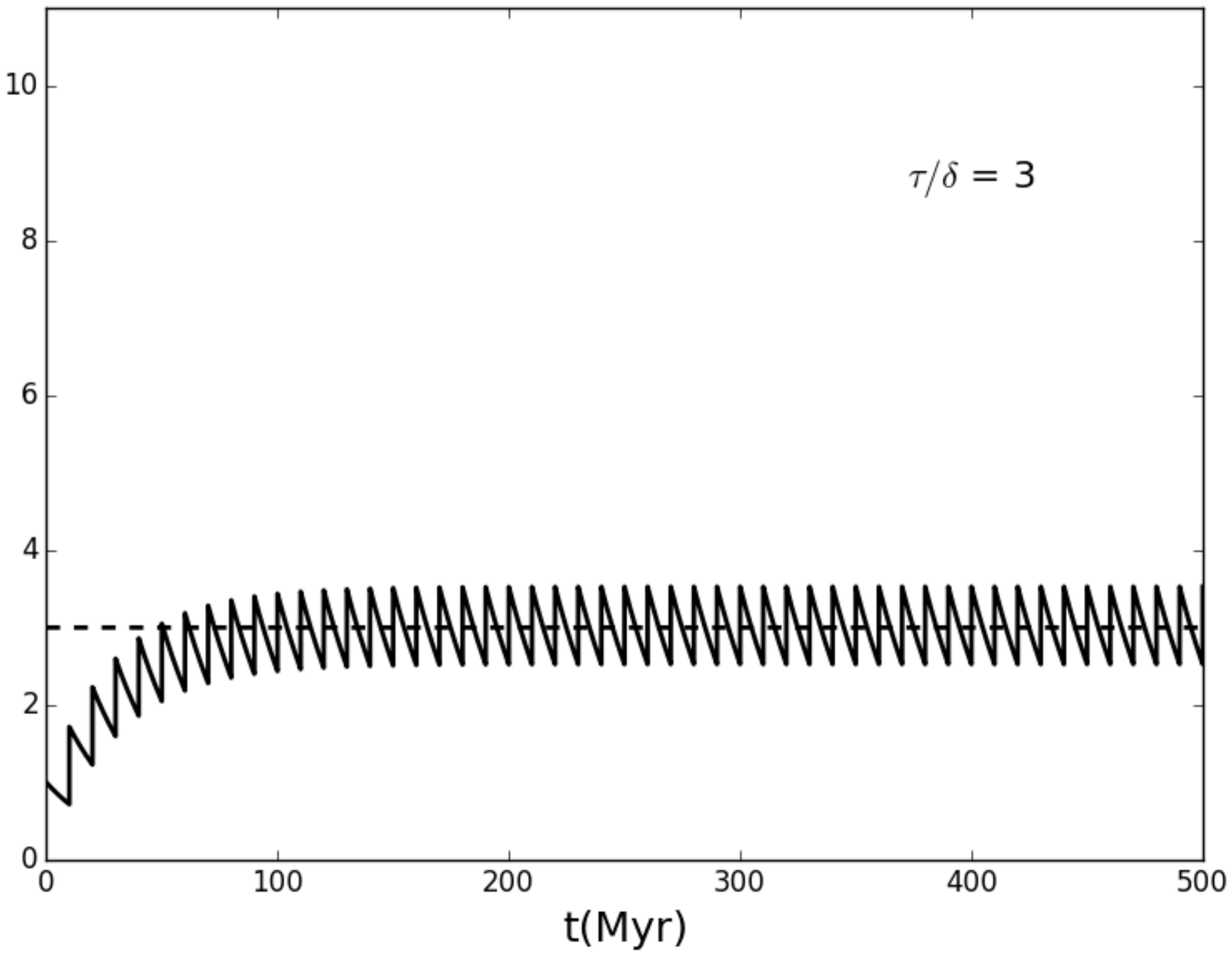} \\
\includegraphics[width=8.5cm,angle=0]{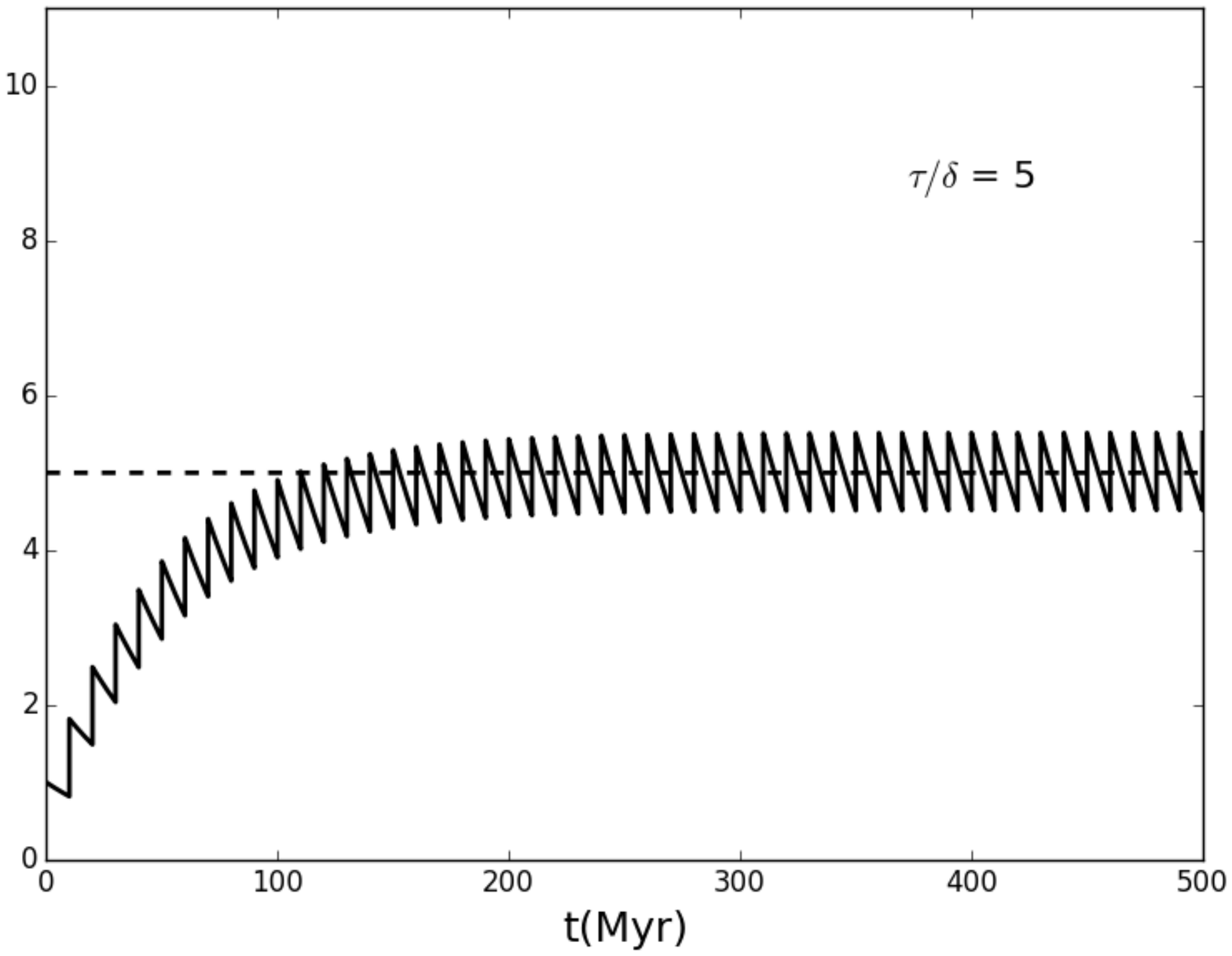} 
\includegraphics[width=8.5cm,angle=0]{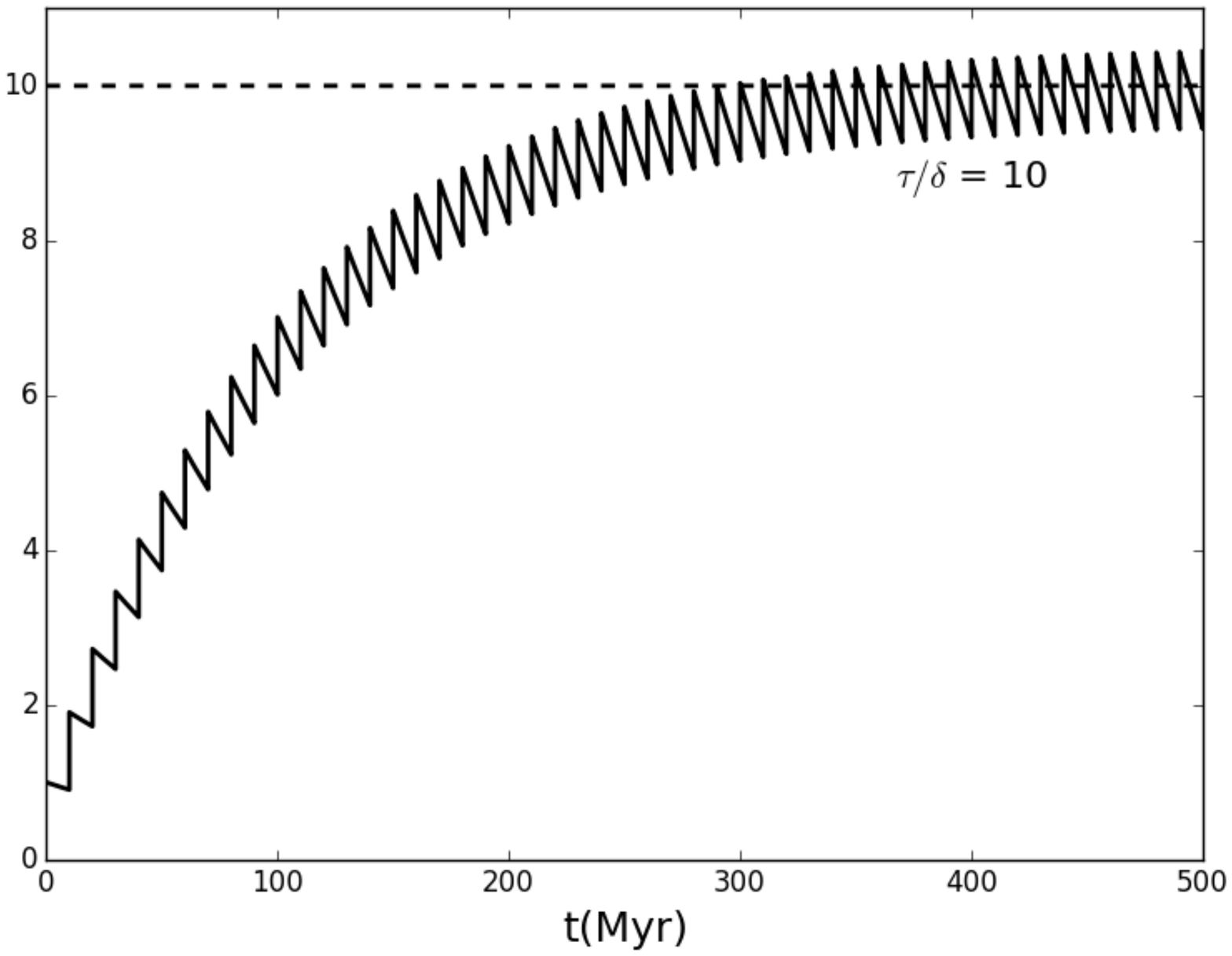} \\
\end{minipage}
\begin{minipage}[t]{16.5 cm}
\caption{Time evolution of Eq.\ref{eq:andres}, assuming $\delta=10$ Myr, and $\tau/\delta$ as indicated in each panel and plotted as the dashed straight line, which also represents the steady-state value. The larger $\tau/\delta$, the more events $N$ it takes to  
reach steady-state. Setting the value of $\tau$ and varying $\delta$ would result in a perfectly equivalent behaviour, except that the time $N \times \delta$ required to reach steady-state would be longer.  
 \label{fig:GCE}}
\end{minipage}
\end{center}
\end{figure}

The behaviour of Eq.~\ref{eq:andres} as function of time for different 
$\tau/\delta$ values is illustrated in Fig.~\ref{fig:GCE}. When $\tau/\delta$ is less than 0.1 the evolution is dominated by peaks representing the discrete stellar additions and the exponential decay in between the peaks. The memory of all the previous events in this case counts for less than 10\% of the total abundance at the peak points. For $\tau/\delta=0.5$ the memory adds 40\% to the peak abundance. Increasing $\tau/\delta$ the memory becomes 
more and more predominant, and the granularity effect can be accounted for as an uncertainty around the steady-state value. The higher the steady-state factor ($\tau/\delta$), the smaller (relatively) becomes the fluctuation around it due to granularity. E.g., for $\tau/\delta =5$ the steady-state factor is $5 \pm 1$, i.e., the relative error bar due to 
granularity is 20\%, for $\tau/\delta =10$, the value is 
$10 \pm 1$ and the relative error is only 10\%. In other words, when $\tau/\delta$ is greater than 10, the abundance can be approximated by the steady-state value within 10\%. 

While the values of $\tau$ are relatively well known, with the exceptions discussed in Sec.~\ref{sec:list}, the values of $\delta$ corresponding to each type of nucleosynthetic event that produced the SLRs are poorly known. 
They can be estimated based on first principle, for example considering how much galactic mass, or volume, is swept by each event, relative to the active star formation area of the galactic disk, and considering the rate in time of the given event in the whole Galaxy. Using this ``snowplow'' approach, Meyer \& Clayton \cite{meyer00} (see their Section 7) estimated, for example the value of $\delta$ for CCSNe to be 5-10 Myr. Following the same reasoning as Meyer \& Clayton, for the typical $s$-process events contributing to the cosmic abundances of the $s$-process elements, i.e., from AGB stars of initial mass between roughly 2 and 4 \msun, one can derive $\delta \approx 50$ Myr. The rate of these AGB events in the Galaxy is higher than that of CCSNe, due to the initial stellar mass function, which favours lower over higher masses, however, $\delta$ is higher for AGB stars than for CCSNe because the slower AGB wind and smaller ejected mass result in the sweeping of a smaller volume of the galactic disk. For NSMs, the potential $r$-process nucleosynthetic events, the velocity is higher but the total mass of the ejecta is lower compared to CCSNe. These effects may balance each other resulting in the same kinetic energy. The event rate is very uncertain, 100-10,000 times lower than the 1-2$\times 10^4$ Myr$^{-1}$ rate for CCSNe, i.e., approximately 1 to 200 Myr$^{-1}$ for NSMs (see discussion in \cite{cote17a}), which covers the rate derived from observation of pulsars in the Milky Way of $\sim$20 Myr$^{-1}$ with large uncertainties \cite{kim15}, Consequently, the value of $\delta$ is also very uncertain, likely $>$ 500 Myr. 

Hotokezaka et al. \cite{hotokezaka15} used an approach based on diffusion instead of snowplowing (see their Eq.~2) and derived average $\delta$ values of roughly 100 Myr for hypothetical $r$-process events with a galactic rate of 300 Myr$^{-1}$, and of roughly 500 Myr for $r$-process events with a galactic rate of 5 Myr$^{-1}$. From analysis of \iso{244}Pu in the ESS and in terrestrial samples they concluded that events with the lower rate are favoured. In this framework, a galactic rate for CCSNe of 20,000 Myr$^{-1}$ would translate into a $\delta$ of roughly 30 Myr, assuming the same mixing length parameter as for the diffusion into the ISM of material from a NSM event. 

Clearly, better determinations of the values of $\delta$ are required from detailed models of GCE, stellar populations and of the dispersion of stellar ejecta in the Galaxy, and individual assessments need to be performed for each SLR. Nevertheless, it is clear that for many of the SLR cases considered here $\tau/\delta$ is likely to be less than unity and the effect resulting from the granularity of the stellar events will need to be considered carefully.

Following from the discussion above, the actual abundance of a given SLR in the ISM material from which the Sun formed can be better described by the following equation, which represents the SLR abundance just after the last stellar addition:


\begin{equation}
\label{eq:andres2}
N_{\rm SLR} \propto p_{\rm SLR}\left(1+\sum_{i=1}^{\infty}e^{-i\delta/\tau}\right) = p_{\rm SLR}\left(1 + \frac{1}{e^{\delta/\tau}-1}\right), 
\end{equation}

\noindent where $p_{\rm SLR}$ is the abundance produced by each single event, 1 represents the last event, and the exponential term represent the memory of all the previous additions. This formula is more general than the steady-state equation, and includes it in the limit $\tau/\delta >> 1$, where the steady-state equation can be recovered by expanding $e^{\delta/\tau}$ into a polynomial series. For the opposite limit, $\tau/\delta << 1$, the memory term goes to zero, and the abundance is simply proportional to how much is produced by the single last event. Note that there is not an equal sign but a proportionality sign in the equation above because 
it does not include the dilution factor representing the distance from the stellar source to the presolar matter. This dilution can be easily factorised by normalising the formula to the abundance of the stable isotope of reference, if it is produced by the same stellar site:

\begin{equation}
\label{eq:stable2}
N_{\rm stable} \propto p_{\rm stable} \frac{T_{\rm gal}}{\delta},
\end{equation}

\noindent where $p_{\rm stable}$ is the abundance produced by each single event (the same type of event that produces the SLRs), and $T_{\rm Gal}/\delta$ the total number of events over the age of the Galaxy before the formation of the Solar System. Combining Eqs.~\ref{eq:andres2} and \ref{eq:stable2}, we obtain:


\begin{equation}
\label{eq:final}
\frac{N_{\rm SLR}}{N_{\rm stable}} = 
K \times \frac{p_{\rm SLR}}{p_{\rm stable}} \times \frac{\delta}{T_{\rm Gal}} \times \left(1 +
\frac{1}{e^{\delta/\tau}-1}\right).
\end{equation}

\noindent Here we have added a parameter $K \geq 1$ representing the potential effects of infall of primordial gas described above in relation to the calculation of the steady-state abundance ratio. 

In summary, Eq.~\ref{eq:final} contains two free parameters: $K$ and $\delta$. Both need to be assessed using detailed models of GCE and stellar populations. Lugaro et al. \cite{lugaro16} attempted at deriving such parameters based on the comparison of the values derived from Eq.~\ref{eq:final} to those derived from the full GCE model of Travaglio et al. \cite{travaglio14} for the four $p$-process nuclei \iso{97,98}Tc, \iso{146}Sm, and \iso{92}Nb,  assuming that they are exclusively produced by SNIa. With four values to match and two free parameters the system in this case is over-determined and self-consistent results were obtained using $\delta=8$ Myr and $K=2$. However, this approach has the intrinsic problem of equating the results from an equation that takes granularity into account to the results of a GCE model that does not. In fact, the derived $\delta=8$ Myr may be too short for the recurrence time of SNIa, since they represent roughly 20\% of all supernovae. Overall, much future work is required to 
assess which $N_{\rm SLR}/N_{\rm stable}$ ratios better represent the galactic contribution to the SLRs in the matter that made up the ESS, as well as most importantly which uncertainties are produced by the granularity effect. 

 \subsection{Deriving timescales}
\label{sec:times}
 
\begin{figure}[tb]
\begin{center}
\includegraphics[width=8.5cm,angle=0]{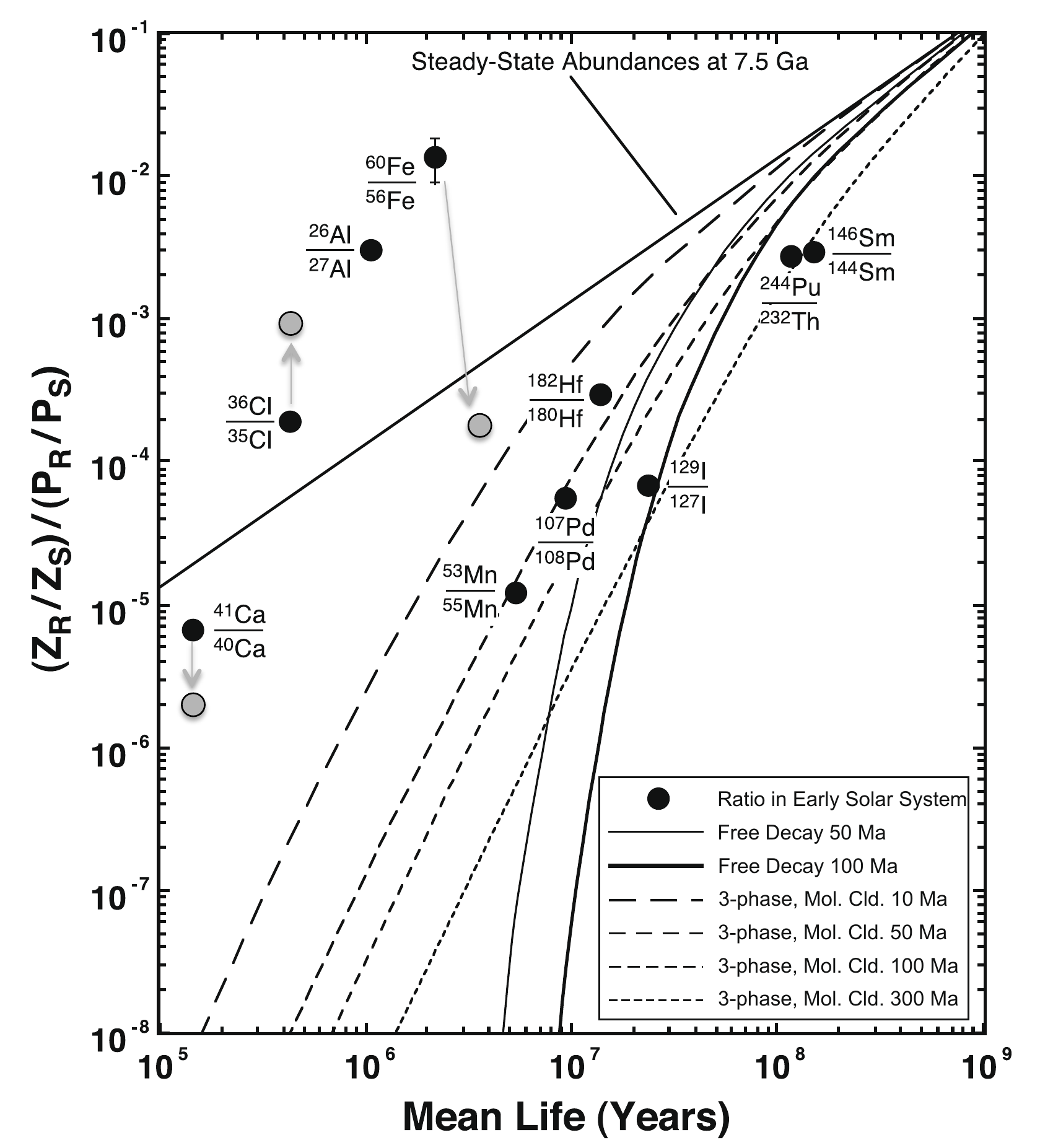} 
\caption{The ESS abundance ratios of SLRs to their stable reference isotope, normalised to their stellar production ratios, are plotted against their mean-lives $\tau$ and compared to models for the steady-state abundances in the ISM (upper solid line). Also shown are model abundances calculated assuming the indicated $T_{\rm isolation}$ (``free decay'', lower solid lines). The dashed curves give steady-state abundances for molecular clouds calculated using the indicated mixing time scales. The figure is updated from Fig.~3 of \cite{huss09}, the grey arrows point to the values obtained using the revised values for the ESS abundances of \iso{36}Cl, \iso{41}Ca, and \iso{60}Fe and the new half-life of \iso{60}Fe, according to Table~\ref{table:SLRs}. \label{fig:times}}
\end{center}
\end{figure} 

\begin{figure}[tb]
\begin{center}
\includegraphics[width=8.5cm,angle=270]{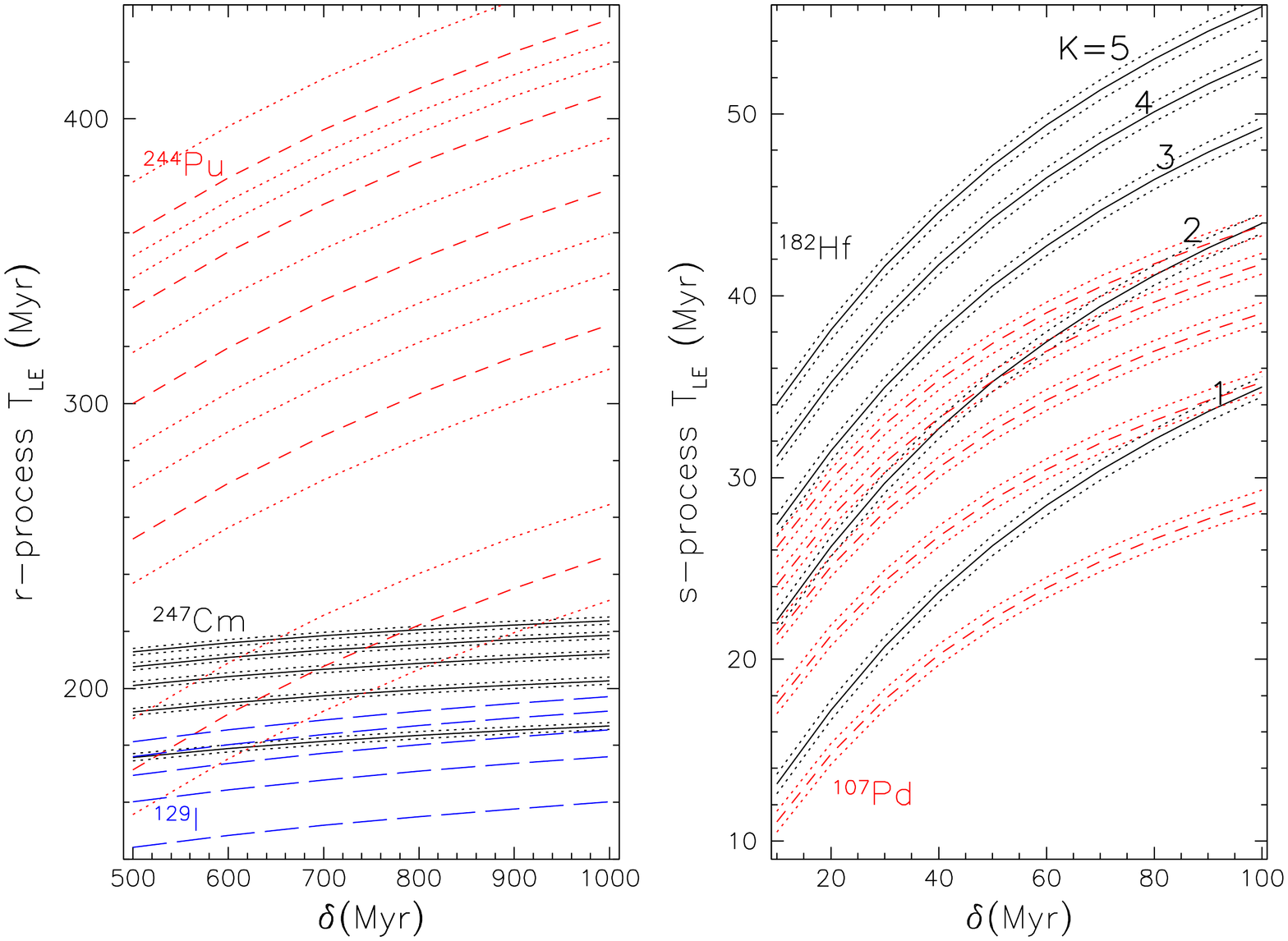} 
\caption{Times from the last $r$-process (left panel) and $s$-process (right panel) events as function of the $\delta$ and 
$K$ values used in Eq.~\ref{eq:final} according to the available SLRs indicated. The dotted lines represent the uncertainty related to the ESS abundance reported in Table~\ref{table:SLRs} (they are not plotted for \iso{129}I because they overlap with the plotted line). \label{fig:times1}}
\end{center}
\end{figure} 

The reason why it is crucial to derive the $N_{\rm SLR}/N_{\rm stable}$ ratios in the galactic parcel of matter that ended up in the stellar nursery where the Sun was born is that by comparing them to those measured in CAIs and other meteoritic materials, using Eq.~\ref{eq:basicdiff} we can derive an estimate of the $isolation$ time, $T_{\rm isolation}$. This time was defined in Sec~\ref{sec:galaxy} as the time interval just prior to the formation of the Sun during which the abundance of the SLR ($N_{\rm SLR}$) was affected by radioactive decay only. This is an implication of the fact that in (G)MCs matter is colder and denser than in the ISM medium preventing mixing between the two (within a $\sim$~100 Myr timescale, \cite{deAvillez02}). The isolation refers to the fact that as mixing was prevented, further stellar additions from the galactic background to the MC where the Sun was born were also prevented. 

One example of the application of such methodology is shown in Fig.~\ref{fig:times}, where the steady-state $N_{\rm SLR}/N_{\rm stable}$ ratio were used to derive the reported time intervals. From results as shown there
it was concluded early on that the isolation times derived from 
several SLRs were inconsistent with each other. Specifically, 
the most short-lived nuclei such as \iso{26}Al and \iso{41}Ca would not 
survive for the isolation times of the order of Myr derived from  less short-lived nuclei like \iso{129}I and \iso{182}Hf. 
Figure~\ref{fig:times} also shows the case when the concept of isolation time is somewhat relaxed by considering slow mixing (with a timescale of the order of 100 Myr) between the warmer ISM and the colder molecular clouds. This moves the lines somewhat upwards in the plot but 
does not change the general conclusions significantly: roughly, the ESS abundances of the SLR nuclei with $T_{1/2} > 5$ Myr appear to be qualitatively compatible with the decay of their ISM abundances, while those with the shorter half-lives (\iso{41}Ca, \iso{36}Cl, \iso{26}Al) are much higher than expected. 
This calls for sources of, e.g., \iso{26}Al much closer in time to the birth of the Sun, as it will be discussed in detail in Sec.~\ref{sec:birth}.
Furthermore, important inconsistencies were also present among 
the less short-lived nuclei themselves, in particular the isolation time derived from \iso{129}I was  much longer than that derived from \iso{182}Hf.
Multiple $r$ processes and/or different local sources for \iso{182}Hf were invoked until, as discussed in Sec.~\ref{sec:107Pd}, it was realised that 
\iso{182}Hf is also produced by the $s$ process in AGB stars.
 
\begin{figure}[tb]
\begin{center}
\begin{minipage}[t]{16.5 cm}
\includegraphics[width=16cm,angle=0]{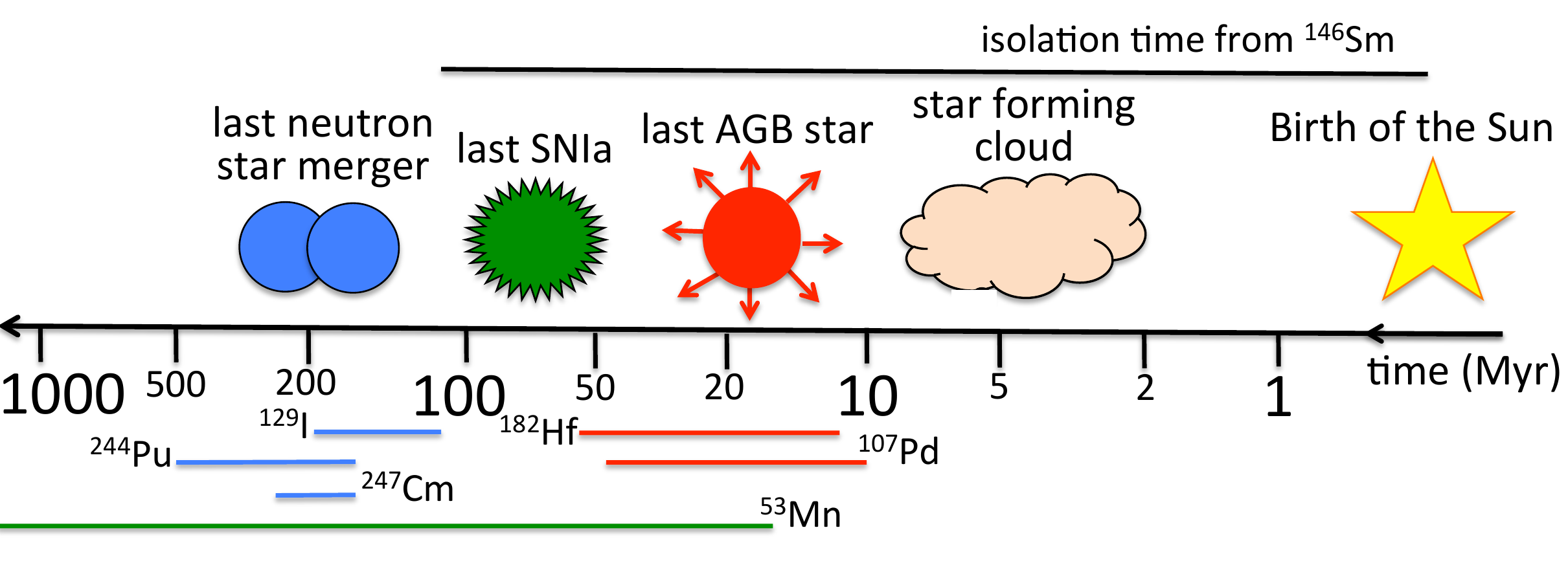} \\
\end{minipage}
\begin{minipage}[t]{16.5 cm}
\caption{Version of Fig.~\ref{fig:Sun} where some of last stellar events to contribute to the Solar System matter have been dated using the specific SLRs indicated and color coded for the different events (red: last AGB star, green: last SNIa, and blue: last neutron star merger). Times are in Myr on a logarithmic scale. The time intervals derive from the error bars on the ESS ratios, and mostly from the uncertainties in the values of $\delta$ (varied between 10 to 100 Myr for the AGB and SNIa nuclei, and between 500 and 1000 Myr for the $r$-process nuclei) and of $K$ (varied between 1 and 5). For $T_{\rm Gal}$ we used 10 Gyr. Possible uncertainties in the stellar production rates were not considered. \label{fig:prehistory}}
\end{minipage}
\end{center}
\end{figure}

In the more general framework of Eq.~\ref{eq:final}, rather than the steady-state equilibrium assumption, depending on the value of $\tau/\delta$ we can in principle provide an estimate for the isolation time only when $\tau/\delta > 3$, with error bars that decrease as $\tau/\delta$ increases (Fig.~\ref{fig:GCE}). Alternatively, for any value of $\tau/\delta$ we can resort instead to estimate the time of the last stellar event that contributed a given SLR and its reference isotope to the Solar System matter ($T_{LE}$). Clearly, the time of a last event represents an upper limit for the isolation time; more precisely, the isolation time cannot be smaller than the time from the last event minus $\delta$. This procedure was used by \cite{lugaro14,lugaro16} and an updated summary of the results is shown in Figs.~\ref{fig:times1} and \ref{fig:prehistory}. A range of $\delta$ values between 10 and 100 Myr is considered there, except for the $r$-process event with $\delta$ between 500 and 1000 Myr. For $K$, a range of values between 1 and 5 were employed, and the error bars in the ESS ratios from Table~\ref{table:SLRs} were taken into account. Only in the case of \iso{244}Pu two values, 4 $\times 10^{-3}$ and 7 $\times 10^{-3}$ were tested in Fig~\ref{fig:prehistory}. Typically, the higher the values of $\delta$ or $K$, the longer the timescale, because the value of $N_{\rm SLR}/N_{\rm stable}$ from Eq.~\ref{eq:final} increases in comparison with the ESS ratio. Clearly better estimates of $\delta$ and $K$ are required to improve this analysis, although Fig.~\ref{fig:times1} already demonstrates that to obtain concordant time estimates for the $r$-process and the $s$-process events the lowest values of $K$ and $\delta$ are favoured. This is because the longer the half-life, the steeper the dependency with $K$ and $\delta$, and the more the lines corresponding to the different isotopes diverge.
We also stress that the stellar production ratios were kept constant (to the specific values reported below), however, these suffer from nuclear and stellar physics uncertainties and could show some variations from event to event, all of which will need to be included for a more complete evaluation of the total uncertainties on the various $T_{LE}$. On top of the SLRs considered in Fig.~\ref{fig:prehistory}, other SLRs may be used to build up the time line of the prehistory of the Sun including \iso{60}Fe, \iso{205}Pb, and \iso{97,98}Tc. The shorter-lived, from \iso{135}Cs to \iso{41}Ca, will also need to be checked in this framework, for consistency.

The relatively most solid results are obtained for the last $s$-process event, a C-rich AGB star\footnote{C-rich AGB stars have C$>$O at their surface due to mixing of material that suffered partial He burning in the deep layers of the star, where also the $s$-process occurs. Models predict that the AGB stars that produce the bulk of $s$-process material in the Galaxy are also C-rich.} of initial mass between 2 and 4 \msun, for which \iso{107}Pd and \iso{182}Hf provide a possible range of intervals in relatively good agreement with each other, for any given values of $K$ and $\delta$. The stellar production factors were taken as typical for the $s$-process: 0.14 and 0.15 for \iso{107}Pd/\iso{108}Pd and \iso{182}Hf/\iso{180}Hf, respectively \cite{lugaro14}. 

The last $r$ process event clearly occurred before the last $s$-process event, as demonstrated by the analysis of \iso{129}I. Although a more detailed statistical analysis of the probability of this specific sequence of events needs to be performed (and may be hampered by our current poor knowledge of the value of $\delta$), the result appears to be consistent with the discussion above that NSMs, the most likely $r$-process production sites are rarer events in the Galaxy than the $s$-process AGB stars in the mass range considered here. The $r$-process production ratio  \iso{129}I/\iso{127}I is well defined by the $r$-process residuals of \iso{129}Xe and \iso{127}I and set to 1.35. On the other hand, the $r$-process production ratios for the two actinides cannot be constrained by the $r$-residual method and are more strongly model dependent \cite{goriely16}. Here, we used 0.3 and 0.4 for the \iso{247}Cm/\iso{235}U and the \iso{244}Pu/\iso{238}U ratio, respectively \cite{goriely16}. Concordance between the two actinides is achieved at the lower limit of the time interval (around 170 Myr), i.e., for values of $K$ and $\delta$ at the low end of the adopted ranges. Applying the same values to the \iso{129}I/\iso{127}I ratio result in a lower time interval of 143 Myr, instead. Concordance could be obtained when using a 22\% longer half-life for \iso{129}I, which has been reported in \cite{timar14}, but is in potential disagreement with current meteoritic data \cite{pravdivtseva17}. It may also be possible that the $r$-process source of \iso{129}I is not the same as that of the actinides; some old halo stars present a so-called ``actinide-boost'', i.e., they are characterised by anomalously high Th/Eu abundance ratios \cite{schatz02,roederer09,ren12,mashonkina14}. Clearly, more investigation is needed.

Further, we consider the production of \iso{53}Mn from SNIa, with a production ratio \iso{53}Mn/\iso{55}Mn of 0.108 \cite{travaglio04}. The derived range in this case is between 15 and 36 My, however, because \iso{53}Mn can also be produced by CCSNe within the stellar nursery of the Sun, it is only possible to give a lower limit for this last event because we may need to add to the $N_{\rm SLR}/N_{\rm stable}$ ratio from GCE also a contribution of these sources within the (G)MC. Note that this is not the case for the $s$- and $r$-process isotopes discussed above, because their sources are all long-lived ($\sim$Gyr) and/or rare stellar objects not expected to be present in shorter-lived ($\sim$Myr) stellar nurseries. 

The case of \iso{146}Sm was considered carefully by \cite{travaglio14} and assumed in that work and here to be produced by the $\gamma$-process in SNIa. Due to its long half-life, this SLR is the closest to potentially providing us with a direct estimate of the isolation time. The corresponding $\tau/\delta$ values for the range of $\delta$ considered in Fig.~\ref{fig:times} are between 1, in which case granularity would give a maximum error of 50\% and 14. Using the upper limit of the estimate of the $N_{\rm SLR}/N_{\rm stable}$ ratio from full GCE models at the time when the Sun was born of 0.01667 \cite{travaglio14}, the resulting upper limit for $T_{\rm isolation}$ is 224 Myr or 112 Myr (the latter shown in Fig.~\ref{fig:times}), using the longer or the shorter possible half-lives, respectively. The lower limit of the $N_{\rm SLR}/N_{\rm stable}$ ratio of $7 \times 10^{-3}$, on the other hand, is below the ratio observed in ESS. This constrain is weak due to the several nuclear physics uncertainties related to both the half-life of this nucleus and its production via the $\gamma$-process. 

Finally, \cite{travaglio14} and \cite{lugaro16} also considered \iso{92}Nb, however we do not include constraints from this nucleus here because its production is still very uncertain. In summary, considering production of \iso{92}Nb in SNIa results in time intervals shorter than those derived from \iso{53}Mn. This indicates that another source of \iso{92}Nb exists in the Galaxy in agreement with the analysis of stable isotopes \cite{travaglio18}. The models by Pignatari et al. \cite{pignatari16a} considered in \cite{lugaro16} found a possible source of \iso{92}Nb in low-mass CCSNe thanks to the $\alpha$-rich freeze-out. However, the stellar uncertainties related to this possibility are very large and other processes may also play a role.

\section{The circumstances of the birth of the Sun}
\label{sec:birth}

Perhaps surprisingly, while we have a good general knowledge of the Sun and of the Solar System, we do not have any consensus on the type of stellar nursery and the 
circumstances in which it was born. A large variety of possibilities and scenarios have been proposed. They are reviewed below in relation to the information derived from SLRs. The role of radionuclides in inferring the circumstances and the environment of the birth of the Sun is twofold: as described in the previous section, information derived from the relatively longer lived (T$_{1/2}$ $>$ 5 Myr) SLRs can allow us to set clocks for the isolation time or the time elapsed between the last nucleosynthetic event and the formation of the Sun. These, in turn, represent precious information on the lifetime of the molecular cloud where the Sun was born. No other means currently exist to derive such information. As mentioned in Sec.~\ref{sec:galaxy}, current observations show that such regions in the Milky Way can live from a few to a few tens of Myr  \cite{hartmann01,murray11}. Lifetimes of 20-30 Myr have been derived from observations of MCs in the Large Magellanic Cloud \cite{kawamura09} and in the galaxy M51 \cite{meidt15}. In the nearby galaxy M33 estimates range from 20-40 Myr  \cite{miura12} to 14 Myr \cite{corbelli17}.

The shortest lived isotopes considered here (T$_{1/2}$ $<$ 2 Myr) most likely did not survive 
significantly throughout the isolation time. This means that a high abundance of them is the 
fingerprint of events that occurred much closer in time to the birth of the Sun. Of special interest 
are SLRs with ``intermediate'' half-lives, 2 Myr $<$ T$_{1/2}$ $<$ 5 Myr, among which are \iso{53}Mn 
and \iso{60}Fe. They may need to be considered both as clocks and as fingerprints. In the case of the 
currently much discussed \iso{60}Fe, its role depends on the difficult determination of its ESS 
value. As discussed in Sec.~\ref{sec:60Fe}, the lowest reported values of \iso{60}Fe/\iso{56}Fe 
($\sim 10^{-8}$) can be explained by the decay of the \iso{60}Fe abundance observed in the ISM via 
$\gamma$-ray astronomy, in which case \iso{60}Fe would act as a clock \cite{tang12}, instead, the 
highest reported values ($\sim 10^{-7}$ - $10^{-6}$) would require an extra source, in which case 
\iso{60}Fe would act as fingerprint.

A vast amount of literature in the past 40 years has been devoted to trying to solve the puzzle of short-lived radioactivity in the ESS. To simplify the complexity of the problem it is useful to separate it into three different questions: 

\begin{itemize}

\item[Q1]{Which stellar sources produce the SLRs in the proportion needed to reproduce their observed ESS abundances?} 

\item[Q2]{Which physical mechanisms allowed such nuclei to be effectively incorporated into the first solids (CAIs, etc) that formed in the ESS?}

\item[Q3]{Which scenarios provide a plausible environment for the birth of the Sun?}

\end{itemize}

While most authors have focused on one or two of these questions, clearly a final answer to the origin of SLRs in the ESS is possible only if all three points are satisfied. To answer the first question the accuracy and precision of the ESS analytical data must be considered carefully against the uncertainties related to the production of SLRs in stars, coming both from nuclear and stellar physics. Furthermore, the contribution discussed above from the GCE to each SLR needs to be accurately established, since only what cannot be explained via the workings of GCE should be attributed to specific events related to the circumstances of the birth of the Sun. In general, as mentioned above, the origin of SLR nuclei with roughly T$_{1/2} > 5$ Myr is dominated by the decay of their abundances in the ISM produced by GCE. However, if the local stellar source invoked to produce the SLRs with shorter life times, e.g. \iso{26}Al, also produces some of the longer living SLRs, then the two components need to be added up. This is particularly relevant for the SLRs with intermediate half-lives mentioned above such as \iso{53}Mn and \iso{60}Fe. The timing of the pollution -- injection into the protosolar cloud or into the already formed disk -- and the form in which the nuclei were transported and incorporated (gas or dust) are relevant to answering the second question. For the third, scenarios for the circumstances of the birth of the Sun need to be considered within the currently established framework for star formation. The likelihood of any such proposed scenario (in terms of probability) also needs to be investigated. Moreover, there are external, independent constraints that should be considered for all three questions. For example, in relation to Q1, the processes that produce SLRs in stars must also be able to explain observations of such nuclei in the Galaxy, on Earth, and in meteoritic stardust grains. In relation to Q2, considering dust as the carriers of SLRs from a given source must be examined in relation to direct observations and theories of the production of such dust. Finally, an answer to Q3 must also account for other properties specific to the Solar System, for example, the stability of the planetary orbits \cite{adams10}, the fact that the disk is truncated at 30 AU \cite{portegies18}, and even its observed spin-orbit misalignment \cite{wijnen17}. 

In the following three sections we summarise past and current attempts at answering the three questions above, their pros and cons, and their implications. We present them in such a structured way for sake of clarity. In reality, however, the three questions also provide constraints to each other and it is always necessary to keep in mind the links between them. 

\subsection{The stellar sources}
\label{sec:Q1}

A simple but quick, effective method to test stellar sources against Q1 is to consider by which factor $f$ the stellar yield of a given SLR ($M_{\rm SLR}^{\star}$) needs to be diluted to match its abundance in mass observed in the ESS ($M_{\rm SLR}^{\rm ESS}$). In other words, which fraction of the total mass of the SLR ejected from the star we need to embed into the presolar material:

\begin{equation}
f = \frac{M_{\rm SLR}^{\rm ESS}}{M_{\rm SLR}^{\star}}.
\label{eq:f}
\end{equation}

\noindent For example, in the case of \iso{26}Al, we can consider the value of the \iso{26}Al/\iso{27}Al ratio from Table~\ref{table:SLRs} and the mass fraction of \iso{27}Al in the Solar System of $6.22 \times 10^{-5}$ \cite{lodders09}. If we assume that the mass of the ESS to be polluted is 1 \msun, then this \iso{27}Al mass fraction corresponds also to its total ESS mass, and $M_{\rm 26Al}^{\rm ESS}=3.1 \times 10^{-9}$ \msun\ in total was present in the ESS. 
Typical values of $f$ needed to match this ESS abundance of \iso{26}Al are of the order of $10^{-3}$-$10^{-5}$, since $M_{\rm 26Al}^{\star}$ is in the range $10^{-4}$-$10^{-6}$ \msun (see, e.g., \cite{lugaro14}). This simple formulation assumes that no contribution from GCE is already present, which may be correct for \iso{26}Al, but probably not for, e.g., \iso{60}Fe or \iso{55}Mn. 

Clearly, a stellar source is a good candidate polluter only if $f$ is the same for all the SLRs under consideration, in which case $f$ corresponds to the overall dilution factor for the whole stellar ejecta: $M_{\rm injected}^{\rm ESS}/M_{\rm total}^{\star}$, where $M_{\rm injected}^{\rm ESS}$ is the mass from the stellar source added to the ESS and $M_{\rm total}^{\star}$ the total mass ejected from the stellar source. At this point, however, we cannot attribute a true physical meaning to the absolute value of $f$ derived from Eq.~\ref{eq:f}, because such a value is meaningful only when considered in comparison to the value derived for different SLRs. For example, as will be discussed in the next section, the value of $M_{\rm SLR}^{\rm ESS}$ assumes a given mass for the ESS. This assumption depends on the scenario considered, 1 \msun\ may be reasonable, for example, in case the protosolar cloud is polluted -- although it probably represents a lower limit since young stellar object{\bf s} may lose up to half the initial mass of the collapsing cloud via jets. On the other hand, if we consider a scenario where the protosolar disk is polluted, much lower ESS masses should be considered down to 0.01 \msun, depending on the evolutionary phase of the disk. In this case, for example, $M_{\rm 26Al}^{\rm ESS}$ would decrease accordingly by a factor of 100 with respect to the value given above, to $3.1 \times 10^{-11}$, and the value of $f$ will also decrease to $10^{-5}$-$10^{-7}$. In any case, we stress that for the purpose of answering Q1 the absolute value of $f$ is not relevant because the aim here is to match the {\it relative} proportions of the SLRs considered. The absolute value of $f$ is important for Q2 because it needs to be compared to the value derived from the physical pollution process: which depends on the distance from the stellar source and the injection efficiency (Sec.~\ref{sec:Q2}). 

The approach of Eq.~\ref{eq:f} represents an oversimplification of the scenario of a single nearby stellar polluter also because it usually considers total stellar yields, i.e., it assumes that the distribution of the SLRs is homogeneous in the ejecta, and it does not account for differences in condensation properties of the different SLRs, which may be relevant in relation to possible differences between gas and dust accretion (Sec.~\ref{sec:Q2}). Still, it is a powerful tool in that it allows to give a quick, quantitative general answer to the question if a given source produces the SLR abundances in the relative distribution required to match the meteoritic data, and to identify problems that may be present.

Two further issues need to be noted. A potential time delay $\Delta t$ between the time the SLR is ejected by the stellar source and its incorporation into the first solids is not included Eq.~\ref{eq:f}. Such a time interval has usually been considered also in the simple analysis above \cite{wasserburg06,lugaro12a,lugaro14}. To do that, the number of SLR nuclei derived after a given dilution $f$ is applied needs to be multiplied by a factor $e^{-\Delta t/\tau}$, where $\Delta t$ represents a second parameter, after $f$. However, since usually the analysis is applied to more than two SLRs, the system is still over-determined. In general, it is found that the value of $\Delta t$ is around 1 Myr and is controlled by the shortest-lived SLR under consideration, i.e., \iso{41}Ca, and that it does not greatly affect the abundances of the other SLRs.

\begin{figure}[tb]
\begin{center}
\begin{minipage}[t]{15 cm}
\includegraphics[width=15cm,angle=0]{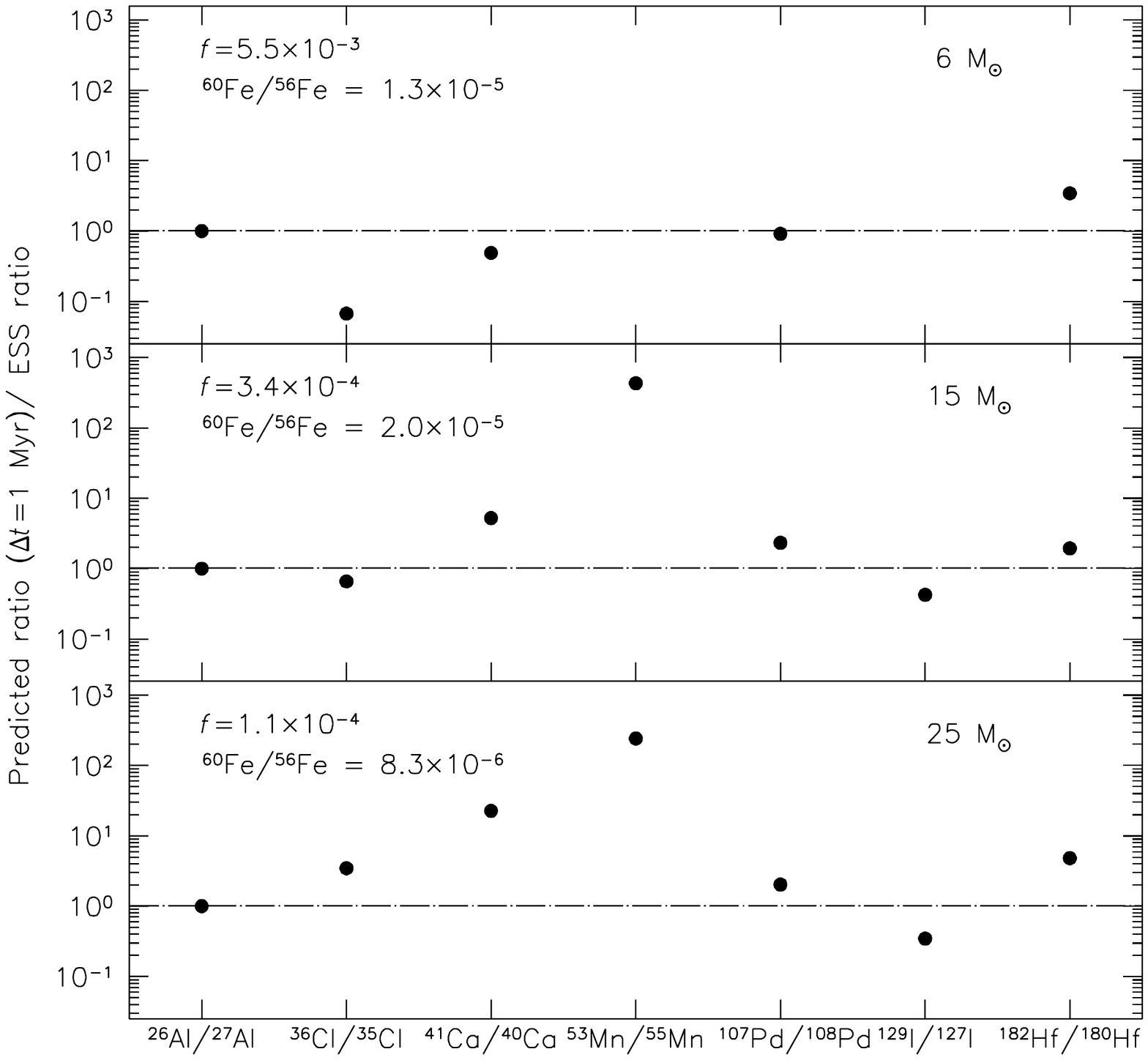}
\caption{Results for selected SLRs from a model that assumes injection from a single stellar source of initial mass 6, 15, and 25 \msun, and a time delay $\Delta t=1$ Myr. The predicted \iso{60}Fe/\iso{56}Fe ratios are indicated in each panel. In the 6 M$_{\odot}$ model, \iso{129}I/\iso{127}I is off scale, many orders of magnitude below unity and  \iso{53}Mn/\iso{55}Mn is zero.  Updated from \cite{lugaro14}. \label{fig:scipaper}}
\end{minipage}
\end{center}
\end{figure}

An example of the exercise described above can be found in Fig.~\ref{fig:scipaper}, where the yields from a typical massive AGB star (of 6 \msun, initially) and two CCSNe (of masses 15 and 25 \msun, initially) are compared to the ESS ratios. As discussed in Sec.~\ref{sec:26Al}, stars of masses below roughly 6 \msun\ have difficulties producing enough \iso{26}Al, at the same time they supply too large amounts of the $s$-process isotopes of the elements heavier the Fe \cite{wasserburg17}. We thus do not consider low-mass AGB stars further, also due to the additional difficulty in envisaging a scenario to answer Q3, i.e. how low-mass AGB could have contributed to the composition of the Sun at its birth. The reason is that these stars are very long lived ($\sim$ 1 Gyr) and are not expected to be associated with star-forming regions \cite{kastner94}.

More massive AGB and Super-AGB stars of mass above 6 \msun\, live much shorter lives ($<$ 50 Myr) and could be envisaged to have been present or have contributed to the composition of the early Solar System \cite{trigo09,lugaro12a,wasserburg17}. These stars produce copious amounts of \iso{26}Al thanks to the proton captures occurring at the base of the convective envelope (see Sec.~\ref{sec:26Al}). Production of all the other SLRs is due to neutron captures and is linked to the dredge-up episodes that carry material into the envelope from the deep He-rich layer, where neutron-capture processes can occur. In the top panel of Fig.~\ref{fig:scipaper} we show a typical example for which a self-consistent solution could be found for a number of SLRs, except that the abundance of \iso{36}Cl is too low, no \iso{53}Mn is produced in these stars, and the \iso{60}Fe/\iso{56}Fe is more than two orders of magnitude higher than the currently recommended value around $10^{-8}$ \cite{wasserburg17}, or a factor of 4 higher than the highest observed values. Similar results can be found for Super-AGB stars of initial masses between 6 and 9 \msun\, \cite{lugaro12a}. However, it should be noted that the efficiency of the dredge-up episodes is one of the main uncertainties in AGB models. For example, if these stars did not experience any dredge-up, no SLRs would be ejected by their winds, except for \iso{26}Al, which is produced directly in the envelope. In this case, it may be plausible to attribute the origin of the other SLRs either to the effect of GCE and/or to local {\it in situ} non-thermal nucleosynthesis together with the Be isotopes. 

The middle and bottom panels of Fig.~\ref{fig:scipaper} present the same dilution exercise for two typical CCSNe. As first proposed by \cite{cameron77}, many authors still consider a nearby CCSN as one of the most promising candidates to have injected the SLR nuclei into the protosolar molecular cloud \cite{Gritschneder12,boss14,pan12}, or into the already formed protoplanetary disk \cite{hester04}. Inferences on and dynamical simulations of the birth environment of the Sun are mostly based on this scenario \cite{looney06,adams14,parker14}. However, as outlined by several authors \cite{wasserburg06} and as seen in Fig.~\ref{fig:scipaper}, many problems are related to the nearby CCSN scenario in terms of abundances. The two features that stand out most in this case are the overproduction of \iso{53}Mn and of \iso{60}Fe. \iso{53}Mn is roughly three orders of magnitude higher than the value observed, \iso{60}Fe between one and three orders of magnitude higher, depending on the choice of the ESS value. This are old-standing problems for CCSNe as candidate sources, and a number of CCSN models have been presented that attempt to resolve them \cite{takigawa08}. One possible traditional solution, at least to the \iso{53}Mn overproduction, is the idea that the inner part of the CCSN ejecta rich in \iso{53}Mn was not incorporated into the ESS, possibly because of extensive fallback of matter onto the CCSN remnant. For this to work it is also required that no significant mixing occurs between the inner and the outer ejecta prior to the fallback, i.e., \iso{53}Mn is not carried to the outer layers. Reducing the amount of \iso{53}Mn via a stronger fallback also reduces the amount of \iso{56}Ni, produced in the same region as \iso{53}Mn, below the value observed in typical CCSNe via its radioactive decay that powers their light curves. In this case, the CCSN that polluted the ESS would have been somewhat fainter than a typical CCSN, which adds to the problem of the low probability for such an event, as discussed in Sec.~\ref{sec:Q3}. Alternatively, it could be considered that the SLRs are required to be in the form of dust in order to be trapped by the protosolar disk and that matter from the CCSN inner layers does not condense into dust grains\footnote{However, meteoritic silicon carbide grains of population X (SiC-X) from CCSNe carry excesses in \iso{44}Ca, the daughter of radioactive \iso{44}Ti (T$_{1/2}$=60 yr) produced in the inner layers of CCSNe.}. Another possible scenario to be investigated is the case of asymmetric CCSN ejecta, potentially powered by different explosion energies. The ESS may have collected the ejecta only from a given direction, with a specific isotopic composition, rather than the bulk average of the CCSN. That CCSNe are asymmetric is demonstrated by observational evidence \cite{wang08} and has been invoked to explain the composition of meteoritic silicon carbide (SiC) stardust of population X \cite{pignatari13c}. The effect on the production of \iso{26}Al, \iso{56}Fe, and \iso{53}Mn has not been investigated yet, however, it may prove difficult to decouple the production of \iso{26}Al and \iso{60}Fe, which are both produced in the more external layers of the CCSNe with respect to \iso{53}Mn \cite{timmes95a,limongi06}. 

No models of CCSNe have been proposed yet that can avoid overproduction of \iso{60}Fe relative to the value recommended in Table~\ref{table:SLRs}. Furthermore, as discussed in Sec.~\ref{sec:60Fe}, the \iso{60}Fe/\iso{26}Al ratio calculated for the ESS is approximately 0.2 - 0.002, while the value derived from the  $\gamma$-ray flux ratio is $\sim$0.55, three times to more than two orders of magnitudes higher than observed in the ESS. Consequently, even the CCSN models that can provide a match to the $\gamma$-ray observations \cite{limongi06} naturally result in overproduction of \iso{60}Fe when compared to the ESS \cite{vasileiadis13}. Note that most CCSN models produce \iso{60}Fe/\iso{26}Al abundance ratios of the order of 3, i.e., 6 times higher than that derived from $\gamma$-ray flux (see, e.g. \cite{austin17}), although recent detailed CCSN models predict values around 1.3, i.e., a discrepancy with the $\gamma$-ray data of less than factor of 3 \cite{sukhbold16}. 

\begin{figure}[tb]
\begin{center}
\begin{minipage}[t]{15 cm}
\includegraphics[width=15cm,angle=0]{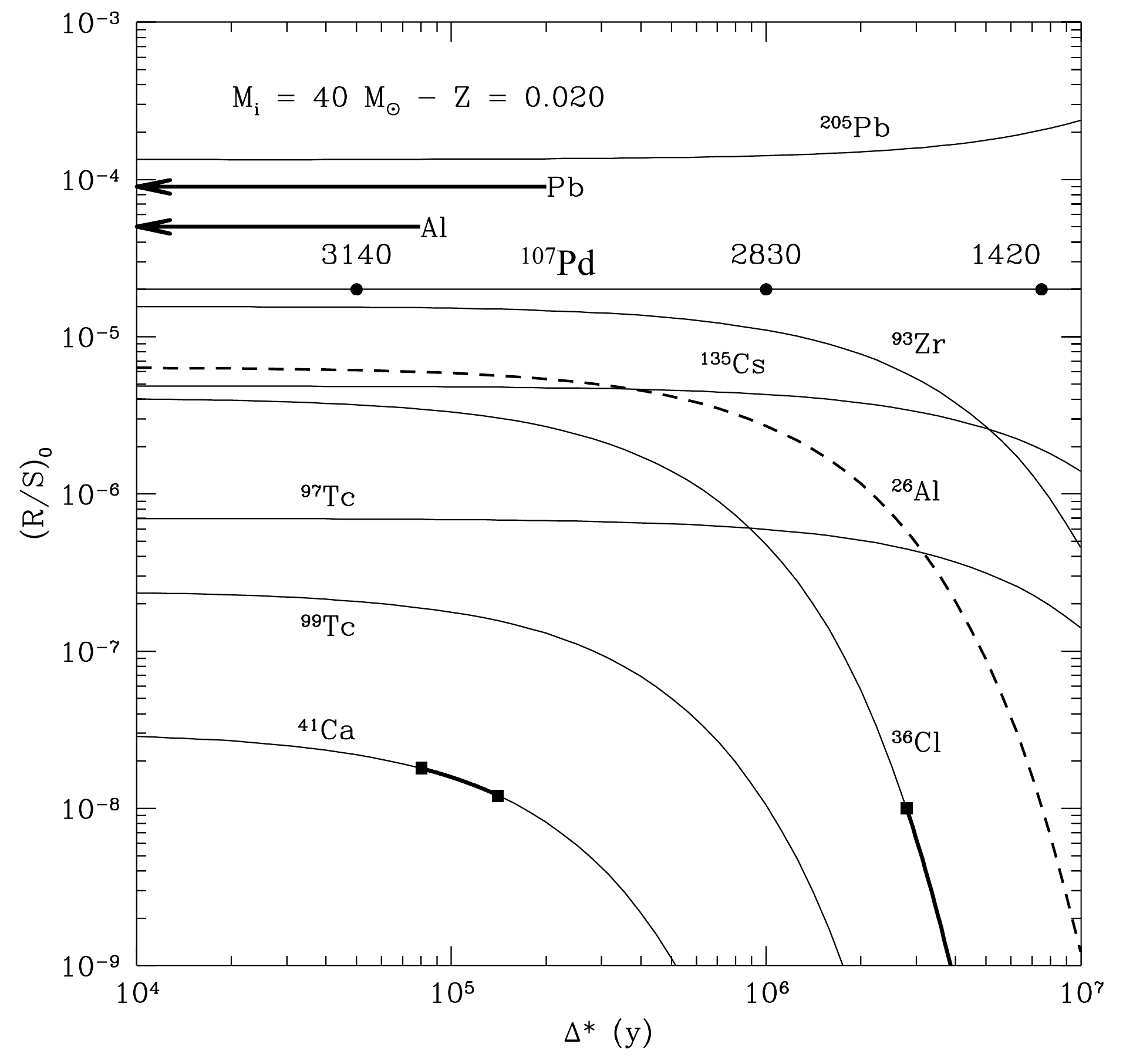}
\caption{Ratios of SLRs to stable isotopes of reference \cite{arnould97} (the labels only indicate the SLR) predicted in the material obtained by mixing matter of Solar System composition with the winds of a WR star of initial mass 40 \msun\, and metallicity 0.02, and let the mixture decay for the time $\Delta^*$ (the x-axis). The dilution factors $1/f$ are calculated as a function of $\Delta^*$, such as the \iso{107}Pd/\iso{108}Pd ratio is set to $2 \times 10^{-5}$. They range between roughly 1500 and 3000, as indicated on the solid flat line at $2 \times 10^{-5}$. The thicker parts on the \iso{41}Ca and \iso{36}Cl lines represent the range of ESS ratios considered by \cite{arnould97}. Credit: Arnould et al., A\&A, 321, 452, 1997, reproduced with permission \textcopyright ESO. \label{fig:arnould}}
\end{minipage}
\end{center}
\end{figure}

Due to the various difficulties for CCSNe in producing the abundances of SLRs in the proportion required to match the ESS values, and mostly due to the potentially prickling issue of the overproduction of \iso{60}Fe relatively to \iso{26}Al, several authors \cite{gaidos09,gounelle12,young14} have turned their attention to the pre-supernova phase of massive stars, in particular to the winds that can be significantly present already during the main sequence phase of WR stars (Sec.~\ref{sec:stars}), These appear to be able carry a number of SLRs, including \iso{26}Al, as well as some amounts of other SLRs such as \iso{36}Cl, \iso{41}Ca, \iso{107}Pd, and \iso{205}Pb, without producing any \iso{53}Mn, which is a typical explosion product, nor overproducing \iso{60}Fe \cite{arnould06,arnould97}. The reason is that the neutron density in these stars is not high enough to efficiently activate the \iso{59}Fe branching point described in Sec.~\ref{sec:60Fe}. 

Figure~\ref{fig:arnould} shows an example of the simple dilution model as applied to the winds of a 40 \msun\, star \cite{arnould97}. The dilution in that study was set so that the \iso{107}Pd/\iso{108}Pd ratio recommended at the time was matched. Assuming that the ESS abundance of \iso{107}Pd is a product of GCE (Fig.~\ref{fig:prehistory}) would relax this constraint and potentially allow a lower value for the dilution factor, leading to an overall increase of the ratios plotted in Fig.~\ref{fig:arnould}, and the possibility to obtain the \iso{26}Al/\iso{27}Al canonical value. Further reports of \iso{26}Al production in WR stars in relation to its ESS abundance can be found in \cite{gounelle12,dwarkadas17}. To our knowledge, following the rotating models presented by \cite{arnould06}, which do not change the picture substantially, there have been no more recent investigations dedicated to analysing the production of the other SLRs on top of \iso{26}Al in these stars in the light of the simple dilution model described above, in particular, not for the new ESS recommended values and with nuclear and stellar physics updates. Such a study is clearly urgently required. Furthermore, it has become evident in recent years that most massive stars experience binary interaction at some point in their life \cite{sana12}. Such an interaction might, for example, result in winds during the pre-supernova phase even in the case of stars of lower masses than in the standard single WR-star scenario. Consequently, it is also urgent to investigate the production of SLRs by binary massive star systems.

\subsection{The injection mechanism}
\label{sec:Q2}

Proceeding from the simple model illustrated above, in relation to the injection mechanism the physical meaning of the dilution factor $f$ effectively includes two aspects: the first is the distance from the source, i.e., the geometric dilution $f_{\rm d}$, the second is the injection efficiency $f_{\rm inj}$, which varies from 0 to 1, and represents the fact that it may not be possible to trap all the available, incoming stellar ejecta into the ESS, depending on the element and the injection mechanism. Ultimately, the final aim of a concerted effort to answer Q1 and Q2 should result in the same value of $f$ calculated as $f = f_{\rm d} \times f_{\rm inj}$ and as by  Eq.~\ref{eq:f}.

Let us first consider the geometric factor only. In relation to the distance the fraction of stellar mass that intercepts the protosolar nebula or disk relative to the total amount of mass ejected is represented by: 

\begin{equation}
\label{eq:intercept}
f_d = \frac{A_{\rm ESS}}{A_{\rm sphere}}=\frac{r^2_{\rm ESS}}{4 d^2}
\label{eq:f2}
\end{equation}

\noindent where $A_{\rm ESS}=\pi r^2_{\rm ESS}$ is the area covered by the ESS material, i.e., the protosolar nebula or disk of radius $r_{\rm ESS}$ and $A_{\rm sphere}=4 \pi d^2$ is the area of the surface of the sphere covered by the stellar ejecta at the distance $d$ between the star and the protosolar nebula or disk. 


In the simple case when $f=f_d$, the distance $d$ can be calculated, by combining Eq.~\ref{eq:f} and \ref{eq:f2}, as:

\begin{equation}
\label{eq:distance1}
d^2 = \frac{r^2_{\rm ESS} \times M_{\rm SLR}^{\star}}{4 \times M_{\rm SLR}^{\rm ESS}}.
\end{equation}


The final value of the distance depends not only on the specific stellar source considered, which controls the value of $M_{\rm SLR}^{\star}$, but also on the particular scenario considered, which controls the values of $r_{\rm ESS}$ and $M_{\rm SLR}^{\rm ESS}$. Depending on the timing of the injection, the value of $r_{\rm ESS}$ can vary from, e.g., 0.5 pc, when considering a diffuse protosolar nebula, down to 0.0005 pc (100 AU), when considering a protosolar disk instead. Also, as discussed in the previous section, $M_{\rm ESS}$ can vary from twice the mass of the Sun, down to 0.01 - 0.5 \msun, when considering the disk. Because $d$ follows $r_{\rm ESS}$ and the square root of $M_{\rm ESS}$, the dependency on $r_{\rm ESS}$ is stronger.

For example, when considering pollution by a CCSN that produces a typical total mass of \iso{26}Al $\simeq 5 \times 10^{-5}$ \msun, $M_{\rm ESS}$ = 0.01 \msun, and a disk of typical radius 100 AU, distances required are of the order of 0.3 pc. We recall that in this case the mass of \iso{26}Al in the ESS is $3.1 \times 10^{-11}$ \msun, i.e. 100 times less than reported at the start of Sec.~\ref{sec:Q1} given that the mass of the ESS to be polluted is 100 times smaller. If instead we imagine pollution of a more dispersed cloud of radius 0.1 pc and mass 1 \msun, inferred distances are of the order of 6 pc. This difference has an impact on the scenarios to be considered, as well as on the consideration that a too close CCSN may disperse the nebula or destroy the disk.

The whole picture of the significance of $f$ must also incorporate the effect of the {\it injection efficiency}. 
In simple words, $f_{\rm inj}$ represents the difficulty of mixing hot, energetic matter coming from the stellar source (the energy depending on the speed of the winds or the supernova ejecta) into the cold protosolar cloud, or the relatively dense disk. Furthermore, if the material was incorporated in the form of dust, chemical fractionation of different elements (and hence different injection efficiencies) could result depending on the likelihood of a given element to be incorporated into dust grains. Many of the SLRs considered here belong to relatively refractory elements and one could assume in first approximation that there should not be huge variations in the injection efficiency for different elements. However, when considering 
rather volatile or moderately volatile elements such as I, Cl, Pb, Sn, Cs (as well as O) the injection efficiency could be different. 

In general, two types of injection mechanisms have been investigated. The original idea was the triggering the formation of the Solar System by the interaction of the cold molecular cloud with hot supernova ejecta ({\it early} injection) \cite{cameron77}. This idea has been investigated in much detail in relation to supernova ejecta as well as  WR-star and AGB winds  \cite{boss08,boss10a,boss10,boss13,boss14,boss15,boss17,Gritschneder12} in order to derive if it is possible to inject material from stellar, hot ejecta into the cold molecular cloud. The injection efficiency has been calculated via detailed hydrodynamical models, which can also predict the potential cloud disruption. Generally, it has been found that is possible for a nearby CCSN to trigger the collapse of the cloud and at the same time inject enough SRLs to account for their ESS abundances. For example, \cite{boss15} reports an injection efficiency of up to 0.1. 
In the same category of {\it early} injection can also be considered the scenario proposed by \cite{tatischeff10}, where the bow shock shell of a runaway WR star allows the injection of the SLRs present in the winds, with the following supernova ejecta as the possible collapse triggering factor. The scenario proposed by \cite{gounelle12,dwarkadas17} also invokes the winds of a massive WR star as the origin of \iso{26}Al but assumes that the formation of the Sun takes place in the gas belonging to the dense shell compacted by such winds. The shell would be enriched by these winds in SLRs to the required levels if the injection efficiency was at least 1\%.

The other type of scenario envisaged instead considers a {\it late} injection of stellar material directly in the already formed disk \cite{hester04}. This scenario is based on observations of protostars located on the edges of bubbles generated by the UV radiation from massive stars, which will subsequently explode as CCSNe and pollute the disk. For the injection of SLRs from the CCSN ejecta material into the disk to be efficient, the atoms must be trapped into dust grains large enough ($\simeq \mu$m) to be captured by the disk \cite{ouellette07,ouellette10,goodson16,close17,portegies18}. Discussion on this idea is also related to the problem mentioned above (Sec.~\ref{sec:26Al}) that we do not observe variations in the O isotopic ratios between, e.g., CAI and micro-corundum grains that are rich in \iso{26}Al and those that are poor in \iso{26}Al. This means that in the first approximation given by the simple dilution models described above, the O isotopic ratios before and after mixing should not be altered to an observable level. All stars from the massive, Super-AGB to the CCSNe  discussed above have problems with matching this constraint within the {\it late} injection scenario \cite{gounelle07,ellinger10,lugaro12a}. A possible way to solve this issue is that O and Al have different injection efficiencies, due to the former not being fully incorporated into dust \cite{ellinger10}. In fact, for example, considering material of solar composition, only roughly 1/3 of all the O abundance would be locked into dust, which is the amount corresponding to the total abundance of all the refractory heavier metals that we can assume to form oxygen-rich dust. 

\subsection{The environment of the birth of the Sun}
\label{sec:Q3}

The main problem with polluting the nascent Solar System with SLRs is that of timescale. The process requires one or more dying star(s) expelling the polluting material into its/their surroundings, to be located in time and space nearby a star just being born. Timescales of stellar lifetimes are relatively long, even a very short-lived 60 \msun\ star needs roughly 4 Myr to reach the end of the main sequence. A typical 25 \msun\ star, whose final CCSN is traditionally invoked as a candidate polluter, needs 7 Myr to evolve. This contrasts with the much shorter timescale of star formation in stellar clusters, of the order of few Myr \cite{elmegreen00}. There have been a number of proposed solutions. One of them is the {\it late} injection idea, which proposes that the protosolar disk (rather than the protosolar cloud) was polluted. This allows the lifetime of the disk, $<$ 10 Myr \cite{ribas15} to be added to the age spread so that stars with ages up to roughly 10 Myr can become potential polluters\footnote{It should be noted, however, that only 5-10\% of stars of mass below 2 \msun\ still present a disk beyond 3 Myr of age \cite{ribas15} and that for the ESS indirect evidence from chondrules indicates that the disk lived less than 5 Myr \cite{bollard15,wang15}.}. In this case, as discussed above, the typical distance between the disk and the source is of the order of 0.03 pc, which allows to put the source within the same cluster as the Sun (see Fig.~\ref{fig:williams}). 

Another solution is to consider the bigger picture of star formation, i.e., the top panel of Fig.~\ref{fig:williams}, where a GMC is shown, whose complex may live up to a few tens of Myr. In this case, the scenario would involve pollution of the presolar cloud ({\it early} injection), given the distances from the source calculated above of the order of 5 pc. This class of scenarios includes the supernova trigger scenario, or the setting of the birth of the Sun in the vicinity of a massive star, such as a runaway WR star \cite{tatischeff10}, or directly inside the shell compacted by the wind of a WR star \cite{gounelle12,gounelle15,dwarkadas17}. In relation to the top panel of Fig.~\ref{fig:williams}, this would be somehow at the border between the green HII region generated by massive stars, and the red star formation cloud.

One common feature of the {\it early} and {\it late} scenarios is that they both involve a relatively {\it local} source for the SLRs, typically one star of a given initial mass and lifetime, located at a given distance. Interestingly, it should be pointed out that while this local scenario is invoked to explain the shortest-lived radionuclides, i.e., those in Fig.~\ref{fig:times} which plot above the steady-state line and cannot be explained by the workings of GCE, in particular \iso{26}Al, it cannot be excluded {\it a priori} that such a proposed single, nearby star also produced some of the longer-lived nuclei \cite{meyer00,wasserburg06}. In this case, due to the vicinity of the source to the ESS, it is not necessary to invoke a major production site, since the dilution with the surrounding matter would be small enough to also allow a minor production site to provide a significant contribution (see Fig.~\ref{fig:scipaper}). This feature of such models actually can create problems of overproduction of some SLRs, which are already inherited from GCE. For example, in the specific cases of \iso{26}Al and \iso{182}Hf there are direct constraints from FUN-CAIs exhibiting the standard ESS \iso{182}Hf/\iso{180}Hf value and no \iso{26}Al \cite{holst13}. At least for this couple it seems that a common origin as predicted by some stellar models (Fig.~\ref{fig:scipaper}) should be excluded. 

Another point common to these {\em local} scenarios is that the likelihood of the required circumstances to happen is relatively small, of the order of some percent or less, depending on the specific model and the specific analysis \cite{williams07,gounelle08,williams10,pan12,gounelle15,lichtenberg16a,nicholson17}. In simple terms, this derives from the fine-tuning of the initial mass of the polluter, leading to constraints on the number of stars born in the same cluster as the Sun and the distance from the stellar source to the Sun. A corollary from the local scenario is that our Solar System would be from rare to relatively uncommon and that most planetary systems in the Galaxy would have a low(er) \iso{26}Al abundance \cite{lichtenberg16a}. 

An alternative to the local scenario has been recently proposed: a {\it global} scenario, which involves the large-scale evolution of a GMC and its potential chemical self-enrichment, which may occur if the lifetime of the cloud is of the order of a few tens of Myr. In this case, massive, short-lived stars born in the regions of the molecular cloud that collapse earlier can provide the short-lived SLRs to stars born in the regions that collapse later. As mentioned above, the lifetimes of molecular clouds can range up to a few tens of Myr, which makes this scenario plausible.  
Interestingly, in the global scenario the injection mechanism related to Q2 is less relevant, since the matter precursor to the presolar cloud is primordially enriched. On the other hand, it is more difficult in this case to explain the existence of FUN-CAIs and micro-corundum grains that show no \iso{26}Al enrichment \cite{kuffmeier16}. These could be explained within the local scenario if they formed prior to the injection of \iso{26}Al \cite{krot12}, and in the global scenario by spatial inhomogeneities in the disk. 

For example, \cite{gounelle09} proposed that \iso{60}Fe was inherited in the molecular cloud where the Sun was born from a previous generation of CCSNe born in a different molecular cloud. Combined with an origin for \iso{26}Al from a {\it local} massive star \cite{gounelle12}, this scenario requires three discrete generations of stars, and two star-formation triggering episodes, which may be unlikely \cite{parker16}. The most sophisticated study to date is that of \cite{vasileiadis13}, who modelled from first principles the production, transport, and admixing of freshly synthesised \iso{26}Al and \iso{60}Fe from CCSNe in star-forming regions within GMC using a multi-dimensional magnetohydrodynamic code. They demonstrated that mixing is efficient and the effect of self-pollution on the composition of newborn stars inside molecular clouds becomes evident relatively quickly, within 10–30 Myr. They concluded that variable presence of \iso{26}Al is a general feature of these regions, however, it is accompanied by an abundance of \iso{60}Fe at or above the upper limit of the currently debated range. This is an inevitable consequence of current CCSN models, which produce comparable yields for \iso{26}Al and \iso{60}Fe, as discussed above in relation to Q1. 
To overcome this problem, \cite{gaidos09} and \cite{young14} speculated that contributions from the winds of massive stars would be enhanced in molecular cloud environments, relative to the galactic ISM, and this would explain the origin of \iso{26}Al. The conundrum here is that, if massive stars evolving in GMC are responsible for the enrichment of SLRs in the ISM, it is not clear why the \iso{60}Fe/\iso{26}Al ratio in the ISM ($\simeq$0.55 from $\gamma$-ray observations) and in GMC ($\simeq$0.2 - 0.002 from the ESS) should be so different. 

One key feature of the global scenario that distinguishes it from the local scenario is that the presence of SLRs would affect many stellar births, since most stars are born in giant molecular clouds. The vast majority of stars in the Galaxy would be born with a high abundance of \iso{26}Al, as in the Solar System. As discussed in Sec.~\ref{sec:26Alhabit} this will have major consequences on the thermo-dynamical evolution of their planetesimals.

\section{The effect of radioactive decay on the evolution of the Solar System solid bodies}
\label{sec:radioheat}

As introduced in Sec~\ref{sec:habit}, the main processes by which radioactivity can affect habitability are radiation (in a direct way) and heat generation (mostly indirectly). In terms of radiation, it should also be noted that the high energetic particles produced by radioactive decay can affect the overall evolution of protoplanetary disks by contributing to the ionisation of matter \cite{cleeves13,adams14}. Considering the effect of different star formation rates in different galaxies, starburst and high-redshift galaxies may be richer in SLRs than the Milky Way Galaxy, which then results in overall higher ionisation levels \cite{lacki14}. Another proposed effect of the decay energy of \iso{26}Al is to drive lightning discharge that could contribute to the melting of chondrule precursors \cite{johansen18}. Here, we will start by considering the heat generated by radioactivity in Solar System solid bodies, and then specifically focus on the relation between the decay of \iso{26}Al and the evolution of early planetesimals, the first rocks of sizes roughly 1 to 200 km from which the rocky planets are believed to have originated via gravitationally controlled accretion.

\subsection{Radioactive heating sources in the Solar System}
\label{sec:heat}

\begin{figure}[tb]
\begin{center}
\begin{minipage}[t]{16.5 cm}
\includegraphics[width=16cm,angle=0]{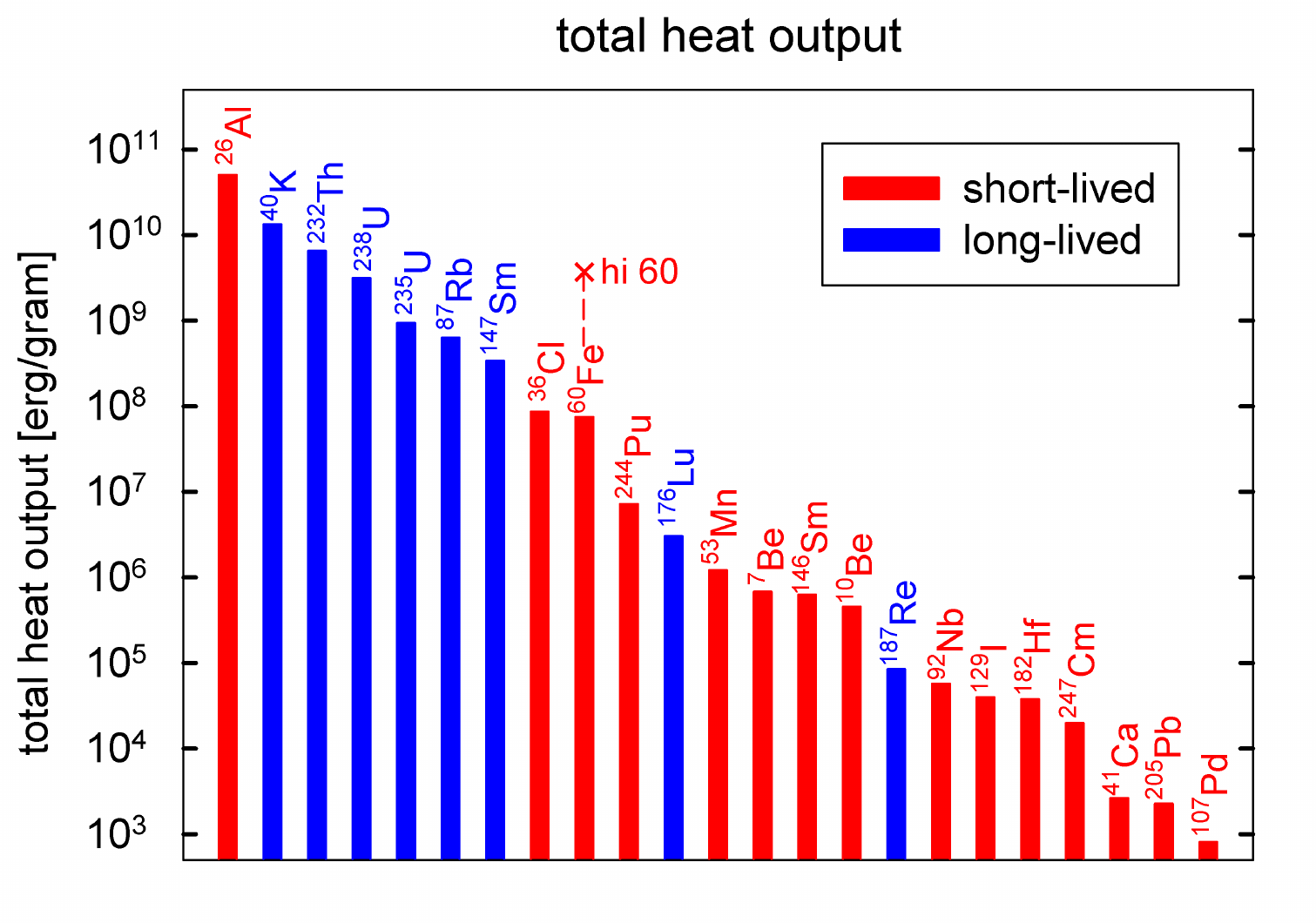} 
\caption{Total energy available for heating from radioactive decay per gram of material in the Solar System (CI meteorite composition) after the initially present isotope has completely decayed. For \iso{10}Be/\iso{9}Be a value of 1 $\times 10^{-3}$ was used. For \iso{60}Fe also the situation for a high abundance (\iso{60}Fe/\iso{56}Fe = 5 $\times 10^{-7}$ as compared to 1.01 $\times 10^{-8}$ given in Table~\ref{table:SLRs}) is indicated.  \label{fig:heat1}}
\end{minipage}
\end{center}
\end{figure}

\begin{figure}[tb]
\begin{center}
\begin{minipage}[t]{16.5 cm}
\includegraphics[width=16cm,angle=0]{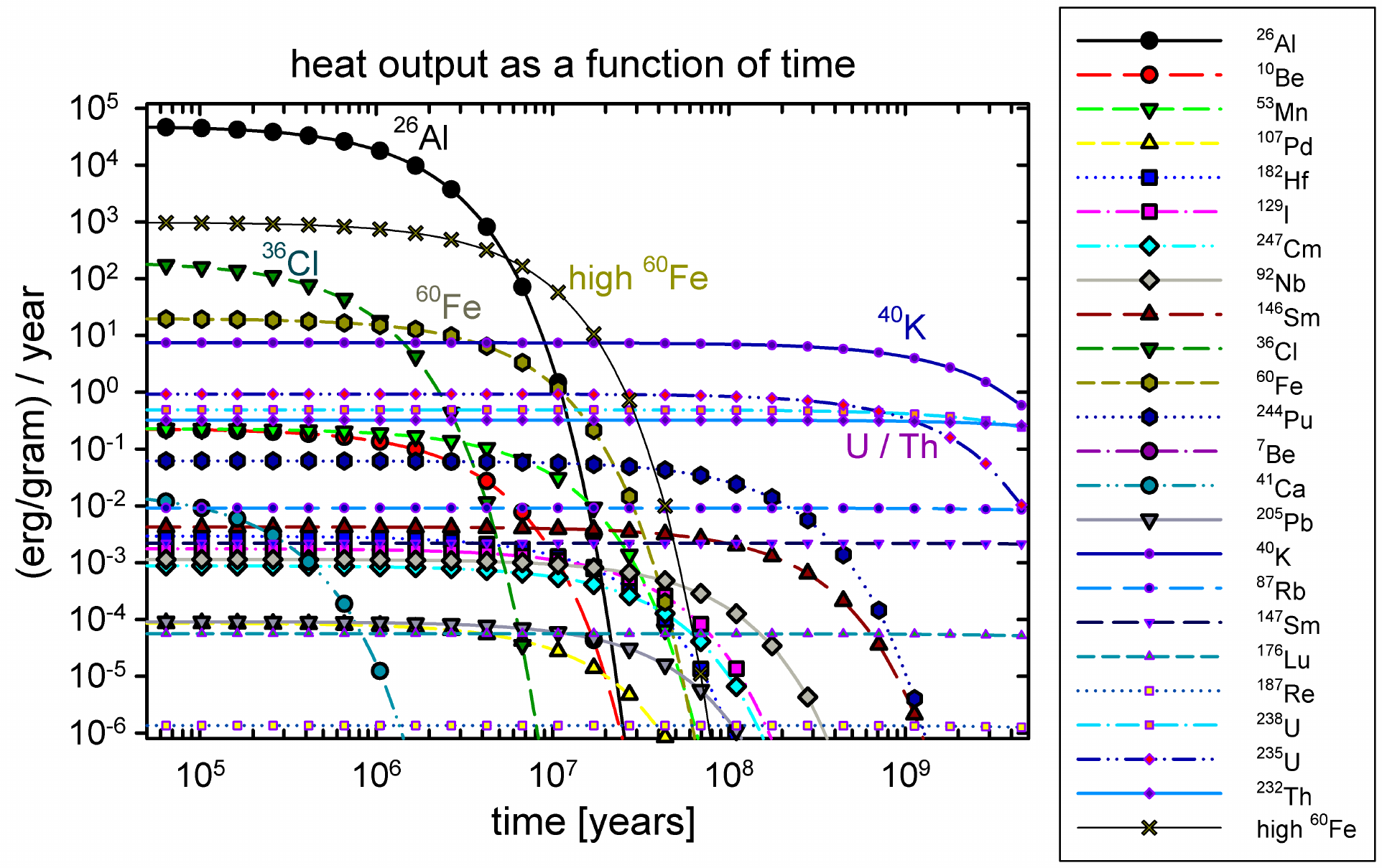} \\
\caption{Net energy output as a function of time, starting 50,000 years after time zero, defined as the time when the nuclei start to decay. For the half-life of \iso{146}Sm we used the 103 Myr value. The initial abundances of \iso{10}Be and \iso{60}Fe are taken as indicated in Fig.~\ref{fig:heat1}.  \label{fig:heat2}}
\end{minipage}
\end{center}
\end{figure}

\begin{table}
\begin{center}
 \centering
 \caption{Decay energy, abundance in CI chondritic meteorites (\cite{lodders09}; normalised to a Si abundance of 10.7 wt\%), and total heat energy available from the decay per gram of rock. The SLRs are listed in the same order as in Table~\ref{table:SLRs}, some long-lived radionuclides are also included at the end of the table, in order of ascending mass.}
\label{tab:heat}
\vspace{0.3cm}
 \begin{tabular}{llll}
 \hline
 \multicolumn{4}{l}{Short-lived radionuclides}\\
 & Energy (erg/decay) & CI abundance (atoms/gr) & Total E (erg/gr) \\ 
\iso{26}Al & $5.00 \times 10^{-6}$ & $1.01 \times 10^{16}$ & $5.07 \times 10^{10}$ \\
\iso{10}Be & $3.24 \times 10^{-7}$ & $1.40 \times 10^{12}$ & $4.56 \times 10^{5}$ \\
\iso{53}Mn & $8.23 \times 10^{-9}$ & $1.48 \times 10^{14}$ & $1.22 \times 10^{6}$ \\
\iso{107}Pd & $1.49 \times 10^{-8}$ & $5.43 \times 10^{10}$ & $8.10 \times 10^{2}$ \\
\iso{182}Hf & $2.96 \times 10^{-6}$ & $1.28 \times 10^{10}$ & $3.78 \times 10^{4}$ \\
\iso{129}I & $1.23 \times 10^{-7}$ & $3.23 \times 10^{11}$ & $3.97 \times 10^{4}$ \\
\iso{247}Cm & $2.67 \times 10^{-5}$ & $7.45 \times 10^{8}$ & $1.99 \times 10^{4}$ \\
\iso{92}Nb & $2.03 \times 10^{-6}$ & $2.81 \times 10^{10}$ & $5.71 \times 10^{4}$ \\
\iso{146}Sm & $4.05 \times 10^{-6}$ & $1.56 \times 10^{11}$ & $6.31 \times 10^{5}$ \\
\iso{36}Cl & $3.95 \times 10^{-7}$ & $2.19 \times 10^{14}$ & $8.67 \times 10^{7}$ \\
\iso{60}Fe & $4.16 \times 10^{-6}$ & $1.80 \times 10^{13}$ & $7.49 \times 10^{7}$ \\
\iso{244}Pu & $2.50 \times 10^{-5}$ & $2.89 \times 10^{11}$ & $7.23 \times 10^{6}$ \\
\iso{7}Be & $7.99 \times 10^{-8}$ & $8.56 \times 10^{12}$ & $6.84 \times 10^{5}$ \\
\iso{41}Ca & $4.31 \times 10^{-9}$ & $6.17 \times 10^{11}$ & $2.66 \times 10^{3}$ \\
\iso{205}Pb & $8.35 \times 10^{-9}$ & $3.03 \times 10^{11}$ & $2.53 \times 10^{3}$ \\
\hline
\multicolumn{4}{l}{Long-lived radionuclides}\\
 & Energy (erg/decay) & CI abundance (atoms/gr) & Total E (erg/gr) \\ 
\iso{40}K & $1.05 \times 10^{-6}$ & $1.27 \times 10^{16}$ & $1.33 \times 10^{10}$ \\
\iso{87}Rb & $1.31 \times 10^{-7}$ & $4.83 \times 10^{15}$ & $6.33 \times 10^{8}$ \\
\iso{147}Sm & $3.70 \times 10^{-6}$ & $9.18 \times 10^{13}$ & $3.40 \times 10^{8}$ \\
\iso{176}Lu & $1.24 \times 10^{-6}$ & $2.46 \times 10^{12}$ & $3.04 \times 10^{6}$ \\
\iso{187}Re & $9.90 \times 10^{-10}$ & $8.58 \times 10^{13}$ & $8.49 \times 10^{4}$ \\
\iso{232}Th & $6.46 \times 10^{-5}$ & $1.01 \times 10^{14}$ & $6.52 \times 10^{9}$ \\
\iso{235}U & $7.10 \times 10^{-5}$ & $1.33 \times 10^{13}$ & $9.45 \times 10^{8}$ \\
\iso{238}U & $7.64 \times 10^{-5}$ & $4.13 \times 10^{13}$ & $3.15 \times 10^{9}$ \\
 \hline
 \hline
 \end{tabular}
\end{center}
\end{table}


In Fig.~\ref{fig:heat1} we plot the total energy available for heating from radioactive decay per gram of material in the Solar System. In Table~\ref{tab:heat} we report the numbers used to derive this energy. We have assumed the ESS ratios for the SLRs reported in Table~\ref{table:SLRs} (choosing $10^{-3}$ for the \iso{10}Be/\iso{9}Be ratio) 
and a rock with the composition of CI chondritic meteorites \cite{lodders09}. This class of meteorites shows the closest composition to the primordial (before Li destruction via H burning) solar photosphere (except for the volatile elements, H, C, N, O, and noble gases). 
The radionuclides are ordered in the figure according to output energy. All SLRs are included, as well as the long-lived 
\iso{40}K, \iso{232}Th, and \iso{235,238}U and a few other long-lived nuclides of interest for geochemical applications. 


For the calculation of the decay energy we took all the required information from the National Nuclear Data Center website (www.nndc.bnl.gov). For the $\alpha$ decays, we used the full Q values, since there is no loss due to  energy carried away by neutrinos and since in this way  the energy in the recoil energy of the daughter nucleus is also included, typically $\sim$2\%. For the $\beta$-decays, we summed the energies carried by various types of electrons and radiation (save neutrinos), but neglected the recoil of the daughter, which is complicated to obtain precisely but typically less than one per mil. In the decay chains branchings of less than one per mil were neglected, as well as the 1.2 per mil fission branch at \iso{244}Pu. The released energies are very close to those calculated by \cite{ruedas17}, for the nuclides reported there (\iso{26}Al,\iso{60}Fe, \iso{40}K and the long-lived Th and U isotopes). Small differences (of at most 2-3\%) may be due to different sources for the decay data and/or different treatment of recoils.

In Fig.~\ref{fig:heat2} we plot the net energy output as a function of time. Obviously, only a portion of the total available energy from the long-lived radionuclei has been expended so far, since they are still alive.
Among the SLRs, the decay energy from \iso{26}Al dominates by far, given its very high abundance. Interestingly, if the high \iso{36}Cl values and low \iso{60}Fe values are correct, \iso{36}Cl would appear to be more important than \iso{60}Fe (still much smaller than \iso{26}Al) as a heat source during the first few Myr of the Solar System. However, heating by \iso{36}Cl may be 
``self-defeating'': within a normal type of meteoritic rock (primarily Mg, Fe silicates) the melting will probably decompose the compound containing Cl and the \iso{36}Cl will be lost to the gas phase, thus removing the possibility of further heating of the rock from this source. 
After 10 Myr or so, \iso{60}Fe becomes the most important SLR, however, already by this time the long-lived \iso{40}K starts to dominate the total energy output. Although \iso{53}Mn is relatively abundant, it provides very little energy because it decays via electron capture only, with no $\gamma$-ray, and essentially all the energy is carried away by the neutrino.

Since the heat scales with the abundance, if, for example, \iso{60}Fe was 50 times higher than assumed here, its contribution to the total heating energy would become roughly 7\% of the \iso{26}Al contribution, and after about 10 Myr its energy output would still be higher than that from \iso{40}K. Alternatively, if the abundance of \iso{26}Al was below the canonical value the evolution of the solids would have been somewhat different \cite{larsen16}. From the discussion above (Sec.~\ref{sec:birth}) it is clear that we cannot assume that other stars are born with the same SLR inventory as the ESS, thus the relative contributions could be enormously different in extra-solar protoplanetary disks. For example, no significant SLR heating source may be present in a system with a \iso{26}Al abundance four orders of magnitudes lower than in the ESS, i.e., if the lower limit of the \iso{26}Al/\iso{27}Al ratio derived from the $\gamma$-ray background of $2 \times 10^{-6}$ decayed for an isolation time of 6 Myr and there was no local production in the stellar nursery. In relation to the long-lived isotopes, we note that unlike the Th and U isotopes, which are pure $r$-process products, \iso{87}Rb and \iso{147}Sm can also be produced by the $s$-process. This implies that their initial abundances relative to Th and U could be very different in other planetary systems, resulting in further possible significant sources of heat. 



 \subsection{Incorporation into minerals}
\label{sec:carriers}

Beyond the characteristic decay energy and occurrence of radioactive isotopes in any hypothetical protoplanetary disk, to produce heat and influence processes relevant to astrobiology, the isotopes have to be incorporated into the solids and finally the planetary bodies, satellites etc. forming there. Here, the related possibilities and expected characteristics are summarised, although we note that serious knowledge gaps exist.

Radioactive isotopes could be trapped in the same way as stable isotopes in any crystalline lattice by two basic processes: during primary mineral condensation in the ISM or especially in the protoplanetary nebula, and by later migration inside already condensed planetesimals\footnote{It is worth mentioning that many isotopes could be simply adsorbed on the porous surfaces of grains, however, substantial accumulation is not expected by this process.}. While the first is more influenced by the conditions inside protoplanetary disks, the second is governed by conditions inside the given planetesimal. Such secondary migration processes could produce environments with enhanced heat production, causing melting, evaporation, gas release, and various chemical reactions. The important radioactive isotopes can be grouped according to their compatibility to various minerals: (i) those which can be accommodated by many (including major) minerals in the long term -- lithophile elements like Al, K, Mn, but also siderophile elements like Pd; (ii) some elements with more complex behaviour like Fe; and (iii) trace elements like U and Th that are incompatible because of their atomic radius and valence. Isotopes accumulated by secondary migration would mostly be relevant for long-lived radionuclides, where this can cause melting, for example, producing mare basalts on the Moon around 3.9-3.5 Ga ago \cite{Morota2011}. 

In relation to primary mineral condensation, if the necessary starting elements exist in reasonable amounts (elevated spatial density) below their condensation temperatures, for SLRs faster accretion could increase the heating effect by early incorporation of isotopes before they decay. In the second case of later migration, the processes of release, migration, and accumulation of isotopes already entrapped inside minerals do matter mainly for long-lived radionuclides because resulting elevated concentrations would cause a net heating effect, while resulting low concentrations would prevent a temperature increase, as the target material loses the heat faster than it is produced. 

In the following we are going to discuss in a simplified manner conditions and mineral types that are relevant to these aspects, i.e., influencing the primary condensation or accumulation of heat producing nuclides inside the first solids. Since the system is complex, it is difficult to quantitatively estimate the distribution and occurrence of certain radioactive isotopes in planetesimals of extra-solar planetary disks, thus only rough estimates on the relevant factors will be summarised.

The chemical environment has an effect due to the presence/absence of the various chemical elements with which the radioactive nuclei can form solids. In this respect the carbon/oxygen ratio in the disks \cite{Helling2014} is an overall dominant factor, allowing the formation of either primarily oxidised (like Al$_2$O$_3$) or reduced (like FeS) components. The C/O ratio is expected to increase with time along with the condensation of H$_2$O in a disk \cite{Fortney2012} and the condensation pathways of refractory solids differ at different C/O ratios \cite{Bond2010}. For the most important SLR \iso{26}Al, oxygen-rich conditions are more favourable for trapping in minerals, relative to a reduced protoplanetary disk – although observational evidence is restricted only to our Solar System, where oxidised phases were dominant. Extra-solar planetary systems more rich in C may behave differently in this respect.

The condensation sequence of the different nuclei follows their different volatility \cite{Ali-Dib2014}. For ESS conditions, the most refractory element among the radioactive isotopes listed in this work is Hf (however, it will have a low spatial concentration), followed by Al (which is much more abundant and forms oxides like corundum, hibonite and gehlenite minerals around 1700 K), followed by the condensation of Nb, Be and Fe around 1500-1300 K (these could be present in enstatite and forsterite silicates formed around 1400 K but only at the trace amount level). Feldspars condense around 1000 K and are carriers of Al and K, while Pb would alloy with Fe at around 700 K \cite{lodders09}.

Condensation speed and the available time for condensation also influence the radioactive heating effect of SLRs as in the case of fast condensation the nuclides are entrapped with still higher abundances. Based on models and observations, the characteristic timescales towards planet formation are the following: 0.01 Myr is required for a Class 0 protostellar object \cite{Murillo2013} to became a Class I object \cite{Harsono2014}, when more than 50\% of the envelope has  fallen onto the central protostar; 0.1 Myr is estimated for the condensation of first solids in the Solar System (CAIs formed at $>$1300 K, containing Ca and Al carrier minerals); and 1 Myr is required for building the first planetesimals. These durations are of the same order as the half-life of \iso{26}Al (0.72 Myr). Thus, in the case of substantial injection of \iso{26}Al into a protoplanetary disk and taking the 0.1 Myr time scale for the condensation of the first solids, large amounts of \iso{26}Al will be encapsulated in solids. Early planetary condensation  may also result in more efficient melting of accumulated material. In the Solar System, the first basaltic angrite meteorites formed around 10 Myr after the formation of the very first CAI. This interval roughly coincides with the age of many debris disks produced by the fragmentation of some already condensed planetesimals \cite{Apai2010}, confirming that 10 Myr is the timescale typical for the formation of km-sized planetesimals, where in turn melting and differentiation can result in secondary accumulation of radionuclides.

The spatial density of certain nuclei will have a strong effect on the  heat production. Of course, the higher the concentration, the higher is the possibility for radioactive decay produced heating. This is particularly relevant for example in the case of heterogeneous distribution of \iso{26}Al \cite{larsen11,larsen16}, where melting of planetesimals may occur in one place and not another, even if they formed at the same time.

\subsection{Implications from the decay of \iso{26}Al decay on planetesimal evolution}
\label{sec:26Alhabit}

Based on model computations and meteorite evidences, the dominant process contributing to the very early melting of planetesimals was the heat generated by the decay of $^{26}$Al (see also Fig.~\ref{fig:heat2}). This decay could melt even relatively small (above 10-20 km diameter) planetesimals or asteroids \cite{Greenwood2005,Kleine2005,lichtenberg16b}. It modified the mineral content, melted ice to liquid water, thus possibly producing a range of different molecules potentially important for habitability both inside these objects and also for other bodies onto which they impacted \cite{monteux17}. Consequently, the most frequently investigated heat source in the ESS is  $^{26}$Al and its role in the thermal evolution of young planetary bodies in the Solar System. 

The direct heating effect of \iso{26}Al depends on its concentration and how early the first planetesimals formed \cite{Beuther2014}. For a given planetesimal size, larger radionuclide concentration and faster growth will, of course, result in higher temperatures.
In case of a sufficient heating effect, the thermo-mechanical evolution of the planetesimal is strongly affected \cite{moskovitz11,lichtenberg16b}. In fact, melting and differentiation is mostly controlled, for a given size planetesimal, by its initial amount of \iso{26}Al. The size controls the heat escape, via the volume to surface ratio, so that larger bodies are able to retain their heat more efficiently.   
By the direct heat effect, co-accreted ice could melt \cite{Metzler1992,brearley06,Elkins-Tanton2017}, and liquid water circulation may result \cite{Clayton1984}. This, in turn, leads to alteration of rocks, including formation of phyllosilicates \cite{Zolensky1989}, which are highly effective for polymerization of smaller organic molecules \cite{Osinski2005}. Phyllosilicates can also stabilise these molecules against other agents by adsorbing them on their mineral surfaces \cite{Hazen2010}. Liquid water circulation inside planetesimals could produce organics also regardless of phyllosilicates, however, this is still poorly explored \cite{Hartman1983,Martins2011}.
Another important feed-back effect coupled to the radioactive heat might increase the temperature even further. If radioactive heating melts at least part of the ice that is mixed with silicates, the so formed liquid water may be able to cause various exothermic hydration reactions -- the most likely among them being serpentinization: the weathering of olivine -- that may become an even more effective heat source than the radioactivity itself \cite{Gobi2017}. Naturally, this process will not start unless the radioactive heat is sufficient to increase the temperature enough to melt the water ice in the first place. 

\begin{figure}[tb]
\begin{center}
\begin{minipage}[t]{15 cm}
\includegraphics[width=15cm,angle=0]{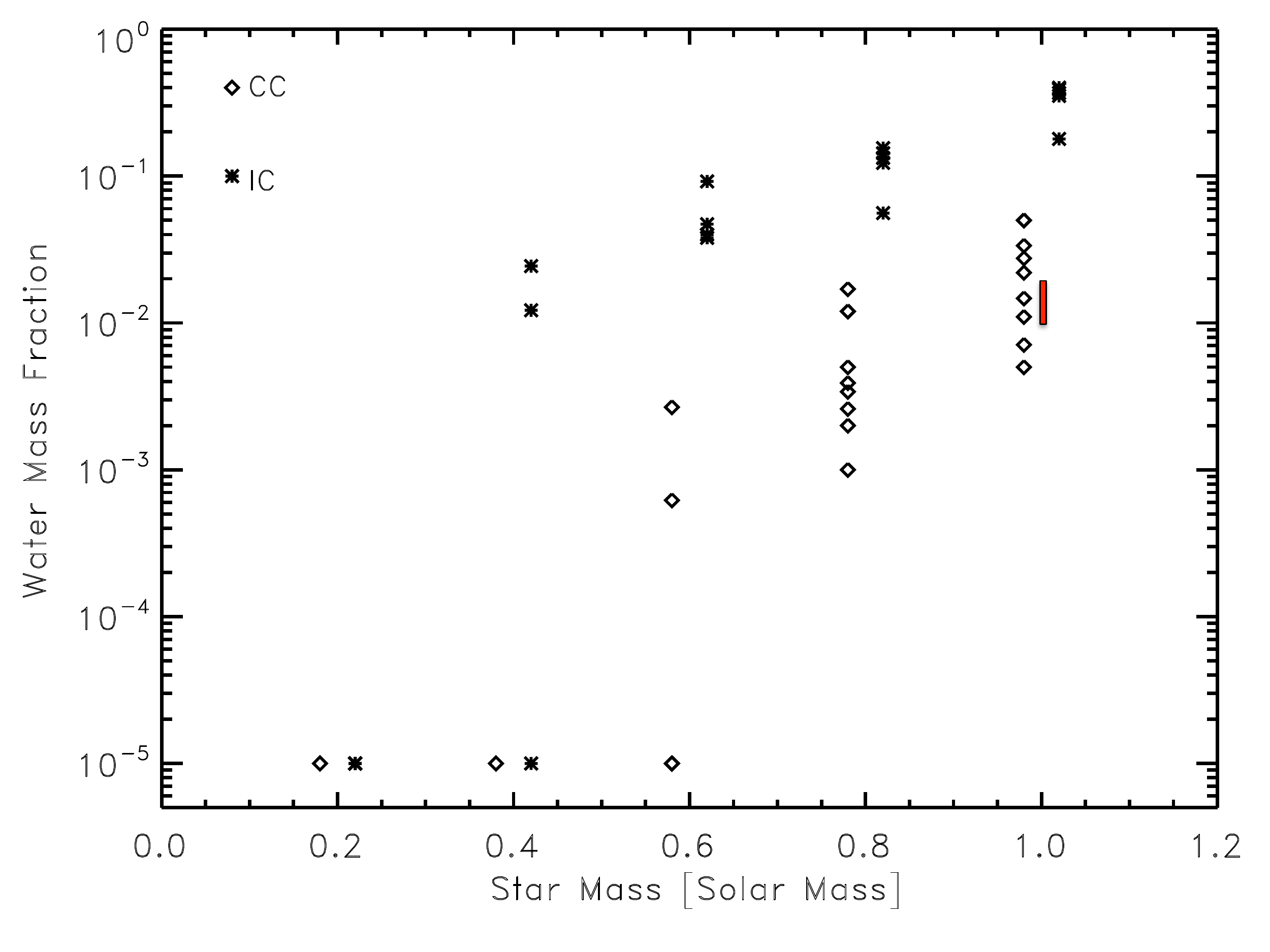}
\caption{The water mass fraction in terrestrial planets in the habitable zone of stars of mass between 0.2 and 1.2 \msun\, as computed by the dynamical models of \cite{ciesla15} starting from planetesimals of different composition. The different data points show the outcomes of different simulation runs. Open symbols represent planets produced in simulations where planetesimals beyond the snow line have the same water content (5\%) as carbonaceous chondrites (CC), which are believed to come from the same type of planetesimals that were the source for most of Earth's water. Asterisks represent planets produced in simulations where planetesimals beyond the snow line have 50\% water content (IC, for icy). The red rectangle represents the estimates for the water mass fraction of the Earth. Figure modified from Ciesla et al. ``Volatile Delivery to Planets from Water-rich Planetesimals around Low Mass Stars'' (https://doi.org/10.1088/0004-637X/804/1/9) \cite{ciesla15} ©AAS. Reproduced with permission. \label{fig:ciesla}}
\end{minipage}
\end{center}
\end{figure}

Because planetesimals are believed to represent the building blocks of rocky planets, as they accrete onto each other via dynamical interaction in the disk, clearly the internal structure and the composition of planetesimals will have an impact on the composition of the planets to which they contribute\footnote{A new model of pebble accretion for the growth of protoplanets has recently been proposed \cite{johansen15}. The role of heating via \iso{26}Al still needs to be investigated in relation to this scenario.}. One important consequence that has received attention in recent years is the delivery of water to terrestrial planets.
In addition to the distance from the central star\footnote{Only beyond the ``snow line'', or ``ice line'', it was cold enough for water to condense into solid ice grains.}, the water budget (H$_{2}$O content) of any planetary body depends on several parameters and conditions, making it difficult to evaluate the whole range of possibilities. However, the dominant effects can be considered and can be grouped into the primordial content acquired during direct accretion and early evolution of the first solid bodies, and later accretion, due to the impact of planetesimals on the surface of larger planetary embryos. In this context, $^{26}$Al may play a role by influencing the temperature inside, and water vapour outgassing from, the first solids. In the Solar System the heat from relatively high $^{26}$Al content combined with ``fast accretion'' ultimately led to a moderately low H$_2$O content of the first accreted bodies. 
Early formed planetesimals beyond the snow line are predicted to initially contain 50\% ice \cite{lodders03}, however, carbonaceous chondrites, which represent objects that formed beyond the snowline, contain only up to 5-10\% water by mass, suggesting that a large fraction of the initial ice was lost. The later water delivery onto growing planets might therefore be low because of a low water content of the impacting bodies. 
According to \cite{morbidelli00} water was probably carried to Earth by planetary embryos coming from the outer asteroid belt, beyond the snow line. In detail, the situation is complex, but according to Ciesla et al. \cite{ciesla15} the increased $^{26}$Al content and the faster accretion in the earliest history of the Solar System led to a low water content in the final planetary bodies. In their simulations, these authors studied  the effect of changing the composition of planetesimals on the amount of water delivered to terrestrial planets. Their results demonstrate that planets formed from ice-rich planetesimals (50\% ice, as compared to the 5\% value in carbonaceous chondrites) would result more likely in having higher water mass fractions than the $\sim$0.1-0.2\% of the Earth (Fig.~\ref{fig:ciesla}). The water degassing process due to the presence of \iso{26}Al and its effect on the water fraction of terrestrial planets is being investigated currently also by Lichtenberg et al.\footnote{https://figshare.com/articles/Desert\_versus\_ocean\_worlds\_a\_planet\_population\_dichotomy\_from\_26Al\_enrichment/5577388/1} with a similar result, namely that systems with larger planetesimals and higher \iso{26}Al will result in planets that are more strongly depleted in water. Considering another extreme, full water coverage of the surface of a planet may be harmful for habitability by inhibiting the carbon cycle \cite{alibert14} and silicate weathering \cite{abbot12}. Furthermore, in the case of deep global oceans, the first organisms forming at the rock-water interface would not have been able to exploit solar radiation as an energy source, and chemical energy sources might not have been sufficient to lead to an advanced evolution as occurred on the Earth.

Because of the deep implications of the presence of \iso{26}Al in  planetary systems it is crucial to understand its distribution in star forming regions. The first step would be to explain its high abundance in the ESS, however, as discussed in Sec.~\ref{sec:birth} we are far from consensus on the answer to this question. In particular, 
the different scenarios discussed in Sec.~\ref{sec:Q3} are characterised by very different probabilities of occurrence, from less than 1 permil (in some cases of local scenarios) to almost 100\% (in the case of global scenarios). 
While it would appear sensible to assume that the birth of the Sun was a typical event, ultimately the question of the probability of \iso{26}Al occurrence could be considered irrelevant to the origin of SLRs in the ESS, since we only have one observation. However, a statistical analysis becomes crucial if we wish to establish the percentage of extra-solar planetary systems born with a significant amount of \iso{26}Al. 

Qualitative and indirect observations tend to support the idea that a high abundance of \iso{26}Al is common in extra-solar planetary systems. Qualitative indication comes from the ability of $\gamma$-ray satellite observatories such as INTEGRAL to map the abundance of \iso{26}Al in the Galaxy. These maps show that a higher abundance of \iso{26}Al is found in the galactic plane, which is consistent with its main production site being massive stars, since these are also  preferentially located closer to the galactic plane. As calculated above, the total mass of \iso{26}Al in the Galaxy, when scaled to the mass of the Galaxy, is very roughly 5 times lower than its abundance in the ESS. However, as mentioned in Sec.~\ref{sec:26Al} its distribution is not homogeneous, with groups of massive stars associated to star-forming regions such as Cygnus and Scorpius-Centaurus clearly more enriched in \iso{26}Al than other regions of the Galaxy. Still, as discussed above, it is also not clear if and how such material can find its way into pre-stellar cold clumps. Further indirect observational support to the idea of a universally high \iso{26}Al in newborn stars may be inferred from the abundances of, among others, Al, Si, and Fe in the atmospheres of white dwarfs that are believed to have accreted some of their own asteroids and appear to carry the sign of early differentiation of these asteroids driven by the heat from the decay of \iso{26}Al \cite{jura13,jura14}. On the other hand, a low probability for the presence of high \iso{26}Al abundances would fit in with anthropic selection for the presence of \iso{26}Al in the ESS, as has been proposed based on its possible implications for the existence of life on Earth \cite{gilmour09}.

%
%

\section{Conclusions}
\label{sec:conclusions}

We have reviewed the meteoric evidence for the presence of SLRs in the ESS, their production in stars, the simple models available to predict their GCE evolution, and the methodology that allows us to use SLRs as clocks to measure the isolation time of presolar matter inside its parent molecular cloud. Most stars produce some kind of SLRs and enrich the ISM with radioactivity at the end of the lives via winds or explosions. We have then considered the origin of the shortest-lived isotopes, such as \iso{26}Al, in the context of different scenarios for the formation of the Sun, and the impact of the heat from radioactivity on the evolution of solid bodies in a planetary system, focusing in particular on the effects resulting from the presence of \iso{26}Al. Depending on the timescale and speed of early condensation and accretion in disks, melting of planetesimals may produce circulating water and related chemical reactions influencing habitability. 
Our main conclusions and future prospects are the following:

\begin{enumerate}

\item The abundance of \iso{60}Fe in the ESS needs to be firmly established before we can proceed to select the possible stellar source responsible for the presence of \iso{26}Al. Among the other SLRs, the determination of the ESS abundances of \iso{36}Cl, \iso{41}Ca, \iso{205}Pb, and \iso{244}Pu require special attention.

\item Accurate and precise predictions for the stellar production of SLRs are still hampered by many uncertainties in the nuclear physics input. These range from the proton capture reactions on \iso{25}Mg (including the feeding factor to the ground state of \iso{26}Al) and \iso{26}Al, to the neutron production rates \iso{13}C($\alpha$,n)\iso{16}O and \iso{22}Ne($\alpha$,n)\iso{25}Mg, the neutron-capture cross sections, as well as the decay rates of the various branching points (related to the production of, e.g., \iso{60}Fe and \iso{182}Hf), the decay rate of \iso{146}Sm as well as the rates of the reactions that lead to its production, and the $r$-process production of isotopes belonging to the actinides. More details have been given for each SLRs in the dedicated subsections of Sec.~\ref{sec:list}. Current and future experimental facilities, among them LUNA, FAIR, and n\_TOF at CERN, will allow to take up the challenge to improve much of the current situation. In the case where theoretical estimates are required, such as the temperature dependence of decay rates, better nuclear models need to be employed and uncertainties evaluated. 

\item Updated and improved stellar models for all the sites of production (Table~\ref{table:stars}) need to be constantly considered in the light of SLR production. For example, the recent models by \cite{sukhbold16} should be examined in relation to all the SLRs considered here, starting with the procedure presented in Sec.~\ref{sec:Q1}. In this context, 
a coherent picture of SLR nucleosynthesis needs to be built that is able to include the interpretation of all the available constraints, each with their own significance. These include stardust grains \cite{groopman15}, likely originating each from a different star or CCSN, $\gamma$-ray observations \cite{diehl13}, for which a grand average of stellar yields in the Galaxy needs to be considered, and measurements of current radioactivity in the Earth's crust and other terrestrial and lunar samples (e.g., \cite{wallner16}), to which probably only one or a few CCSNe contributed. Observations of live \iso{26}Al of CCSN origin in the same terrestrial samples that show live \iso{60}Fe, for example, and comparison of the \iso{60}Fe/\iso{26}Al ratio with that derived from $\gamma$-ray observations can provide new constraints for the build up of this coherent picture. 
At the same time proposed future space telescopes like eASTROGAM \cite{deangelis17} and GRIPS \cite{greiner12} will provide a new, improved understanding of the occurrence of \iso{26}Al and \iso{60}Fe in the Galaxy and in star forming regions. 

\item In order to properly describe the effect of discrete stellar additions of SLRs to the presolar matter and derive information about the lifespan of its stellar nursery, the free parameters $K$ and $\delta$ in Eq.~\ref{eq:final} need to be constrained both by using full GCE and stellar population models, and an improved understanding of the propagation of different types of stellar ejecta in the ISM. This will allow us to derive a better estimate of the isolation time, which in turn may help us to clarify the environment of the birth of the Sun. Hydrodynamical model calculations of the propagation of SLRs in the Galaxy from their stellar sources to the location of star formation need to be carried out also to address the issue of the origin of \iso{26}Al in the ESS. 

\item Models of planetesimal evolution and planet formation including the effect of the heat of \iso{26}Al need to be developed further, including volatile degassing and different scenarios such as pebble accretion. Different abundances for other potential heat sources like \iso{60}Fe for short term heating, or \iso{87}Rb, \iso{147}Sm, and \iso{232}Th for long term heating should be also considered, as under specific conditions they may have elevated abundances compared to those that were present in our Solar System and thus may have a substantial effect on habitability in other planetary systems.

\item Based on a clearer picture obtained from the points above, a statistical analysis of the presence of \iso{26}Al in extra-solar planetary systems will need to be developed. Further steps might be such model approaches that connect the spatial distribution of $^{26}$Al in star-forming regions and the condensation speed of planetesimals in disks there. This might help to evaluate differences between forming planetary systems in their potential of generating liquid water and various chemicals inside their early formed solid bodies. Results may be then compared to independent constraints from future observations of the composition of extra-solar planets, and particularly of their water content as a possible signature of the ultimate effect of \iso{26}Al decay. Missions that aim to discover and characterise exoplanets include the NASA missions K2 and TESS and the upcoming ESA missions CHEOPS and PLATO, in conjunction with more detailed, spectroscopic input from the NASA James Webb Space Telescope (JWST), the prospective ESA ARIEL space mission, and the ground-based Extremely Large Telescopes (ELTs) that are currently in development. The combination of these data sets will give us information on both the bulk and the atmospheric or surface compositions of extra-solar planets. Such further examples of planets beyond the Solar System may eventually provide an independent estimate of the presence of \iso{26}Al and the consequence of its integration to solid condensates.

\end{enumerate}

These tasks lying ahead are challenging, but feasible, and carrying the promise to provide us with a clearer view of where our Solar System and the life within it stands in relation to the vast population of extra-solar planetary systems in the Galaxy. Connection of improved models and further observational data on the occurrence of radionuclides could provide new approaches to estimate the habitability potential of the growing number of recently discovered exoplanetary systems. 

\section{Acknowledgments}

We thank the referees for their comments, which have helped us to improve the discussion and the presentation of the paper. We are thankful for many discussions with Benoit Côté, Tim Lichtenberg, M\'aria Pet\H{o}, Andr\'es Yag\"ue (also for the GCE plotting routine), Marco Pignatari, Maria Sch\"onbachler, L\'aszl\'o Moln\'ar, P\'eter \'Abrah\'am, Martin Bizzarro, Roland Diehl, Francois Tissot, Amanda Karakas, Alexander Heger, and Brett Hennig. This work is supported by the European Research Council (ERC) under the EU Horizon 2020 research and innovation programme (grant agreement No 724560) through ERC Consolidator Grant “RADIOSTAR” to M.L. and the Lend\"ulet grant (LP17-2014) of the Hungarian Academy of Sciences to M.L. \'A.K. thanks the COST TD1308 ORIGINS Action, and M.L. the COST CA16117 ChETEC Action.

\section{References}
 
\noindent [1]
    Diehl R., Hartmann D.H., Prantzos N. (Eds.), Astronomy with Radioactivities, vol. 812, Lecture Notes in Physics, Springer Verlag, Berlin (2011)

\noindent [2]
    Clayton R.N., Onuma N., Grossman L., Mayeda T.K.
    Earth Planet. Sci. Lett., 34 (1977), pp. 209-224

\noindent [3]
    Marks N.E., Borg L.E., Hutcheon I.D., Jacobsen B., Clayton R.N.
    Earth Planet. Sci. Lett., 405 (2014), pp. 15-24

\noindent [4]
    Cayrel R., Hill V., Beers T.C., Barbuy B., Spite M., Spite F., Plez B., Andersen J., Bonifacio P., François P., Molaro P., Nordström B., Primas F.
    Nature, 409 (2001), pp. 691-692

\noindent [5]
    Frebel A., Christlieb N., Norris J.E., Thom C., Beers T.C., Rhee J.
    Astrophys. J., 660 (2007), pp. L117-L120

\noindent [6]
    Dauphas N., Chaussidon M.
    Annu. Rev. Earth Planet. Sci., 39 (2011), pp. 351-386

\noindent [7]
    Borg L.E., Connelly J.N., Boyet M., Carlson R.W.
    Nature, 477 (2011), pp. 70-72

\noindent [8]
    Tissot F.L.H., Dauphas N., Grove T.L.
    Geochim. Cosmochim. Acta, 213 (2017), pp. 593-617

\noindent [9]
    Gargaud M., Amils R., Quintanilla J.C., Cleaves H.J., Irvine W.M., Pinti D.L., Viso M.
    Encyclopedia of Astrobiology
    Springer-Verlag, Berlin, Heidelberg (2011)

\noindent [10]
    Lee T., Papanastassiou D.A., Wasserburg G.J.
    Astrophys. J., 211 (1977), pp. L107-L110

\noindent [11]
    Liu M.-C., Chaussidon M., Srinivasan G., McKeegan K.D.
    Astrophys. J., 761 (2012), p. 137

\noindent [12]
    Tissot F.L.H., Dauphas N., Grossman L.
    Sci. Adv., 2 (2016), Article e1501400

\noindent [13]
    Tang H., Liu M.-C., McKeegan K.D., Tissot F.L.H., Dauphas N.
    Geochim. Cosmochim. Acta, 207 (2017), pp. 1-18

\noindent [14]
    Connelly J.N., Bizzarro M., Krot A.N., Nordlund A., Wielandt D., Ivanova M.A.
    Science, 338 (2012), pp. 651-655

\noindent [15]
    Wallner A., Feige J., Kinoshita N., Paul M., Fifield L.K., Golser R., Honda M., Linnemann U., Matsuzaki H., Merchel S., Rugel G., Tims S.G., Steier P., Yamagata T., Winkler S.R.
    Nature, 532 (2016), pp. 69-72

\noindent [16]
    Wallner A., Faestermann T., Feige J., Feldstein C., Knie K., Korschinek G., Kutschera W., Ofan A., Paul M., Quinto F., Rugel G., Steier P.
    Nature Commun., 6 (2015), p. 5956

\noindent [17]
    Ludwig P., Bishop S., Egli R., Chernenko V., Deneva B., Faestermann T., Famulok N., Fimiani L., Gómez-Guzmán J.M., Hain K., Korschinek G., Hanzlik M., Merchel S., Rugel G.
    Proc. Natl. Acad. Sci., 113 (2016), pp. 9232-9237

\noindent [18]
    Fimiani L., Cook D.L., Faestermann T., Gómez-Guzmán J.M., Hain K., Herzog G., Knie K., Korschinek G., Ludwig P., Park J., Reedy R.C., Rugel G.
    Phys. Rev. Lett., 116 (15) (2016), Article 151104

\noindent [19]
    Breitschwerdt D., Feige J., Schulreich M.M., Avillez M.A.D., Dettbarn C., Fuchs B.
    Nature, 532 (2016), pp. 73-76

\noindent [20]
    Binns W.R., Israel M.H., Christian E.R., Cummings A.C., de Nolfo G.A., Lave K.A., Leske R.A., Mewaldt R.A., Stone E.C., von Rosenvinge T.T., Wiedenbeck M.E.
    Science, 352 (2016), pp. 677-680

\noindent [21]
    Dauphas N., Schauble E.A.
    Annu. Rev. Earth Planet. Sci., 44 (2016), pp. 709-783

\noindent [22]
    G.A. Brennecka, T. Kleine, Lunar and Planetary Science Conference, vol. 47, 2016, p. 1135.

\noindent [23]
    Lugmair G.W., Marti K.
    Earth Planet. Sci. Lett., 35 (1977), pp. 273-284

\noindent [24]
    G.B. Hudson, B.M. Kennedy, F.A. Podosek, C.M. Hohenberg, in: G. Ryder, V.L. Sharpton (Eds.), Lunar and Planetary Science Conference Proceedings, vol. 19, 1989, pp. 547–557.

\noindent [25]
    Brennecka G.A., Kleine T.
    Astrophys. J., 837 (2017), p. L9

\noindent [26]
    Lodders K., Palme H., Gail H.-P.
    Abundances of the Elements in the Solar System
    Landolt-Börnstein, New Series VI/4B, 34, Chapter 4.4. (2009)

\noindent [27]
    Burkhardt C., Kleine T., Oberli F., Pack A., Bourdon B., Wieler R.
    Earth Planet. Sci. Lett., 312 (2011), pp. 390-400

\noindent [28]
    Matthes M., Fischer-Gödde M., Kruijer T.S., Kleine T.
    Geochim. Cosmochim. Acta, 220 (2018), pp. 82-95

\noindent [29]
    Ott U.
    Chem. Der Erde Geochem., 74 (2014), pp. 519-544

\noindent [30]
    Becker H., Walker R.J.
    Chem. Geol., 196 (2003), pp. 43-56

\noindent [31]
    Jacobsen B., Yin Q.-Z., Moynier F., Amelin Y., Krot A.N., Nagashima K., Hutcheon I.D., Palme H.
    Earth Planet. Sci. Lett., 272 (2008), pp. 353-364

\noindent [32]
    Tatischeff V., Duprat J., de Séréville N.
    Astrophys. J., 796 (2014), p. 124

\noindent [33]
    Kruijer T.S., Kleine T., Fischer-Gödde M., Burkhardt C., Wieler R.
    Earth Planet. Sci. Lett., 403 (2014), pp. 317-327

\noindent [34]
    Ott U.
    Alsabti A.W., Murdin P. (Eds.), Handbook of Supernovae, Springer International Publishing Switzerland (2016)

\noindent [35]
    M.K. Haba, Y.-J. Lai, J.F. Wotzlaw, A. Yamaguchi, A. von Quadt, M. Schönbächler, Lunar and Planetary Science Conference, vol. 48, 2017, p. 1739.

\noindent [36]
    Tang H., Dauphas N.
    Astrophys. J., 802 (2015), p. 22

\noindent [37]
    Chaussidon M., Robert F., McKeegan K.D.
    Geochim. Cosmochim. Acta, 70 (2006), pp. 224-245

\noindent [38]
    Liu M.-C.
    Geochim. Cosmochim. Acta, 201 (2017), pp. 123-135

\noindent [39]
    Palk A.G.W., Andreasen R., Rehkämper M., Stunt A., Kreissig K., Schönbächler M., Smith C.
    Meteorit. Planet. Sci., 53 (2018), pp. 167-186

\noindent [40]
    Brennecka G.A., Borg L.E., Romaniello S.J., Souders A.K., Shollenberger Q.R., Marks N.E., Wadhwa M.
    Geochim. Cosmochim. Acta, 201 (2017), pp. 331-344

\noindent [41]
    Chmeleff J., von Blanckenburg F., Kossert K., Jakob D.
    Nucl. Instrum. Methods Phys. Res., 268 (2010), pp. 192-199

\noindent [42]
    Gounelle M., Chaussidon M., Rollion-Bard C.
    Astrophys. J., 763 (2013), p. L33

\noindent [43]
    Kinoshita N., Paul M., Kashiv Y., Collon P., Deibel C.M., DiGiovine B., Greene J.P., Henderson D.J., Jiang C.L., Marley S.T., Nakanishi T., Pardo R.C., Rehm K.E., Robertson D., Scott R., Schmitt C., Tang X.D., Vondrasek R., Yokoyama A.
    Science, 335 (2012), pp. 1614-1617

\noindent [44]
    Mishra R.K., Chaussidon M.
    Earth Planet. Sci. Lett., 398 (2014), pp. 90-100

\noindent [45]
    Telus M., Huss G.R., Nagashima K., Ogliore R.C., Tachibana S.
    Geochim. Cosmochim. Acta, 221 (2018), pp. 342-357

\noindent [46]
    Merrill S.P.W.
    Astrophys. J., 116 (1952), pp. 21-26

\noindent [47]
    Woosley S.E., Heger A., Weaver T.A.
    Rev. Modern Phys., 74 (2002), pp. 1015-1071

\noindent [48]
    Langer N.
    Ann. Rev. Astron. Astrophys., 50 (2012), pp. 107-164

\noindent [49]
    Karakas A.I., Lattanzio J.C.
    Publ. Astron. Soc. Aust., 31 (2014), p. e030

\noindent [50]
    De Marco O., Izzard R.G.
    Publ. Astron. Soc. Aust., 34 (2017), p. e001

\noindent [51]
    Janka H.-T.
    Annu. Rev. Nucl. Part. Sci., 62 (2012), pp. 407-451

\noindent [52]
    Riess A.G., Filippenko A.V., Challis P., Clocchiatti A., Diercks A., Garnavich P.M., Gilliland R.L., Hogan C.J., Jha S., Kirshner R.P., Leibundgut B., Phillips M.M., Reiss D., Schmidt B.P., Schommer R.A., Smith R.C., Spyromilio J., Stubbs C., Suntzeff N.B., Tonry J.
    Astron. J., 116 (1998), pp. 1009-1038

\noindent [53]
    Cameron A.G.W.
    Publ. Astron. Soc. Pac., 69 (1957), p. 201

\noindent [54]
    Burbidge E.M., Burbidge G.R., Fowler W.A., Hoyle F.
    Rev. Modern Phys., 29 (1957), pp. 547-650

\noindent [55]
    Wallerstein G., Iben I. Jr., Parker P., Boesgaard A.M., Hale G.M., Champagne A.E., Barnes C.A., Käppeler F., Smith V.V., Hoffman R.D., Timmes F.X., Sneden C., Boyd R.N., Meyer B.S., Lambert D.L.
    Rev. Modern Phys., 69 (1997), pp. 995-1084

\noindent [56]
    Käppeler F., Gallino R., Bisterzo S., Aoki W.
    Rev. Modern Phys., 83 (2011), pp. 157-194

\noindent [57]
    Thielemann F.-K., Arcones A., Käppeli R., Liebendörfer M., Rauscher T., Winteler C., Fröhlich C., Dillmann I., Fischer T., Martinez-Pinedo G., Langanke K., Farouqi K., Kratz K.-L., Panov I., Korneev I.K.
    Prog. Part. Nucl. Phys., 66 (2011), pp. 346-353

\noindent [58]
    Herwig F., Pignatari M., Woodward P.R., Porter D.H., Rockefeller G., Fryer C.L., Bennett M., Hirschi R.
    Astrophys. J., 727 (2011), p. 89

\noindent [59]
    Hampel M., Stancliffe R.J., Lugaro M., Meyer B.S.
    Astrophys. J., 831 (2016), p. 171

\noindent [60]
    Kilpatrick C.D., Foley R.J., Kasen D., Murguia-Berthier A., Ramirez-Ruiz E., Coulter D.A., Drout M.R., Piro A.L., Shappee B.J., Boutsia K., Contreras C., Di Mille F., Madore B.F., Morrell N., Pan Y.-C., Prochaska J.X., Rest A., Rojas-Bravo C., Siebert M.R., Simon J.D., Ulloa N.
    Science, 358 (2017), pp. 1583-1587

\noindent [61]
    Côté B., Fryer C.L., Belczynski K., Korobkin O., Chruślińska M., Vassh N., Mumpower M.R., Lippuner J., Sprouse T.M., Surman R., Wollaeger R.
    Astrophys. J., 855 (2018), p. 99

\noindent [62]
    Hotokezaka K., Piran T., Paul M.
    Nat. Phys., 11 (2015), pp. 1042-1044

\noindent [63]
    Pignatari M., Göbel K., Reifarth R., Travaglio C.
    Internat. J. Modern Phys. E, 25 (2016), pp. 1630003-1630232

\noindent [64]
    Travaglio C., Randich S., Galli D., Lattanzio J., Elliott L.M., Forestini M., Ferrini F.
    Astrophys. J., 559 (2001), pp. 909-924

\noindent [65]
    Tinsley B.M.
    Fundam. Cosm. Phys., 5 (1980), pp. 287-388

\noindent [66]
    Matteucci F.
    Chemical Evolution of Galaxies, Astronomy and Astrophysics Library, 978-3-642-22490-4, Springer-Verlag, Berlin, Heidelberg (2012)

\noindent [67]
    Asplund M., Grevesse N., Sauval A.J., Scott P.
    Ann. Rev. Astron. Astrophys., 47 (2009), pp. 481-522

\noindent [68]
    Casagrande L., Schönrich R., Asplund M., Cassisi S., Ramírez I., Meléndez J., Bensby T., Feltzing S.
    Astron. Astrophys., 530 (2011), p. A138

\noindent [69]
    Bensby T., Feltzing S., Oey M.S.
    Astron. Astrophys., 562 (2014), p. A71

\noindent [70]
    Spitoni E., Romano D., Matteucci F., Ciotti L.
    Astrophys. J., 802 (2015), p. 129

\noindent [71]
    Williams J.
    Contemp. Phys., 51 (2010), pp. 381-396

\noindent [72]
    Hartmann L., Ballesteros-Paredes J., Bergin E.A.
    Astrophys. J., 562 (2001), pp. 852-868

\noindent [73]
    Murray N.
    Astrophys. J., 729 (2011), p. 133

\noindent [74]
    Dib S., Gutkin J., Brandner W., Basu S.
    Mon. Not. R. Astron. Soc., 436 (2013), pp. 3727-3740

\noindent [75]
    Haisch K.E. Jr., Lada E.A., Lada C.J.
    Astrophys. J., 553 (2001), pp. L153-L156

\noindent [76]
    Williams J.P., Cieza L.A.
    Ann. Rev. Astron. Astrophys., 49 (2011), pp. 67-117

\noindent [77]
    Lodders K.
    Astrophys. Space Sci. Proc., 16 (2010), p. 379

\noindent [78]
    de Avillez M.A., Mac Low M.-M.
    Astrophys. J., 581 (2002), pp. 1047-1060

\noindent [79]
    Gounelle M., Meibom A.
    Astrophys. J., 664 (2007), pp. L123-L125

\noindent [80]
    Lugaro M., Doherty C.L., Karakas A.I., Maddison S.T., Liffman K., García-Hernández D.A., Siess L., Lattanzio J.C.
    Meteorit. Planet. Sci., 47 (2012), pp. 1998-2012

\noindent [81]
    Herzberg C., Condie K., Korenaga J.
    Earth Planet. Sci. Lett., 292 (2010), pp. 79-88

\noindent [82]
    Jaupart C., Labrosse S., Lucazeau F., Mareschal J.-C.
    Schubert G. (Ed.), Treatise on Geophysics, vol. 7 (second ed.), Elsevier, Oxford (2015), p. 223270

\noindent [83]
    Andrault D., Monteux J., Le Bars M., Samuel H.
    Earth Planet. Sci. Lett., 443 (2016), pp. 195-203

\noindent [84]
    Araki T., Enomoto S., Furuno K., Gando Y., Ichimura K., Ikeda H., Inoue K., Kishimoto Y., Koga M., Koseki Y., Maeda T., Mitsui T., Motoki M., Nakajima K., Ogawa H., Ogawa M., Owada K., Ricol J.-S., Shimizu I., Shirai J., Suekane F., Suzuki A., Tada K., Takeuchi S., Tamae K., Tsuda Y., Watanabe H., Busenitz J., Classen T., Djurcic Z., Keefer G., Leonard D., Piepke A., Yakushev E., Berger B.E., Chan Y.D., Decowski M.P., Dwyer D.A., Freedman S.J., Fujikawa B.K., Goldman J., Gray F., Heeger K.M., Hsu L., Lesko K.T., Luk K.-B., Murayama H., O’Donnell T., Poon A.W.P., Steiner H.M., Winslow L.A., Mauger C., McKeown R.D., Vogel P., Lane C.E., Miletic T., Guillian G., Learned J.G., Maricic J., Matsuno S., Pakvasa S., Horton-Smith G.A., Dazeley S., Hatakeyama S., Rojas A., Svoboda R., Dieterle B.D., Detwiler J., Gratta G., Ishii K., Tolich N., Uchida Y., Batygov M., Bugg W., Efremenko Y., Kamyshkov Y., Kozlov A., Nakamura Y., Karwowski H.J., Markoff D.M., Nakamura K., Rohm R.M., Tornow W., Wendell R., Chen M.-J., Wang Y.-F., Piquemal F.
    Nature, 436 (2005), pp. 499-503

\noindent [85]
    Hazen R., Papineau D., Bleeker W., Downs R., Ferry J., McCoy T., Svejensky D., Yang H.
    Am. Mineral., 93 (2008), pp. 1693-1720

\noindent [86]
    Lazio T.J.W., Shkolnik E., Hallinan G., Planetary Habitability Study Team
    Planetary Magnetic Fields: Planetary Interiors and Habitability
    Planetary Magnetic Fields: Planetary Interiors and Habitability, Tech. Rep. Final Report Funded By the Keck Institute for Space Studies (2016), p. 147

\noindent [87]
    Fisk M.R., Giovannoni S.J.
    J. Geophys. Res., 104 (1999), pp. 11805-11816

\noindent [88]
    Trevors J.
    Res. Microbiol., 153 (2002), p. 487491

\noindent [89]
    Turcotte D.L., Schubert G.
    Geodynamics (second ed.), Cambridge University Press (2002), p. 472

\noindent [90]
    Unterborn C.T., Johnson J.A., Panero W.R.
    Astrophys. J., 806 (2015), p. 139

\noindent [91]
    Wehmeyer B., Pignatari M., Thielemann F.-K.
    Mon. Not. R. Astron. Soc., 452 (2015), pp. 1970-1981

\noindent [92]
    Lodders K.
    Astrophys. J., 591 (2003), pp. 1220-1247

\noindent [93]
    Wasserburg G.J., Busso M., Gallino R., Nollett K.M.
    Nuclear Phys. A, 777 (2006), pp. 5-69

\noindent [94]
    Lugaro M., Heger A., Osrin D., Goriely S., Zuber K., Karakas A.I., Gibson B.K., Doherty C.L., Lattanzio J.C., Ott U.
    Science, 345 (2014), pp. 650-653

\noindent [95]
    Trigo-Rodríguez J.M., García-Hernández D.A., Lugaro M., Karakas A.I., van Raai M., García Lario P., Manchado A.
    Meteorit. Planet. Sci., 44 (2009), pp. 627-641

\noindent [96]
    Wasserburg G.J., Karakas A.I., Lugaro M.
    Astrophys. J., 836 (2017), p. 126

\noindent [97]
    Arnould M., Paulus G., Meynet G.
    Astron. Astrophys., 321 (1997), pp. 452-464

\noindent [98]
    Arnould M., Goriely S., Meynet G.
    Astron. Astrophys., 453 (2006), pp. 653-659

\noindent [99]
    Limongi M., Chieffi A.
    Astrophys. J., 647 (2006), pp. 483-500

\noindent [100]
    Takigawa A., Miki J., Tachibana S., Huss G.R., Tominaga N., Umeda H., Nomoto K.
    Astrophys. J., 688 (2008), pp. 1382-1387

\noindent [101]
    Rauscher T., Heger A., Hoffman R.D., Woosley S.E.
    Astrophys. J., 576 (2002), pp. 323-348

\noindent [102]
    Meyer B.S., Clayton D.D.
    Space Sci. Rev., 92 (2000), pp. 133-152

\noindent [103]
    Lugaro M., Pignatari M., Ott U., Zuber K., Travaglio C., Gyürky G., Fülöp Z.
    Proc. Natl. Acad. Sci., 113 (2016), pp. 907-912

\noindent [104]
    Rauscher T., Dauphas N., Dillmann I., Fröhlich C., Fülöp Z., Gyürky G.
    Rep. Progr. Phys., 76 (6) (2013), Article 066201

\noindent [105]
    Banerjee P., Qian Y.-Z., Heger A., Haxton W.C.
    Nature Commun., 7 (2016), p. 13639

\noindent [106]
    Hayakawa T., Nakamura K., Kajino T., Chiba S., Iwamoto N., Cheoun M.K., Mathews G.J.
    Astrophys. J., 779 (2013), p. L9

\noindent [107]
    Travaglio C., Hillebrandt W., Reinecke M., Thielemann F.-K.
    Astron. Astrophys., 425 (2004), pp. 1029-1040

\noindent [108]
    Travaglio C., Gallino R., Rauscher T., Dauphas N., Röpke F.K., Hillebrandt W.
    Astrophys. J., 795 (2014), p. 141

\noindent [109]
    Bisterzo S., Gallino R., Straniero O., Cristallo S., Käppeler F.
    Mon. Not. R. Astron. Soc., 418 (2011), pp. 284-319

\noindent [110]
    Goriely S., Arnould M.
    Astron. Astrophys., 379 (2001), pp. 1113-1122

\noindent [111]
    Goriely S., Janka H.-T.
    Mon. Not. R. Astron. Soc., 459 (2016), pp. 4174-4182

\noindent [112]
    José J., Hernanz M.
    J. Phys. G: Nucl. Part. Phys., 34 (2007), pp. R431-R458

\noindent [113]
    Gounelle M., Shu F.H., Shang H., Glassgold A.E., Rehm K.E., Lee T.
    Astrophys. J., 640 (2006), pp. 1163-1170

\noindent [114]
    Desch S.J., Connolly H.C. Jr., Srinivasan G.
    Astrophys. J., 602 (2004), pp. 528-542

\noindent [115]
    Tatischeff V., Duprat J., de Séréville N.
    Astrophys. J., 714 (2010), pp. L26-L30

\noindent [116]
    Desch S.J., Morris M.A., Connolly H.C. Jr., Boss A.P.
    Astrophys. J., 725 (2010), pp. 692-711

\noindent [117]
    Duprat J., Tatischeff V.
    Astrophys. J., 671 (2007), pp. L69-L72

\noindent [118]
    Anderson T., Skulski M., Clark A., Nelson A., Ostdiek K., Collon P., Chmiel G., Woodruff T., Caffee M.
    Phys. Rev. C, 96 (1) (2017), Article 015803

\noindent [119]
    Fitoussi C., Duprat J., Tatischeff V., Kiener J., Naulin F., Raisbeck G., Assunção M., Bourgeois C., Chabot M., Coc A., Engrand C., Gounelle M., Hammache F., Lefebvre A., Porquet M.-G., Scarpaci J.-A., de Séréville N., Thibaud J.-P., Yiou F.
    Phys. Rev. C, 78 (4) (2008), Article 044613

\noindent [120]
    Wielandt D., Nagashima K., Krot A.N., Huss G.R., Ivanova M.A., Bizzarro M.
    Astrophys. J., 748 (2012), p. L25

\noindent [121]
    Cameron A.G.W., Fowler W.A.
    Astrophys. J., 164 (1971), p. 111

\noindent [122]
    Urey H.C.
    Proc. Natl. Acad. Sci., 41 (1955), pp. 127-144

\noindent [123]
    Cameron A.G.W., Truran J.W.
    Icarus, 30 (1977), pp. 447-461

\noindent [124]
    Villeneuve J., Chaussidon M., Libourel G.
    Science, 325 (2009), p. 985

\noindent [125]
    Park C., Nagashima K., Krot A.N., Huss G.R., Davis A.M., Bizzarro M.
    Geochim. Cosmochim. Acta, 201 (2017), pp. 6-24

\noindent [126]
    Makide K., Nagashima K., Krot A.N., Huss G.R., Ciesla F.J., Hellebrand E., Gaidos E., Yang L.
    Astrophys. J., 733 (2011), p. L31

\noindent [127]
    Makide K., Nagashima K., Krot A.N., Huss G.R., Hutcheon I.D., Hellebrand E., Petaev M.I.
    Geochim. Cosmochim. Acta, 110 (2013), pp. 190-215

\noindent [128]
    Ireland T.R.
    Aust. J. Earth Sci., 59 (2012), pp. 225-236

\noindent [129]
    Larsen K.K., Trinquier A., Paton C., Schiller M., Wielandt D., Ivanova M.A., Connelly J.N., Nordlund A., Krot A.N., Bizzarro M.
    Astrophys. J. Lett., 735 (2011), p. L37

\noindent [130]
    Schiller M., Connelly J.N., Glad A.C., Mikouchi T., Bizzarro M.
    Earth Planet. Sci. Lett., 420 (2015), pp. 45-54

\noindent [131]
    Budde G., Kruijer T.S., Kleine T.
    Geochim. Cosmochim. Acta, 222 (2018), pp. 284-304

\noindent [132]
    Larsen K.K., Schiller M., Bizzarro M.
    Geochim. Cosmochim. Acta, 176 (2016), pp. 295-315

\noindent [133]
    Straniero O., Imbriani G., Strieder F., Bemmerer D., Broggini C., Caciolli A., Corvisiero P., Costantini H., Cristallo S., DiLeva A., Formicola A., Elekes Z., Fülöp Z., Gervino G., Guglielmetti A., Gustavino C., Gyürky G., Junker M., Lemut A., Limata B., Marta M., Mazzocchi C., Menegazzo R., Piersanti L., Prati P., Roca V., Rolfs C., Rossi Alvarez C., Somorjai E., Terrasi F., Trautvetter H.-P.
    Astrophys. J., 763 (2013), p. 100

\noindent [134]
    Nittler L.R., Alexander O., Gao X., Walker R.M., Zinner E.
    Astrophys. J., 483 (1997), pp. 475-495

\noindent [135]
    Palmerini S., La Cognata M., Cristallo S., Busso M.
    Astrophys. J., 729 (2011), p. 3

\noindent [136]
    Bruno C.G., Scott D.A., Aliotta M., Formicola A., Best A., Boeltzig A., Bemmerer D., Broggini C., Caciolli A., Cavanna F., Ciani G.F., Corvisiero P., Davinson T., Depalo R., Di Leva A., Elekes Z., Ferraro F., Fülöp Z., Gervino G., Guglielmetti A., Gustavino C., Gyürky G., Imbriani G., Junker M., Menegazzo R., Mossa V., Pantaleo F.R., Piatti D., Prati P., Somorjai E., Straniero O., Strieder F., Szücs T., Takács M.P., Trezzi D., LUNA Collaboration
    Phys. Rev. Lett., 117 (14) (2016), Article 142502

\noindent [137]
    Iliadis C., Longland R., Champagne A.E., Coc A.
    Nuclear Phys. A, 841 (2010), pp. 323-388

\noindent [138]
    Lugaro M., Karakas A.I., Bruno C.G., Aliotta M., Nittler L.R., Bemmerer D., Best A., Boeltzig A., Broggini C., Caciolli A., Cavanna F., Ciani G.F., Corvisiero P., Davinson T., Depalo R., di Leva A., Elekes Z., Ferraro F., Formicola A., Fülöp Z., Gervino G., Guglielmetti A., Gustavino C., Gyürky G., Imbriani G., Junker M., Menegazzo R., Mossa V., Pantaleo F.R., Piatti D., Prati P., Scott D.A., Straniero O., Strieder F., Szücs T., Takács M.P., Trezzi D.
    Nat. Astron., 1 (2017), p. 0027

\noindent [139]
    Abia C., Hedrosa R.P., Domínguez I., Straniero O.
    Astron. Astrophys., 599 (2017), p. A39

\noindent [140]
    Timmes F.X., Woosley S.E., Hartmann D.H., Hoffman R.D., Weaver T.A., Matteucci F.
    Astrophys. J., 449 (1995), p. 204

\noindent [141]
    de Smet L., Wagemans C., Wagemans J., Heyse J., van Gils J.
    Phys. Rev. C, 76 (4) (2007), Article 045804

\noindent [142]
    Oginni B.M., Iliadis C., Champagne A.E.
    Phys. Rev. C, 83 (2) (2011), Article 025802

\noindent [143]
    Dillmann I., Heil M., Käppeler F., Plag R., Rauscher T., Thielemann F.
    Woehr A., Aprahamian A. (Eds.), Capture Gamma-Ray Spectroscopy and Related Topics, American Institute of Physics Conference Series, 819 (2006), pp. 123-127

\noindent [144]
    Iliadis C., Champagne A., Chieffi A., Limongi M.
    Astrophys. J. Suppl., 193 (2011), p. 16

\noindent [145]
    Diehl R.
    Rep. Progr. Phys., 76 (2) (2013), Article 026301

\noindent [146]
    Kubryk M., Prantzos N., Athanassoula E.
    Astron. Astrophys., 580 (2015), p. A126

\noindent [147]
    Wang W., Lang M.G., Diehl R., Halloin H., Jean P., Knödlseder J., Kretschmer K., Martin P., Roques J.P., Strong A.W., Winkler C., Zhang X.L.
    Astron. Astrophys., 496 (2009), pp. 713-724

\noindent [148]
    Martin P., Knödlseder J., Diehl R., Meynet G.
    Astron. Astrophys., 506 (2009), pp. 703-710

\noindent [149]
    Diehl R., Lang M.G., Martin P., Ohlendorf H., Preibisch T., Voss R., Jean P., Roques J.-P., von Ballmoos P., Wang W.
    Astron. Astrophys., 522 (2010), p. A51

\noindent [150]
    Zinner E.
    Davis A.M., Holland H.D., Turekian K.K. (Eds.), Treatise on Geochemistry, vol. 1 (second ed.), Meteorites and Cosmochemical Processes, Elsevier, Oxford (2014), pp. 181-213

\noindent [151]
    Groopman E., Zinner E., Amari S., Gyngard F., Hoppe P., Jadhav M., Lin Y., Xu Y., Marhas K., Nittler L.R.
    Astrophys. J., 809 (2015), p. 31

\noindent [152]
    Pignatari M., Wiescher M., Timmes F.X., de Boer R.J., Thielemann F.-K., Fryer C., Heger A., Herwig F., Hirschi R.
    Astrophys. J., 767 (2013), p. L22

\noindent [153]
    van Raai M.A., Lugaro M., Karakas A.I., Iliadis C.
    Astron. Astrophys., 478 (2008), pp. 521-526

\noindent [154]
    Hsu W., Guan Y., Leshin L.A., Ushikubo T., Wasserburg G.J.
    Astrophys. J., 640 (2006), pp. 525-529

\noindent [155]
    Nakashima D., Ott U., Hoppe P., El Goresy A.
    Geochim. Cosmochim. Acta, 72 (2008), pp. 6141-6153

\noindent [156]
    de Smet L., Wagemans C., Goeminne G., Heyse J., van Gils J.
    Phys. Rev. C, 75 (3) (2007), Article 034617

\noindent [157]
    Trinquier A., Birck J.-L., Allègre C.J., Göpel C., Ulfbeck D.
    Geochim. Cosmochim. Acta, 72 (2008), pp. 5146-5163

\noindent [158]
    Göpel C., Birck J.-L., Galy A., Barrat J.-A., Zanda B.
    Geochim. Cosmochim. Acta, 156 (2015), pp. 1-24

\noindent [159]
    Dressler R., Ayranov M., Bemmerer D., Bunka M., Dai Y., Lederer C., Fallis J., StJ Murphy A., Pignatari M., Schumann D., Stora T., Stowasser T., Thielemann F.-K., Woods P.J.
    J. Phys. G: Nucl. Phys., 39 (2012), Article 105201

\noindent [160]
    Nyquist L.E., Kleine T., Shih C.-Y., Reese Y.D.
    Geochim. Cosmochim. Acta, 73 (2009), pp. 5115-5136

\noindent [161]
    Seitenzahl I.R., Cescutti G., Röpke F.K., Ruiter A.J., Pakmor R.
    Astron. Astrophys., 559 (2013), p. L5

\noindent [162]
    Hitomi Collaboration
    Nature, 551 (2017), p. 478

\noindent [163]
    Parikh A., José J., Seitenzahl I.R., Röpke F.K.
    Astron. Astrophys., 557 (2013), p. A3

\noindent [164]
    Seitenzahl I.R., Ciaraldi-Schoolmann F., Röpke F.K., Fink M., Hillebrandt W., Kromer M., Pakmor R., Ruiter A.J., Sim S.A., Taubenberger S.
    Mon. Not. R. Astron. Soc., 429 (2013), pp. 1156-1172

\noindent [165]
    Mishra R., Chaussidon M., Marhas K.
    Nuclei in the Cosmos (NIC XII) (2012), p. 85
    Published online at http://pos.sissa.it/cgi-bin/reader/conf.cgi?confid=146

\noindent [166]
    Tang H., Dauphas N.
    Earth Planet. Sci. Lett., 359–360 (2012), pp. 248-263

\noindent [167]
    P. Boehnke, K.D. McKeegan, T. Stephan, R.C.J. Steele, R. Trappitsch, A.M. Davis, M. Pellin, M-C. Liu, Met. Planet. Sci., LPI Contri. No 1987, 2017.

\noindent [168]
    Wallner A., Bichler M., Buczak K., Dressler R., Fifield L.K., Schumann D., Sterba J.H., Tims S.G., Wallner G., Kutschera W.
    Phys. Rev. Lett., 114 (4) (2015), Article 041101

\noindent [169]
    Ostdiek K.M., Anderson T.S., Bauder W.K., Bowers M.R., Clark A.M., Collon P., Lu W., Nelson A.D., Robertson D., Skulski M., Dressler R., Schumann D., Greene J.P., Kutschera W., Paul M.
    Phys. Rev. C, 95 (5) (2017), Article 055809

\noindent [170]
    Rugel G., Faestermann T., Knie K., Korschinek G., Poutivtsev M., Schumann D., Kivel N., Günther-Leopold I., Weinreich R., Wohlmuther M.
    Phys. Rev. Lett., 103 (7) (2009), Article 072502

\noindent [171]
    Uberseder E., Adachi T., Aumann T., Beceiro-Novo S., Boretzky K., Caesar C., Dillmann I., Ershova O., Estrade A., Farinon F., Hagdahl J., Heftrich T., Heil M., Heine M., Holl M., Ignatov A., Johansson H.T., Kalantar N., Langer C., Le Bleis T., Litvinov Y.A., Marganiec J., Movsesyan A., Najafi M.A., Nilsson T., Nociforo C., Panin V., Pietri S., Plag R., Prochazka A., Rastrepina G., Reifarth R., Ricciardi V., Rigollet C., Rossi D.M., Savran D., Simon H., Sonnabend K., Streicher B., Terashima S., Thies R., Togano Y., Volkov V., Wamers F., Weick H., Weigand M., Wiescher M., Wimmer C., Winckler N., Woods P.J.
    Phys. Rev. Lett., 112 (21) (2014), Article 211101

\noindent [172]
    Uberseder E., Reifarth R., Schumann D., Dillmann I., Pardo C.D., Görres J., Heil M., Käppeler F., Marganiec J., Neuhausen J., Pignatari M., Voss F., Walter S., Wiescher M.
    Phys. Rev. Lett., 102 (15) (2009), Article 151101

\noindent [173]
    Li K.A., Lam Y.H., Qi C., Tang X.D., Zhang N.T.
    Phys. Rev. C, 94 (6) (2016), Article 065807

\noindent [174]
    Wanajo S., Janka H.-T., Müller B.
    Astrophys. J. Lett., 774 (2013), p. L6

\noindent [175]
    Woosley S.E.
    Astrophys. J., 476 (1997), pp. 801-810

\noindent [176]
    Reynolds J.H.
    Phys. Rev. Lett., 4 (1960), pp. 8-10

\noindent [177]
    Gilmour J.D., Pravdivtseva O.V., Busfield A., Hohenberg C.M.
    Meteorit. Planet. Sci., 41 (2006), pp. 19-31

\noindent [178]
    Turner G., Busfield A., Crowther S.A., Harrison M., Mojzsis S.J., Gilmour J.
    Earth Planet. Sci. Lett., 261 (2007), pp. 491-499

\noindent [179]
    Hagee B., Bernatowicz T.J., Podosek F.A., Johnson M.L., Burnett D.S.
    Geochim. Cosmochim. Acta, 54 (1990), pp. 2847-2858

\noindent [180]
    M. Peto, S.B. Jacobsen, Planetary Science Conference, vol. 39, 2008, p. 2499.

\noindent [181]
    K.M. Pető, Accretion: Building New Worlds, Houston, TX, USA. Abstract 2038.

\noindent [182]
    O. Pravdivtseva, A. Meshik, C.M. Hohenberg, in: S. Kubono, T. Kajino, S. Nishimura, T. Isobe, S. Nagataki, T. Shima, Y. Takeda (Eds.), 14th International Symposium on Nuclei in the Cosmos, NIC2016, JPS Conf. Proc. 14, 2017, 011005.

\noindent [183]
    Pravdivtseva O., Meshik A., Hohenberg C.M., Krot A.N.
    Geochim. Cosmochim. Acta, 201 (2017), pp. 320-330

\noindent [184]
    Ratzel U., Arlandini C., Käppeler F., Couture A., Wiescher M., Reifarth R., Gallino R., Mengoni A., Travaglio C.
    Phys. Rev. C, 70 (6) (2004), Article 065803

\noindent [185]
    Takahashi K., Yokoi K.
    At. Data Nucl. Data Tables, 36 (1987), p. 375

\noindent [186]
    Reifarth R., Käppeler F., Voss F., Wisshak K., Gallino R., Pignatari M., Straniero O.
    Astrophys. J., 614 (2004), pp. 363-370

\noindent [187]
    Rauscher T.
    Astrophys. J., 755 (2012), p. L10

\noindent [188]
    Arlandini C., Käppeler F., Wisshak K., Gallino R., Lugaro M., Busso M., Straniero O.
    Astrophys. J., 525 (1999), pp. 886-900

\noindent [189]
    Burkhardt C., Kleine T., Bourdon B., Palme H., Zipfel J., Friedrich J.M., Ebel D.S.
    Geochim. Cosmochim. Acta, 72 (2008), pp. 6177-6197

\noindent [190]
    Schönbächler M., Carlson R.W., Horan M.F., Mock T.D., Hauri E.H.
    Geochim. Cosmochim. Acta, 72 (2008), pp. 5330-5341

\noindent [191]
    R. Andreasen, M. Rehkämper, G.K. Benedix, K.J. Theis, M. Schönbächler, C.L. Smith, Lunar and Planetary Science Conference, vol. 43, 2012, p. 2902.

\noindent [192]
    Baker R.G.A., Schönbächler M., Rehkämper M., Williams H.M., Halliday A.N.
    Earth Planet. Sci. Lett., 291 (2010), pp. 39-47

\noindent [193]
    Nielsen S.G., Rehkämper M., Halliday A.N.
    Geochim. Cosmochim. Acta, 70 (2006), pp. 2643-2657

\noindent [194]
    Wasserburg G.J., Busso M., Gallino R.
    Astrophys. J., 466 (1996), p. L109

\noindent [195]
    Ott U., Kratz K.-L.
    New Astron. Rev., 52 (2008), pp. 396-400

\noindent [196]
    Bondarenko V., Berzins J., Prokofjevs P., Simonova L., von Egidy T., Honzátko J., Tomandl I., Alexa P., Wirth H.-F., Köster U., Eisermann Y., Metz A., Graw G., Hertenberger R., Rubacek L.
    Nuclear Phys. A, 709 (2002), pp. 3-59

\noindent [197]
    Goriely S.
    Astron. Astrophys., 342 (1999), pp. 881-891

\noindent [198]
    Yokoi K., Takahashi K., Arnould M.
    Astron. Astrophys., 145 (1985), pp. 339-346

\noindent [199]
    Iizuka T., Lai Y.-J., Akram W., Amelin Y., Schönbächler M.
    Earth Planet. Sci. Lett., 439 (2016), pp. 172-181

\noindent [200]
    Travaglio C., Rauscher T., Heger A., Pignatari M., West C.
    Astrophys. J., 854 (2018), p. 18

\noindent [201]
    Fehr M.A., Rehkämper M., Halliday A.N., Hattendorf B., Günther D.
    Meteorit. Planet. Sci., 44 (2009), pp. 971-984

\noindent [202]
    Bermingham K.R., Mezger K., Desch S.J., Scherer E.E., Horstmann M.
    Geochim. Cosmochim. Acta, 133 (2014), pp. 463-478

\noindent [203]
    Patronis N., Dababneh S., Assimakopoulos P.A., Gallino R., Heil M., Käppeler F., Karamanis D., Koehler P.E., Mengoni A., Plag R.
    Phys. Rev. C, 69 (2) (2004), Article 025803

\noindent [204]
    Bisterzo S., Gallino R., Käppeler F., Wiescher M., Imbriani G., Straniero O., Cristallo S., Görres J., deBoer R.J.
    Mon. Not. R. Astron. Soc., 449 (2015), pp. 506-527

\noindent [205]
    Liu N., Savina M.R., Gallino R., Davis A.M., Bisterzo S., Gyngard F., Käppeler F., Cristallo S., Dauphas N., Pellin M.J., Dillmann I.
    Astrophys. J., 803 (2015), p. 12

\noindent [206]
    Clayton D.D.
    Arnett W.D., Truran J.W. (Eds.), Nucleosynthesis: Challenges and New Developments (1985), p. 65

\noindent [207]
    Huss G.R., Meyer B.S., Srinivasan G., Goswami J.N., Sahijpal S.
    Geochim. Cosmochim. Acta, 73 (2009), pp. 4922-4945

\noindent [208]
    Wang W., Harris M.J., Diehl R., Halloin H., Cordier B., Strong A.W., Kretschmer K., Knödlseder J., Jean P., Lichti G.G., Roques J.P., Schanne S., von Kienlin A., Weidenspointner G., Wunderer C.
    Astron. Astrophys., 469 (2007), pp. 1005-1012

\noindent [209]
    Côté B., Belczynski K., Fryer C.L., Ritter C., Paul A., Wehmeyer B., O’Shea B.W.
    Astrophys. J., 836 (2017), p. 230

\noindent [210]
    Kim C., Perera B.B.P., McLaughlin M.A.
    Mon. Not. R. Astron. Soc., 448 (2015), pp. 928-938

\noindent [211]
    M. A. de Avillez, M.-M. Mac Low, Astrophys. J 2002, 581, 1047–1060.

\noindent [212]    
    Timar J., Elekes Z., Singh B.
    Nucl. Data Sheets, 121 (2014), pp. 143-394

\noindent [213]
    Schatz H., Toenjes R., Pfeiffer B., Beers T.C., Cowan J.J., Hill V., Kratz K.-L.
    Astrophys. J., 579 (2002), pp. 626-638

\noindent [214]
    Roederer I.U., Kratz K.-L., Frebel A., Christlieb N., Pfeiffer B., Cowan J.J., Sneden C.
    Astrophys. J., 698 (2009), pp. 1963-1980

\noindent [215]
    Ren J., Christlieb N., Zhao G.
    Astron. Astrophys., 537 (2012), p. A118

\noindent [216]
    Mashonkina L., Christlieb N., Eriksson K.
    Astron. Astrophys., 569 (2014), p. A43

\noindent [217]
    Pignatari M., Herwig F., Hirschi R., Bennett M., Rockefeller G., Fryer C., Timmes F.X., Ritter C., Heger A., Jones S., Battino U., Dotter A., Trappitsch R., Diehl S., Frischknecht U., Hungerford A., Magkotsios G., Travaglio C., Young P.
    Astrophys. J. Suppl., 225 (2016), p. 24

\noindent [218]
    Kawamura A., Mizuno Y., Minamidani T., Filipović M.D., Staveley-Smith L., Kim S., Mizuno N., Onishi T., Mizuno A., Fukui Y.
    Astrophys. J. Suppl., 184 (2009), pp. 1-17

\noindent [219]
    Meidt S.E., Hughes A., Dobbs C.L., Pety J., Thompson T.A., García-Burillo S., Leroy A.K., Schinnerer E., Colombo D., Querejeta M., Kramer C., Schuster K.F., Dumas G.
    Astrophys. J., 806 (2015), p. 72

\noindent [220]
    Miura R.E., Kohno K., Tosaki T., Espada D., Hwang N., Kuno N., Okumura S.K., Hirota A., Muraoka K., Onodera S., Minamidani T., Komugi S., Nakanishi K., Sawada T., Kaneko H., Kawabe R.
    Astrophys. J., 761 (2012), p. 37

\noindent [220]
    Corbelli E., Braine J., Bandiera R., Brouillet N., Combes F., Druard C., Gratier P., Mata J., Schuster K., Xilouris M., Palla F.
    Astron. Astrophys., 601 (2017), p. A146

\noindent [222]
    Adams F.C.
    Ann. Rev. Astron. Astrophys., 48 (2010), pp. 47-85

\noindent [223]
    Portegies Zwart S., Pelupessy I., van Elteren A., Wijnen T.P.G., Lugaro L.M.
    Astron. Astrophys. (2018)
    accepted, arXiv:1802.04360

\noindent [224]
    Wijnen T.P.G., Pelupessy F.I., Pols O.R., Portegies Zwart S.
    Astron. Astrophys., 604 (2017), p. A88

\noindent [225]
    Kastner J.H., Myers P.C.
    Astrophys. J., 421 (1994), pp. 605-615

\noindent [226]
    Gritschneder M., Lin D.N.C., Murray S.D., Yin Q.-Z., Gong M.-N.
    Astrophys. J., 745 (2012), p. 22

\noindent [227]
    Boss A.P., Keiser S.A.
    Astrophys. J., 788 (2014), p. 20

\noindent [228]
    Pan L., Desch S.J., Scannapieco E., Timmes F.X.
    Astrophys. J., 756 (2012), p. 102

\noindent [229]
    Hester J.J., Desch S.J., Healy K.R., Leshin L.A.
    Science, 304 (2004)

\noindent [230]
    Looney L.W., Tobin J.J., Fields B.D.
    Astrophys. J., 652 (2006), pp. 1755-1762

\noindent [231]
    Adams F.C., Fatuzzo M., Holden L.
    Astrophys. J., 789 (2014), p. 86

\noindent [232]
    Parker R.J., Church R.P., Davies M.B., Meyer M.R.
    Mon. Not. R. Astron. Soc., 437 (2014), pp. 946-958

\noindent [233]
    Wang L., Wheeler J.C.
    Ann. Rev. Astron. Astrophys., 46 (2008), pp. 433-474

\noindent [234]
    Vasileiadis A., Nordlund A., Bizzarro M.
    Astrophys. J., 769 (2013), p. L8

\noindent [235]
    Austin S.M., West C., Heger A.
    Astrophys. J., 839 (2017), p. L9

\noindent [236]
    Sukhbold T., Ertl T., Woosley S.E., Brown J.M., Janka H.-T.
    Astrophys. J., 821 (2016), p. 38

\noindent [237]
    Gaidos E., Krot A.N., Williams J.P., Raymond S.N.
    Astrophys. J., 696 (2009), pp. 1854-1863

\noindent [238]
    Gounelle M., Meynet G.
    Astron. Astrophys., 545 (2012), p. A4

\noindent [239]
    Young E.D.
    Earth Planet. Sci. Lett., 392 (2014), pp. 16-27

\noindent [240]
    Dwarkadas V.V., Dauphas N., Meyer B., Boyajian P., Bojazi M.
    Astrophys. J., 851 (2017), p. 147

\noindent [241]
    Sana H., de Mink S.E., de Koter A., Langer N., Evans C.J., Gieles M., Gosset E., Izzard R.G., Le Bouquin J.-B., Schneider F.R.N.
    Science, 337 (2012), p. 444

\noindent [242]
    Boss A.P., Ipatov S.I., Keiser S.A., Myhill E.A., Vanhala H.A.T.
    Astrophys. J., 686 (2008), p. L119

\noindent [243]
    Boss A.P., Keiser S.A.
    Astrophys. J., 717 (2010), pp. L1-L5

\noindent [244]
    Boss A.P., Keiser S.A., Ipatov S.I., Myhill E.A., Vanhala H.A.T.
    Astrophys. J., 708 (2010), pp. 1268-1280

\noindent [245]
    Boss A.P., Keiser S.A.
    Astrophys. J., 770 (2013), p. 51

\noindent [246]
    Boss A.P., Keiser S.A.
    Astrophys. J., 809 (2015), p. 103

\noindent [247]
    Boss A.P.
    Astrophys. J., 844 (2017), p. 113

\noindent [248]
    Ouellette N., Desch S.J., Hester J.J.
    Astrophys. J., 662 (2007), pp. 1268-1281

\noindent [249]
    Ouellette N., Desch S.J., Hester J.J.
    Astrophys. J., 711 (2010), pp. 597-612

\noindent [250]
    Goodson M.D., Luebbers I., Heitsch F., Frazer C.C.
    Mon. Not. R. Astron. Soc., 462 (2016), pp. 2777-2791

\noindent [251]
    Close J.L., Pittard J.M.
    Mon. Not. R. Astron. Soc., 469 (2017), pp. 1117-1130

\noindent [252]
    Ellinger C.I., Young P.A., Desch S.J.
    Astrophys. J., 725 (2010), pp. 1495-1506

\noindent [253]
    Elmegreen B.G.
    Astrophys. J., 530 (2000), pp. 277-281

\noindent [254]
    Ribas Á., Bouy H., Merín B.
    Astron. Astrophys., 576 (2015), p. A52

\noindent [255]
    Bollard J., Connelly J.N., Bizzarro M.
    Meteorit. Planet. Sci., 50 (2015), pp. 1197-1216

\noindent [256]
    H. Wang, B.P. Weiss, B.G. Downey, J. Wang, Y.K. Chen-Wiegart, J. Wang, C.R. Suavet, R.R. Fu, E.A. Lima, M.E. Zucolotto, Lunar and Planetary Science Conference, vol. 46, 2015, p. 2516.

\noindent [257]
    Gounelle M.
    Astron. Astrophys., 582 (2015), p. A26

\noindent [258]
    Holst J.C., Olsen M.B., Paton C., Nagashima K., Schiller M., Wielandt D., Larsen K.K., Connelly J.N., Jorgensen J.K., Krot A.N., Nordlund A., Bizzarro M.
    Proc. Natl. Acad. Sci. USA, 110 (2013), p. 88198823

\noindent [259]
    Williams J.P., Gaidos E.
    Astrophys. J., 663 (2007), pp. L33-L36

\noindent [260]
    Gounelle M., Meibom A.
    Astrophys. J., 680 (2008), pp. 781-792

\noindent [261]
    Lichtenberg T., Parker R.J., Meyer M.R.
    Mon. Not. R. Astron. Soc., 462 (2016), pp. 3979-3992

\noindent [262]
    Nicholson R.B., Parker R.J.
    Mon. Not. R. Astron. Soc., 464 (2017), pp. 4318-4324

\noindent [263]
    Kuffmeier M., Frostholm Mogensen T., Haugbølle T., Bizzarro M., Nordlund A.
    Astrophys. J., 826 (2016), p. 22

\noindent [264]
    Krot A.N., Makide K., Nagashima K., Huss G.R., Ogliore R.C., Ciesla F.J., Yang L., Hellebrand E., Gaidos E.
    Meteorit. Planet. Sci., 47 (2012), pp. 1948-1979

\noindent [265]
    Gounelle M., Meibom A., Hennebelle P., Inutsuka S.-i.
    Astrophys. J., 694 (2009), pp. L1-L5

\noindent [266]
    Parker R.J., Dale J.E.
    Mon. Not. R. Astron. Soc., 456 (2016), pp. 1066-1072

\noindent [267]
    Cleeves L.I., Adams F.C., Bergin E.A., Visser R.
    Astrophys. J., 777 (2013), p. 28

\noindent [268]
    Lacki B.C.
    Mon. Not. R. Astron. Soc., 440 (2014), pp. 3738-3748

\noindent [269]
    Johansen A., Okuzumi S.
    Astron. Astrophys., 609 (2018), p. A31

\noindent [270]
    Ruedas T.
    Geochem. Geophys. Geosyst., 18 (2017), p. 3530

\noindent [271]
    Morota T., Haruyama J., Ohtake M., Matsunaga T., Honda C., Yokota Y., Kimura J., Ogawa Y., Hirata N., Demura H., Iwasaki A., Sugihara T., Saiki K., Nakamura R., Kobayashi S., Ishihara Y., Takeda H., Hiesinger H.
    Earth Planet. Sci. Lett., 302 (2011), pp. 255-266

\noindent [272]
    Helling C., Woitke P., Rimmer P.B., Kamp I., Thi W.-F., Meijerink R.
    Life, 4 (2014)

\noindent [273]
    Fortney J.J.
    Astrophys. J., 747 (2012), p. L27

\noindent [274]
    Bond J.C., O’Brien D.P., Lauretta D.S.
    Astrophys. J., 715 (2010), pp. 1050-1070

\noindent [275]
    Ali-Dib M., Mousis O., Pekmezci G.S., Lunine J.I., Madhusudhan N., Petit J.-M.
    Astron. Astrophys., 561 (2014), p. A60

\noindent [276]
    Murillo N.M., Lai S.-P., Bruderer S., Harsono D., van Dishoeck E.F.
    Astron. Astrophys., 560 (2013), p. A103

\noindent [277]
    Harsono D., Jørgensen J.K., van Dishoeck E.F., Hogerheijde M.R., Bruderer S., Persson M.V., Mottram J.C.
    Astron. Astrophys., 562 (2014), p. A77

\noindent [278]
    Apai D.A., Lauretta D.S.
    Protoplanetary Dust: Astrophysical and Cosmochemical Perspectives (2010), pp. 128-160

\noindent [279]
    Greenwood R.C., Franchi I.A., Jambon A., Buchanan P.C.
    Nature, 435 (2005), pp. 916-918

\noindent [280]
    Kleine T., Mezger K., Palme H., Scherer E., Münker C.
    Geochim. Cosmochim. Acta, 69 (2005), pp. 5805-5818

\noindent [281]
    Lichtenberg T., Golabek G.J., Gerya T.V., Meyer M.R.
    Icarus, 274 (2016), pp. 350-365

\noindent [282]
    Monteux J., Golabek G.J., Rubie D.C., Tobie G., Young E.D.
    Space Sci. Rev., 214 (2017), Article 39

\noindent [283]
    Beuther H., Klessen R.S., Dullemond C.P., Henning T.
    Protostars and Planets VI (2014)

\noindent [284]
    Moskovitz N., Gaidos E.
    Meteorit. Planet. Sci., 46 (2011), pp. 903-918

\noindent [285]
    Metzler K., Bischoff A., Stoeffler D.
    Geochim. Cosmochim. Acta, 56 (1992), pp. 2873-2897

\noindent [286]
    Brearley A.
    Meteorites and the Early Solar System II (eds. D. Lauretta, H.Y. McSween Jr.) (2016), pp. 587-624

\noindent [287]
    Elkins-Tanton L.T., Weiss B.P. (Eds.), Planetesimals: Early Differentiation and Consequences for Planets (Cambridge Planetary Science) (2017)

\noindent [288]
    Clayton R.N., Mayeda T.K.
    Earth Planet. Sci. Lett., 67 (1984), pp. 151-161

\noindent [289]
    Zolensky M.E., Bourcier W.L., Gooding J.L.
    Icarus, 78 (1989), pp. 411-425

\noindent [290]
    Osinski G.R., Kring D.
    Large Meteorite Impacts and Planetary Evolution
    (2005)

\noindent [291]
    Hazen R.M., Sverjensky D.
    Cold Spring Harb Perspect. Biol., 2 (2010), p. a00216

\noindent [292]
    H. Hartman, B. Fegley, R.G. Prinn, J.S. Lewis, Lunar and Planetary Science Conference, vol. 14, 1983. pp. 279–280.

\noindent [293]
    Martins Z.
    Elements, 7 (2011), pp. 35-40

\noindent [294]
    Góbi S., Kereszturi Á.
    Mon. Not. R. Astron. Soc., 466 (2017), pp. 2099-2110

\noindent [295]
    Ciesla F.J., Mulders G.D., Pascucci I., Apai D.
    Astrophys. J., 804 (2015), p. 9

\noindent [296]
    Johansen A., Mac Low M.-M., Lacerda P., Bizzarro M.
    Sci. Adv., 1 (2015), Article 1500109

\noindent [297]
    Morbidelli A., Chambers J., Lunine J.I., Petit J.M., Robert F., Valsecchi G.B., Cyr K.E.
    Meteorit. Planet. Sci., 35 (2000), pp. 1309-1320

\noindent [298]
    Alibert Y.
    Astron. Astrophys., 561 (2014), p. A41

\noindent [299]
    Abbot D.S., Cowan N.B., Ciesla F.J.
    Astrophys. J., 756 (2012), p. 178

\noindent [300]
    Jura M., Xu S., Young E.D.
    Astrophys. J., 775 (2013), p. L41

\noindent [301]
    Jura M., Young E.D.
    Annu. Rev. Earth Planet. Sci., 42 (2014), pp. 45-67

\noindent [302]
    Gilmour J.D., Middleton C.A.
    Icarus, 201 (2009), pp. 821-823

\noindent [303]
    De Angelis A., Tatischeff V., Tavani M., Oberlack U., Grenier I., Hanlon L., Walter R., Argan A., von Ballmoos P., Bulgarelli A., Donnarumma I., Hernanz M., Kuvvetli I., Pearce M., Zdziarski A., Aboudan A., Ajello M., Ambrosi G., Bernard D., Bernardini E., Bonvicini V., Brogna A., Branchesi M., Budtz-Jorgensen C., Bykov A., Campana R., Cardillo M., Coppi P., De Martino D., Diehl R., Doro M., Fioretti V., Funk S., Ghisellini G., Grove E., Hamadache C., Hartmann D.H., Hayashida M., Isern J., Kanbach G., Kiener J., Knödlseder J., Labanti C., Laurent P., Limousin O., Longo F., Mannheim K., Marisaldi M., Martinez M., Mazziotta M.N., McEnery J., Mereghetti S., Minervini G., Moiseev A., Morselli A., Nakazawa K., Orleanski P., Paredes J.M., Patricelli B., Peyré J., Piano G., Pohl M., Ramarijaona H., Rando R., Reichardt I., Roncadelli M., Silva R., Tavecchio F., Thompson D.J., Turolla R., Ulyanov A., Vacchi A., Wu X., Zoglauer A.
    Exp. Astron., 44 (2017), pp. 25-82

\noindent [304]
    Greiner J., Mannheim K., Aharonian F., Ajello M., Balasz L.G., Barbiellini G., Bellazzini R., Bishop S., Bisnovatij-Kogan G.S., Boggs S., Bykov A., DiCocco G., Diehl R., Elsässer D., Foley S., Fransson C., Gehrels N., Hanlon L., Hartmann D., Hermsen W., Hillebrandt W., Hudec R., Iyudin A., Jose J., Kadler M., Kanbach G., Klamra W., Kiener J., Klose S., Kreykenbohm I., Kuiper L.M., Kylafis N., Labanti C., Langanke K., Langer N., Larsson S., Leibundgut B., Laux U., Longo F., Maeda K., Marcinkowski R., Marisaldi M., McBreen B., McBreen S., Meszaros A., Nomoto K., Pearce M., Peer A., Pian E., Prantzos N., Raffelt G., Reimer O., Rhode W., Ryde F., Schmidt C., Silk J., Shustov B.M., Strong A., Tanvir N., Thielemann F.-K., Tibolla O., Tierney D., Trümper J., Varshalovich D.A., Wilms J., Wrochna G., Zdziarski A., Zoglauer A.
    Exp. Astron., 34 (2012), pp. 551-582




\end{document}